\pdfoutput=1 %

\documentclass[11pt,letterpaper]{scrartcl}

\makeatletter
\DeclareOldFontCommand{\rm}{\normalfont\rmfamily}{\mathrm}
\DeclareOldFontCommand{\sf}{\normalfont\sffamily}{\mathsf}
\DeclareOldFontCommand{\tt}{\normalfont\ttfamily}{\mathtt}
\DeclareOldFontCommand{\bf}{\normalfont\bfseries}{\mathbf}
\DeclareOldFontCommand{\it}{\normalfont\itshape}{\mathit}
\DeclareOldFontCommand{\sl}{\normalfont\slshape}{\@nomath\sl}
\DeclareOldFontCommand{\sc}{\normalfont\scshape}{\@nomath\sc}
\makeatother

\usepackage[top=2.4cm, bottom=2.4cm, left=3.2cm, right=3cm,footskip=0.8cm]{geometry}

\usepackage[T1]{fontenc}
\usepackage[utf8]{inputenc}
\usepackage{abstract}

\usepackage[hidelinks]{hyperref}
\usepackage[parfill]{parskip}
\usepackage{slashed}
\usepackage{microtype}
\usepackage{csquotes}
\usepackage{graphicx}
\usepackage{amsmath}
\usepackage{mathtools}
\usepackage{amssymb}
\usepackage{braket}
\usepackage{array}
\usepackage{caption}
\usepackage{subcaption}
\usepackage{verbatim}
\usepackage{xcolor}

\usepackage[numbers,sort&compress]{natbib}

\usepackage[automark]{scrlayer-scrpage}
\clearpairofpagestyles
\ohead{\leftmark\ifstr{\rightmark}{\leftmark}{}{ -- \rightmark}}
\ofoot*{\centering\pagemark}
\usepackage{cleveref}

\usepackage[auth-sc,affil-it]{authblk}

\usepackage{scalefnt}
\newcommand{\abbrev}{\scalefont{.9}}
\newcommand{\NNLO}{\text{\abbrev NNLO}}
\newcommand{\NNNLO}{\text{\abbrev N$^3$LO}}

\newcommand{\IR}{\text{\abbrev IR}}
\newcommand{\SU}{\text{\abbrev SU}}

\newcommand{\UV}{\text{\abbrev UV}}
\newcommand{\QCD}{\text{\abbrev QCD}}

\newcommand{\PDF}{\text{\abbrev PDF}}

\newcommand{\IBP}{\text{\abbrev IBP}}
\newcommand{\LaMET}{\text{\abbrev LaMET}}
\newcommand{\qgraf}{\text{\abbrev QGRAF}}
\newcommand{\DGLAP}{\text{\abbrev DGLAP}}

\newcommand{\GPD}{\text{\abbrev GPD}}
\newcommand{\MSBAR}{\text{\abbrev $\overline{\text{MS}}$}}
\newcommand{\LmuP}{\ensuremath{L_\mu^\mathcal{P}}}
\newcommand{\LmuQ}{\ensuremath{L_\mu^\mathcal{Q}}}

\newcommand{\diffmaster}{\ensuremath{I^\eta(1,0,0,1)}}
\newcommand{\diffmastertwo}{\ensuremath{I^\eta(0,1,1,0)}}

\newcommand{\zt}{\tilde{z}}
\newcommand{\alphas}{\ensuremath{\alpha_\text{s}}}

\newcounter{notecount}

\makeatletter
\renewcommand\maketitle{
	\begin{center}
		{\huge\bfseries\@title\par\vspace{0.3em}}
		{\scshape\@author, \@date}
	\end{center}
}
\makeatother

\newpairofpagestyles{firstpage}{
	\ohead*{\normalfont SMU-PHY-26-05}
	\ofoot*{\centering\pagemark}
}

\begin{document}
	\thispagestyle{firstpage}
	\title{\Large {Unpolarized quasi- and pseudo-distributions at one loop: \\gluon correlator decomposition, matching, and the region $|x|>1$
	}}
	
	\author[1,2]{Christopher Monahan}
	\author[1,4]{Tobias Neumann}
	
	\affil[1]{Department of Physics, William \& Mary, Williamsburg, Virginia, USA}
	\affil[2]{Department of Physics, Colorado College, Colorado Springs, Colorado, USA}
	\affil[4]{Department of Physics, Southern Methodist University, Texas, USA}
	
	\date{}
	\maketitle
	
	\vspace{0.5cm}

	\begin{onecolabstract}
		\vspace{0.5cm}
		Lattice \QCD{} determinations of parton distribution functions (\PDF{}s) match matrix elements of space-like separated fields onto light-cone \PDF{}s. For gluons this matching depends on how the two field-strength tensors are contracted, and existing one-loop calculations each fix one index combination. We compute the one-loop matrix elements of the unpolarized quark and gluon quasi- and pseudo-distributions in the \MSBAR{} scheme and their matching to the quark and gluon \PDF{}s. Decomposing the gluon correlator into six covariant form factors, we derive the $6\times6$ renormalization mixing matrix. The eigenvectors of this matrix classify the multiplicatively renormalizable combinations, and any lattice operator choice can be obtained from our results without a further loop calculation.

		Evaluating all Fourier transforms in general dimension $d$ before expanding in $\epsilon$, we resolve the structure of the exterior region $\lvert x\rvert>1$ and reconcile reported discrepancies between coordinate- and momentum-space kernels. By the Paley--Wiener theorem, the analytic part of the one-loop equal-time correlator transforms into $\lvert x\rvert\leq1$, so the exterior region is generated by the branch cut of the short-distance scale logarithm at vanishing Ioffe time. At leading power, the exterior quasi-distribution carries no independent non-perturbative physics, but is fixed pointwise by the light-cone distribution through the renormalization group. At large $\lvert x\rvert$, the gluon-in-gluon quasi-distribution has a power-law tail with one of three coefficients, fixed by the operator's eigenvalue sector.

		In the gluon-in-gluon channel, we find that the coordinate-space matching kernel of Balitsky, Morris and Radyushkin differs from our \MSBAR{} calculation by a finite polynomial, identical in each of their index combinations. We trace this difference to a single step of their own appendix, where a loop-generated index contracted between the two field strengths is summed over two transverse directions instead of all $d-2$. The omitted evanescent operator generates exactly the missing finite polynomial when it multiplies the collinear pole. Using our form-factor basis, we find agreement with the independent calculation of Yao, Ji and Zhang, which employs a different gluon operator, and we show that the same polynomial separates their result from that of Balitsky, Morris and Radyushkin.

		In current unpolarized gluon pseudo-distribution extractions, which implement the ratio scheme, the normalization at zero Ioffe time protects the total momentum fraction $\langle x\rangle_g$ from this correction. We estimate that, at $\alphas=0.3$, this finite difference suppresses the extracted normalized second and third gluon moments by approximately $11\%$ and $13\%$ relative to the \MSBAR{} values. This corresponds to a systematically harder \MSBAR{} gluon distribution than previously reported.

		\vspace{0.5cm}
	\end{onecolabstract}
	
	\newpage
	\tableofcontents

	\pagestyle{scrheadings}

	\newpage
	\section{Introduction}\label{sec:intro}
	
	Understanding the internal structure of protons is a challenge that brings together particle 
	and nuclear physics. Protons are one of the basic building blocks of nuclei and 
	a vital experimental probe of phenomena at the energy frontier. Quantum chromodynamics (\QCD{}) 
	captures the dynamics of the strong nuclear force that binds quarks and gluons together to form 
	protons, but it has only recently become possible to determine the internal structure of 
	protons directly from \QCD{}. 
	
	The three-dimensional structure of protons is described by generalized parton distributions
	(\GPD{}s) and transverse momentum dependent distributions, which characterize the correlations 
	between the longitudinal momentum of the internal constituents and their transverse position 
	or momentum, respectively~\cite{Diehl:2003ny}. Parton distribution functions (\PDF{}s) capture the longitudinal
	momentum structure of protons, and can be interpreted at leading order as the probability of 
	finding a specific quark or gluon with a particular fraction of the momentum of the proton. 
	\PDF{}s are important theoretical inputs for many experimental analyses at the Large Hadron 
	Collider~\cite{Gao:2017yyd,PDF4LHCWorkingGroup:2022cjn}.
	
	Both \GPD{}s and \PDF{}s are defined through matrix elements of fields at light-like 
	separations~\cite{Collins:1981uw,Muller:1994ses,Ji:1998pc,Radyushkin:1996nd,Radyushkin:1997ki}. This definition precludes direct calculation via lattice \QCD{}, the formulation 
	of 
	\QCD{} on a Euclidean hypercubic lattice. To circumvent this difficulty, \PDF{}s can be extracted from matrix
	elements of fields with space-like separations, which can be determined using Euclidean lattice 
	\QCD{}~\cite{Briceno:2017cpo}. Large-momentum effective theory (\LaMET{})
	\cite{Ji:2013fga,Ji:2013dva,Ji:2014gla,Ji:2020ect} and short-distance factorization 
	\cite{Radyushkin:2016hsy,Radyushkin:2017cyf,Ma:2014jla,Ma:2017pxb,Izubuchi:2018srq} provide 
	complementary approaches to extracting \PDF{}s from lattice calculations of spatially-extended 
	operators.
	
	 In the context of large-momentum effective theory, quasi-distributions~\cite{Ji:2013dva}, which were inspired by the original 
	formulation of \PDF{}s in the infinite momentum frame, can be related to light-cone \PDF{}s 
	through a factorization formula with power corrections that vanish in the infinite proton 
	momentum limit. Pseudo-distributions~\cite{Radyushkin:2016hsy,Radyushkin:2017cyf}, on the other 
	hand, are formulated within short-distance factorization and can be related to light-cone 
	\PDF{}s via an alternative 
	factorization relation. Both quasi- and pseudo-distributions are examples of the more general 
	class of factorizable matrix elements, or ``good lattice cross-sections'' 
	\cite{Ma:2014jla,Ma:2017pxb}, that can be analyzed in the spirit of the global fitting paradigm 
	used to extract \PDF{}s from experimental cross-sections. Indeed, the results of lattice 
	calculations can be treated as ``data'' to be combined with experimental data in
	global analyses 
	\cite{Bringewatt:2020ixn,DelDebbio:2020rgv,JeffersonLabAngularMomentumJAM:2022aix,Karpie:2023nyg}.
	
	The space-like extended
	operators relevant to quasi- and pseudo-\PDF{}s and \GPD{}s have both
    logarithmic \UV{} divergences and power \UV{} divergences. In refs.~\cite{Ishikawa:2017faj,Ji:2017oey,Green:2017xeu} the
	renormalizability of extended quark operators was studied in detail to all orders in 
	perturbation theory. These extended quark operators generate a linear power divergence that must be removed nonperturbatively and a variety of methods have been proposed to tackle this problem~\cite{Monahan:2016bvm,Chen:2016fxx,Orginos:2017kos,Ishikawa:2017faj,Chen:2017mzz,Alexandrou:2017huk,Ji:2020brr,LatticePartonCollaborationLPC:2021xdx,Constantinou:2022aij}.

	Explicit perturbative calculations have also been performed at one-loop~\cite{Xiong:2013bka,Constantinou:2017sej,Izubuchi:2018srq,Chou:2022drv} and two-loop~\cite{Ji:2015jwa,Li:2020xml,Chen:2020ody,Chen:2020iqi}. Axial gauge removes the need to calculate Feynman diagrams containing gauge fields from the Wilson line, but the gauge propagators are considerably more complicated than in $R_\xi$ gauges. Coulomb gauge has also been proposed~\cite{Gao:2023lny} as a method to remove the linear divergence and enable the straightforward use of off-axis momenta, which otherwise generate corner divergences~\cite{Musch:2010ka}.
			
	Gluon distributions were first studied at one-loop (in a cutoff scheme) in 
	ref.~\cite{Wang:2017qyg}. It was found that linear divergences exist even in diagrams without 
	the Wilson line, but the cutoff scheme breaks gauge invariance, which complicates the mixing 
	pattern of the operators under renormalization. This effect was studied further in 
	ref.~\cite{Wang:2017eel} using the auxiliary field 
	formalism~\cite{Ji:2017oey,Green:2017xeu}.
	In principle, the spatially-extended gluon operator encodes 36 different operators that could 
	mix 
	under renormalization.
	However, using the auxiliary field formalism, multiplicative renormalization was demonstrated 
	in ref.~\cite{Zhang:2018diq} for certain combinations of operators (Lorentz contractions). In 
	ref.~\cite{Li:2018tpe} this was extended to all possible operators, based on a diagrammatic 
	analysis.
		
	The one-loop gluon correlator matching to both quark and gluon distributions has been calculated in
	ref.~\cite{Wang:2019tgg} in a momentum subtraction scheme and using the auxiliary field 
	formalism. The unpolarized and polarized gluon distribution matching were  calculated at one-loop in the 
	\MSBAR{}-scheme in refs.~\cite{Balitsky:2019krf} and \cite{Balitsky:2021cwr}, respectively, with particular attention paid to the choice of Lorentz indices to minimize 
	contamination of structures that do not directly contribute to the matching with the \PDF{}. 
	In the auxiliary field formalism, methods from heavy-quark effective theory
	can be applied, e.g. as demonstrated to compute three-loop vacuum results~\cite{Braun:2020ymy}.

	Ref.~\cite{Yao:2022vtp} computed the complete set of one-loop matching kernels for spacelike correlators onto light-cone distributions, $C_{qq}$, $C_{qg}$, $C_{gq}$ and $C_{gg}$, taking \GPD{}s as the primary object. The kernels are given in coordinate, pseudo and momentum space in the \MSBAR{} scheme, together with their conversion to the ratio and hybrid schemes, and \PDF{}s and distribution amplitudes are recovered as limits, with those kernels written out in coordinate and momentum space only. The one-loop matching for gluon quasi \GPD{}s was obtained independently in ref.~\cite{Ma:2022gty} using a momentum-space tensor decomposition. Both calculations are carried out in Feynman gauge. The two results initially appeared to differ in the off-diagonal gluon-in-quark channel $C_{gq}$. This channel is subtle because the off-diagonal matching kernel is singular at vanishing operator separation, a consequence of the mass-dimension mismatch between the gluon and quark operators. Ref.~\cite{Ji:2025cbb} has since shown that, once this singularity is consistently treated in dimensional regularization, the two calculations agree at the level of physical (convolution) kernels.
	
    Nonperturbative studies of unpolarized gluon distributions appear in refs.~\cite{Fan:2018dxu,Fan:2020cpa,HadStruc:2021wmh,Fan:2022kcb,Delmar:2023agv} for the nucleon and in refs.~\cite{Fan:2021bcr,Good:2023ecp} for mesons. The first analyses of the polarized gluon distribution appear in refs.~\cite{HadStruc:2022yaw,Khan:2022vot}. Mixing occurs at the factorization stage and, for the gluon distributions, contributions from 
	the iso-scalar quark matrix elements must also be included for a controlled extraction of the 
	gluon \PDF{}s from lattice \QCD{}. The computation of iso-scalar quark matrix elements requires 
	the determination of computationally expensive disconnected contributions and in many lattice 
	calculations these effects have been 
	neglected~\cite{Fan:2018dxu,Fan:2020cpa,Fan:2021bcr,HadStruc:2021wmh,HadStruc:2022yaw}. The 
	first determination of the unpolarized gluon distribution that incorporated this mixing 
	appeared in ref.~\cite{Delmar:2023agv}.
	
	In this paper we present a one-loop determination of the gluon quasi- and pseudo-distribution matrix elements and their matching to the flavor-singlet quark and gluon \PDF{}s. We organize the results using the covariant six-form-factor decomposition of the gluon correlator, $\{\mathcal{M}_{pp},\mathcal{M}_{zz},\mathcal{M}_{zp},\mathcal{M}_{pz},\mathcal{M}_{ppzz},\mathcal{M}_{gg}\}$, introduced in ref.~\cite{Balitsky:2019krf}, and derive the corresponding mixing matrix under renormalization. Any extended gluon operator can therefore be obtained from our results without further loop calculations. The matrix elements are computed in both a covariant ($R_\xi$) and an axial gauge, and their agreement provides a stringent test of the gauge invariance of our result. The amplitudes are generated in an automated pipeline from \qgraf{}~\cite{Nogueira:1991ex} to master integrals. Taken together, this consolidates the forward (\PDF{}) limit of the recent complete-kernel results of ref.~\cite{Yao:2022vtp} and provides an independent, multi-gauge cross-check. In the pseudo-distribution sector, where that reference presents kernels only at the level of \GPD{}s, we supply the explicit forward-limit gluon expressions.

	We obtain the matching relation of ref.~\cite{Balitsky:2019krf} by keeping only the leading-twist projection of the one-loop correction. In addition, we give explicit expressions for the five auxiliary form factors, $\mathcal{M}_{zz}$, $\mathcal{M}_{zp}$, $\mathcal{M}_{pz}$, $\mathcal{M}_{ppzz}$, and $\mathcal{M}_{gg}$, in \cref{sec:matrixelems}. We additionally reproduce the \MSBAR{} gluon results of refs.~\cite{Balitsky:2019krf,Balitsky:2021qsr} for all three index combinations they consider, including their gluon splitting kernel, the ultraviolet coefficients, plus distributions, and endpoint contact terms, and the auxiliary form factors $\mathcal{M}_{zp}$, $\mathcal{M}_{pz}$ and $\mathcal{M}_{ppzz}$. Our matching kernels differ from theirs by a single finite polynomial in the coefficient of $\mathcal{M}_{pp}$, common to the three combinations. We discuss this discrepancy in more detail around \cref{eq:bmr_residual} and note that the independent coordinate-space gluon kernel of ref.~\cite{Yao:2022vtp} agrees with our result, for both of the gluon operator projections given in that reference. The appendix of ref.~\cite{Balitsky:2021qsr} presents operator-level results that show the source of the discrepancy. The quoted matching kernels are obtained from the operator-level results by contracting a loop-generated index between the two field strength tensors over two instead of $d-2$ transverse directions. Restricting this sum leaves out the contributions from an evanescent operator that multiplies the collinear pole and generates exactly the finite discrepancy.

	We also recover the quark calculations of refs.~\cite{Izubuchi:2018srq,Chou:2022drv}. Moreover, since we perform all Fourier transforms in general $d$, our gluon-in-quark results implement by construction the regularization prescription of ref.~\cite{Ji:2025cbb}, and we verify exact agreement of our results with the quasi kernel for the transverse-trace gluon operator of ref.~\cite{Ji:2025cbb} (\cref{sec:matrixelems}).

	We give a careful treatment of the $x\to\pm\infty$ boundary terms that arise in the matching procedure, consolidating the quark-sector analyses of refs.~\cite{Izubuchi:2018srq,Chou:2022drv} and extending them to the gluon. In the convolution these terms evaluate the light-cone distribution at vanishing parton momentum. For quarks the boundaries at $+\infty$ and $-\infty$ contribute with equal weight and pick out the quark and the antiquark distribution respectively, so the sea quark contribution cancels between them and only the valence distribution remains, with its finite net quark number. For gluons no such cancellation exists, but the gluon correlator carries one additional power of the momentum fraction, so the boundary terms involve the momentum density $y f_g(y)$ instead of $f_g(y)$, and, in the gluon-in-quark channel, the quark momentum density $y f_q(y)$; their integrals, the gluon and quark momentum fractions, are bounded by the momentum sum rule. Thus, in both quark and gluon channels a sum rule, and not a fitted small-$x$ exponent, controls the boundary terms. Moreover, independently of the sum rules or the small-$x$ behavior, the boundary terms probe only partons of vanishing momentum, outside the reach of the perturbative matching.

	Beyond the boundary points, the exterior region $\lvert x\rvert>1$ is fixed pointwise at leading power by the scale evolution of the light-cone distribution and the running of the lattice operator (\cref{sec:symmetries}). At fixed $P^z$, the equal-time correlator is analytic in the Ioffe time $\nu$ up to powers of the short-distance scale logarithm $\log(\zt^2\mu^2) = \log(\nu^2\mu^2/(P^z)^2)$. By the Paley--Wiener theorem, all analytic contributions transform into distributions supported within $\lvert x\rvert\leq1$, so that the exterior support at $\lvert x\rvert>1$ is generated entirely by the branch point of the logarithm at $\nu=0$, whose coefficients are fixed by the renormalization group; at one loop the coefficient is the scale derivative of the correlator at fixed Ioffe time. Consequently, at leading power the exterior quasi-distribution has no non-perturbative content of its own and follows from the light-cone distribution through the matching relation. For the reconstruction of quasi-distributions from lattice data~\cite{Chen:2025cxr,Dutrieux:2025jed,Dutrieux:2025axb,Xiong:2025obq} this means that the exterior of the reconstructed distribution has to agree with the matching relation up to power corrections, so the exterior can be used to test the reconstruction.
	
	In \cref{sec:definitions} we introduce the theoretical framework for quasi- and pseudo-distributions and formulate the covariant six-form-factor decomposition of the gluon correlator. In \cref{sec:renormalization} we analyze the ultraviolet renormalization of the bilocal gluon operators, deriving the $6\times6$ form-factor mixing matrix and classifying its three eigenvalue sectors. In \cref{sec:matching} we establish the matching framework, explain the short-distance origin of the exterior region $\lvert x\rvert>1$ and what it implies for quasi-distributions obtained from lattice data, and show that sum rules control the boundary terms at $x=\pm\infty$. We collect the preliminaries of our calculation in \cref{sec:preliminaries}, including our multi-gauge setup, tensor projection operators, and the reduction to three master integrals. In \cref{sec:matrixelems} we present our one-loop results for matrix elements and matching kernels across all four partonic channels, detail our comparisons with existing calculations in the literature, and evaluate the numerical impact of the matching kernel correction on extracted gluon moments. We conclude in \cref{sec:conclusion} and provide technical details in the appendices: Feynman rules and gauge-fixing conventions in \cref{app:feynman}, master Fourier transforms in general dimension in \cref{app:FT}, and the two-loop vacuum calculation of the gluon renormalization constants in \cref{app:vacuum}.
	
	\section{Theoretical Framework of Quasi- and Pseudo-distributions\label{sec:definitions}}
	The starting point for the extraction of distributions from lattice \QCD{} are the bare 
	matrix elements
	\begin{equation}
		\label{eq:Qmatelem}
		\mathcal{Q}^{\text{bare}}_\Gamma = \braket{H(P) \mid \overline{\Psi}(n) \tau_f \Gamma \,W(n,0)
			\Psi(0) \mid 	H(P)\,}\,,
	\end{equation}
	and
	\begin{equation}
		\label{eq:matelem}
		\mathcal{G}^{\text{bare}}_{\mu\nu\rho\sigma} = \braket{H(P) \mid G_{\mu\nu}(n) W(n,0)
			G_{\rho\sigma}(0) \mid 	H(P)\,}\,,
	\end{equation}
	for the quark and gluon distributions of hadron $H$, respectively. Here
	the nucleon momentum is $P=(P^0,0,0,P^z)$, $\Gamma$ represents a suitably-chosen gamma matrix that depends on the distribution of interest, and $\tau_f$ the quark $SU(2)$-flavor projector, chosen to isolate the iso-vector or iso-scalar quark contributions. In this work we focus on unpolarized quark and unpolarized gluon distributions. The Wilson line, $W(n,0)$, lies in direction $n^\mu$ and is given by
	\begin{equation}
    W(n,0) = \mathcal{P}\exp\left( i\eta_sg_s \int_0^n \mathrm{d}\lambda\, n\cdot A^c(\lambda)\,
		T^c  
		\right)\,.
	\end{equation}
	For quark matrix elements, the Wilson line is in the fundamental representation of $SU(3)$-color, while for the gluon matrix element, the Wilson line is in the adjoint representation. The factor $\eta_s=\pm1$ allows for a sign choice 
	in the gauge transformation that drops out in physical results \cite{Romao:2012pq}.\footnote{In 
	practice one can switch from one choice to the other by the replacement $g\to-g$.}  The Wilson-line direction is usually taken to be $n^\mu=z^\mu=(0,0,0,\zt)$ in
	lattice calculations, with $\zt>0$ the scalar separation as distinct from the four-vector $z$, whereas light-cone
	\PDF{}s are defined with $n^2 =0$.
	In the following we use the Minkowski mostly-minus metric, 
	$g^{\mu\nu}=\mathrm{diag}(1,-1,-1,-1)$, and have $z^2=-\zt^2$, whereas with  Euclidean 
	metric, we have $z_E^2=\zt^2$.

	\subsection{Light-cone parton distributions}
    The bare light-cone \PDF{}s are defined via the Fourier transform with respect to $\xi^-$ of 
    the bare matrix element with fields at light-like separations, where
    $\xi^\pm = (t\pm \tilde{z})/\sqrt{2}$. The quark and gluon matrix elements read
	\begin{equation}
	\label{eq:QmatelemLC}
	\mathcal{Q}_{\Gamma,\text{LC}}^{\text{bare}}(\xi^-) = \braket{H(P) \mid \overline{\Psi}(\xi^-)\tau_f\Gamma
		W(\xi^-,0) \Psi(0) \mid H(P)\,}\,,
    \end{equation}
    and
	\begin{equation}
	\label{eq:matelemLC}
	\mathcal{G}_{\mu\nu\rho\sigma,\text{LC}}^{\text{bare}}(\xi^-) = \braket{H(P) \mid G_{\mu
	\nu}(\xi^-)
		W(\xi^-,0) G_{\rho\sigma}(0) \mid H(P)\,}\,,
    \end{equation}
    with
    \begin{equation}
	W(\xi^-,0) = \mathcal{P}\exp\left( i\eta_sg_s \int_0^{\xi^-} \mathrm{d}\lambda^- \, A^+(\lambda^-)\,
	\right)\,.
    \end{equation}
    The bare unpolarized \PDF{}s depend only on the parton momentum fraction $x$, which enters through the 
	Fourier	transform as the conjugate of the light-cone separation:
    \begin{align}
	f_{q/H}^\text{bare}(x) & = \int_{-\infty}^{\infty}\frac{\mathrm{d}{\xi^-}}{4\pi} e^{-i 	\xi^- x P^+}\, \mathcal{Q}_{\gamma^+,\text{LC}}^{\text{bare}}(\xi^-)\,\label{eq:quarkPDF} \\
	x f_{g/H}^\text{bare}(x) & = \int_{-\infty}^{\infty}\frac{\mathrm{d}{\xi^-}}{2\pi P^+} e^{-i 	\xi^- x P^+}
	g^{\nu\sigma} \mathcal{G}^{\text{bare}}_{+\nu\sigma+,\text{LC}}(\xi^-)\,. \label{eq:gluonPDF}
	\end{align}	
	The choice of normalization in this definition is convention dependent and varies in the literature. We follow the conventions in
	refs.~\cite{Collins:1981uw,CTEQ:1993hwr,Collins:2011zzd}. Throughout, the first subscript labels the observed parton and the second the target, hadronic or partonic: $f_{q/g}$, for example, is the quark distribution in a gluon.

	The bare \PDF{}s are renormalized, for example in the \MSBAR{} scheme, through a Mellin convolution
	\begin{gather}
		f_{j/H}(x,\mu^2) = (Z^{\MSBAR}(\mu^2) \otimes f_{j/H}^\text{bare} )(x)\,,
	\end{gather}
	where
	\begin{equation}
		(f\otimes g)(x) = \int_0^1 \mathrm{d}\alpha \int_0^1 \mathrm{d} \beta\, f(\alpha) g(\beta)
		\delta(\alpha\beta - x)
		= \int_x^1 \frac{\mathrm{d}\beta}{\beta}\, f(x/\beta) g(\beta)\,.
		\label{eq:mellinconv}
	\end{equation}
	
	\subsection{Equal-time correlators for lattice \QCD{}}
	
	Quasi- and pseudo-\PDF{}s are derived from equal-time spatial correlators of quark and gluon fields, which are the matrix elements that Euclidean lattice \QCD{} can compute:
	\begin{equation}
	\label{eq:QmatelemET}
	\mathcal{Q}_{\Gamma,\text{ET}}^{\text{bare}}(\zt,P^z) = \braket{H(P) \mid \overline{\Psi}(z)\tau_f\Gamma
		W(z,0) \Psi(0) \mid H(P)\,}\,,
    \end{equation}
    and
	\begin{equation}
	\label{eq:matelemET}
	\mathcal{G}_{\mu\nu\rho\sigma,\text{ET}}^{\text{bare}}(\zt,P^z) = \braket{H(P) \mid G_{\mu\nu}(z)
		W(z,0) G_{\rho\sigma}(0) \mid H(P)\,}\,,
    \end{equation}
    with
    \begin{equation}
	W(z,0) = \mathcal{P}\exp\left( ig\eta_s \int_0^{\zt} \mathrm{d}\lambda \, A^z(\lambda)\,
	\right)\,.
    \end{equation}
	
	Both quasi- and pseudo-\PDF{}s are derived from the same underlying correlator, via Fourier transforms of the spatial correlator $\mathcal Q(\zt,P^z)$ or $\mathcal{G}(\zt,P^z)$. The two distributions are therefore equivalent leading-power Fourier representations and can be
	obtained from each other through double (inverse) Fourier transforms
	\cite{Izubuchi:2018srq}.

	Quasi-\PDF{}s are obtained from a Fourier transform with respect to the spatial separation $z$ at fixed hadron momentum $P^z$. The conjugate variable, the quasi-\PDF{} momentum fraction $x$, has support on the entire real line $x\in(-\infty,+\infty)$.
	Pseudo-\PDF{}s are obtained from a Fourier transform with respect to the Ioffe time $\nu=\zt P^z=-P\cdot z$~\cite{Ioffe:1969kf,Braun:1994jq} at fixed spatial separation $z^2$. The pseudo-\PDF{} is matched to the light-cone \PDF{} via a short-distance operator product expansion \cite{Izubuchi:2018srq}, so the perturbative matching kernel inherits the strict compact support $-1 \leq \alpha \leq 1$ of the light-cone \PDF{}. This bounded support is a consequence of the $\alpha$-parameter representation of the corresponding Feynman diagrams, where the momentum fraction is strictly bounded by the positivity of the propagator polynomials \cite{Radyushkin:2016hsy,Radyushkin:2017cyf,Radyushkin:2017lvu}. Consequently, the pseudo-\PDF{} itself possesses strictly physical support $x \in [-1, 1]$. In our calculation, we explicitly confirm this compact support at the one-loop level.

	\subsection{Quasi-\PDF{}s}
	The unpolarized quasi-\PDF{}s are given by
	\begin{align}
    \tilde{f}^\text{quasi,bare}_{q/H}(x,P^z) & = \int_{-\infty}^{\infty}
		\frac{\mathrm{d}\zt}{4\pi} e^{i\zt x P^z} \mathcal{Q}_{\gamma^0,\text{ET}}^{\text{bare}}(\zt,P^z)\,,\\
    x \tilde{f}^\text{quasi,bare}_{g/H}(x,P^z) & = -\int_{-\infty}^{\infty}
		\frac{\mathrm{d}\zt}{2\pi P^z} e^{i\zt x P^z}\mathcal{G}_{g,\text{ET}}^{\text{bare}}(\zt,P^z)\,.\label{eq:fgquasi}
	\end{align}
	The gluon quasi-\PDF{} can be defined through different operators, corresponding to different choices of Lorentz indices, provided those operators are in the correct universality class. The minus sign has been introduced so that the leading-twist contribution in any index combination comes with the correct sign, leading to a proper normalization of the tree-level result.

	Typically it is advantageous to choose operators that multiplicatively renormalize \cite{Zhang:2018diq,Li:2018tpe}, such as
	\begin{equation}
    \mathcal{G}^{\text{bare}}_{g,\text{ET}}(\zt,P^z) = \sum_{i = 1}^{2}\left[\mathcal{G}_{i00i,\text{ET}}^{\text{bare}}(\zt,P^z) + \sum_{j=1}^{2}\mathcal{G}_{ijji,\text{ET}}^{\text{bare}}(\zt,P^z)\right]\,,
	\end{equation}
    although this choice is not unique. The normalization also varies in the literature (compare, for example, refs.~\cite{Ji:2013dva}, \cite{Ji:2020ect}, and \cite{Wang:2017qyg}). Here we follow \cite{Wang:2017qyg}.

    \paragraph{Matching.} A matching relation to the light-cone \PDF{} (factorization theorem) is given in a large momentum expansion in the limit $P^z\to \infty$. Assuming a nucleon mass
	$M$, the leading-power factorization theorem is~\cite{Ji:2013dva,Xiong:2013bka,Izubuchi:2018srq,Ji:2020ect}
	\begin{align}
		\tilde f_{i/H}^\text{quasi}(x,P^z,\mu^2)  = & {} \int_{-1}^{1} \frac{\mathrm{d}y}{|y|}
		\sum_jC_{ij}\left(\frac{x}{y},\frac{\mu}{|x|P^z}\right) 
		f_{j/H}(y,\mu^2)  \nonumber \\
		{} & \qquad + \mathcal{O} \left( \frac{M^2}{(P^z)^2}, \frac{\Lambda_\text{QCD}^2}{x^2(P^z)^2}, 
		\frac{\Lambda_\text{QCD}^2}{(1-x)^2(P^z)^2}  \right)\,,
		\label{eq:quasimatching}
	\end{align}
	where the indices $i$ and $j$ run over parton species. Throughout this work $x$ denotes the momentum
	fraction of the distribution under discussion, quasi, pseudo or light-cone, while $y$ denotes the
	light-cone momentum fraction integrated over in a matching convolution. Where we compare directly
	with a reference we retain that reference's own variable for the integrated fraction, so that its
	kernels can be quoted unchanged: $u$ for refs.~\cite{Balitsky:2019krf,Balitsky:2021qsr} and $\alpha$
	for refs.~\cite{Radyushkin:2017cyf,Izubuchi:2018srq,Yao:2022vtp}.
	The $M^2/(P^z)^2$ corrections include target-mass corrections, which multiply the same twist-2 matrix elements as the leading-power term and are therefore fixed by the light-cone \PDF{} itself. For the quark quasi-\PDF{} they have been resummed to all orders in $M^2/(P^z)^2$ at tree level~\cite{Chen:2016utp} and derived to $\mathcal{O}(M^2/(P^z)^2)$ within the light-ray operator product expansion~\cite{Braun:2018brg}. We are not aware of a corresponding result for the gluon operators. The $x$-dependence of the other two power-correction terms follows the counting of ref.~\cite{Ji:2020ect}, in which the corrections are suppressed by the parton momentum $xP^z$ and by the spectator momentum $(1-x)P^z$. For the quark quasi-\PDF{}, the renormalon analysis of ref.~\cite{Braun:2018brg} finds the same enhancement $1/x^2$ at small $x$ and a single power $1/(1-x)$ at large $x$, that is, corrections of order $\Lambda_\text{QCD}^2/(x^2(1-x)(P^z)^2)$. 

The matching coefficients $C_{ij}$ depend on the parton species and on the
	renormalization scheme. For the gluon quasi-\PDF{}, the coefficients also depend on the combination of the Lorentz indices $(\mu,\nu,\rho,\sigma)$. Here the kernel $C_{ij}$ is defined in the form in which the factorization theorem is proven. In practical \LaMET{} extractions the light-cone \PDF{} is instead computed from the lattice quasi-\PDF{} through the inverse relation, whose kernel follows from $C_{ij}$ order by order in $\alphas$~\cite{Chen:2025cxr}. At one loop the inverse kernel is the tree-level kernel minus the one-loop term of $C_{ij}$, so that our one-loop kernels determine both forms.

	The matching calculation is performed with parton states of momentum $p^z \neq P^z$.
	In 	ref.~\cite{Izubuchi:2018srq} it was shown that $p^z = |y| P^z$ should be taken in the 
	matching coefficient. We follow ref.~\cite{Su:2022fiu} and express the coefficient as a 
	function of the momentum of the quasi parton $|x|P^z$. In fact, a renormalon analysis suggests 
	that the natural scale associated with the quasi-\PDF{} is $2xP^z$~\cite{Su:2022fiu}. Earlier studies of power corrections through renormalon analyses appear in \cite{Braun:2018brg,Liu:2020rqi}.
	
	\subsection{Pseudo-\PDF{}s}
	Pseudo-\PDF{}s are given by:
	\begin{align}
    \tilde{f}^\text{pseudo,bare}_{q/H}(x,z^2) & {} = \int_{-\infty}^{\infty}
		\frac{\mathrm{d}\nu}{2\pi} e^{ix\nu} \mathcal{M}_{pp}^{q,\text{bare}}(\nu,z^2)\,,\label{eq:fqpseudo}\\
    x \tilde{f}^\text{pseudo,bare}_{g/H}(x,z^2) & {} = {-}2\int_{-\infty}^{\infty}
		\frac{\mathrm{d}\nu}{2\pi} e^{ix\nu}\mathcal{M}_{pp}^{g,\text{bare}}(\nu,z^2)\,.\label{eq:fgpseudo}
	\end{align}
	Here the quark form factor or pseudo Ioffe-time distribution, $\mathcal{M}_{pp}^{q,\text{bare}}(\nu,z^2)$, can be extracted from the decomposition
	\begin{equation}\label{eq:quarkFF}
	\mathcal{Q}_{\gamma^\alpha,\text{ET}}^{\text{bare}}(\zt,P^z) = \eta_{pp}P^\alpha  \mathcal{M}_{pp}^{q,\text{bare}}(\nu,z^2) + z^\alpha \mathcal{M}_z^{q,\text{bare}}(\nu,z^2)\,.
	\end{equation}
    Here $\eta_{pp} = \frac{1}{2}\operatorname{Tr}\mathbf{1}$ is the spin-averaged tree-level normalization. The four-dimensional convention $\operatorname{Tr}\mathbf{1} = 4$ gives $\eta_{pp} = 2$, while continuing the Dirac trace as $\operatorname{Tr}\mathbf{1} = d$ gives $\eta_{pp} = d/2$. In \cref{sec:matrixelems} we keep the factor $d/4 = 1-\epsilon/2$ explicit and unexpanded in the bare quark distributions; in the matching relation this factor cancels whenever the target parton is a quark, and setting $d/4\to1$ recovers the standard four-dimensional \MSBAR{} normalization.
	The gluon pseudo Ioffe-time distribution, $\mathcal{M}_{pp}^{g,\text{bare}}(\nu,z^2)$, can be obtained from
	various combinations of $\mathcal{G}^{\mu\nu\rho\sigma}$~\cite{Balitsky:2019krf}, as we discuss in more detail in the next section. 
	
	In this paper we generalize the pseudo-\PDF{} definition for gluons to take into account the auxiliary form factors (\cref{sec:formfactors}) that enter for arbitrary operator index combinations. The generalization is similar to the definition of quasi-\PDF{}s:
	\begin{align}
		x \tilde{f}^\text{pseudo,bare}_{g/H}(x,z^2) & {} = {-}\int_{-\infty}^{\infty}
		\frac{\mathrm{d}\nu}{2\pi (P^z)^2} e^{ix\nu} \mathcal{G}_{g,\text{ET}}^{\text{bare}}(\zt,P^z)\,,\label{eq:fgpseudogen}
	\end{align}
	where the division by $(P^z)^2$ ensures consistency with \cref{eq:fgpseudo} for the leading-twist combination of \cref{eq:Gcombination_Mpp}. %
	
	\paragraph{Matching.} Pseudo-\PDF{}s have a matching relation to light-cone \PDF{}s in the short-distance expansion limit $\zt \to 0$.
	The factorization theorem in this case is~\cite{Radyushkin:2017cyf,Radyushkin:2017lvu}
	\begin{align}\label{eq:pseudomatching}
    \tilde f_{i/H}^\text{pseudo} (x,z^2,\mu^2)   = & {}
		\int_{-1}^{1} \frac{\mathrm{d}y}{|y|}
		\sum_jC_{ij}\left(\frac{x}{y},z^2,\mu^2\right) f_{j/H}(y,\mu^2) +
		\mathcal{O} \left( z^2\Lambda_\text{QCD}^2, z^2 M^2 \right)\,.
	\end{align}
	This factorization theorem has been proven via an operator product expansion in
	ref.~\cite{Izubuchi:2018srq}.

	In many cases the \PDF{} is extracted directly from the pseudo Ioffe-time distribution through a kernel relation~\cite{Balitsky:2019krf,Radyushkin:2019mye}
	\begin{align}
    \mathcal{M}_{pp}^i(\nu,z^2) = & {} \int_{-1}^1 \mathrm{d}y\,\sum_jR_{ij}(y\nu,z^2)f_{j/H}(y,\mu^2) +
		\mathcal{O} \left( z^2\Lambda_\text{QCD}^2, z^2 M^2 \right)\,.
	\end{align}
	The kernels $R_{ij}$ can be obtained from a combination of \cref{eq:fqpseudo,eq:fgpseudo,eq:pseudomatching} and are given explicitly in ref.~\cite{Balitsky:2019krf}. Ref.~\cite{Balitsky:2021qsr} provides the equivalent matching relations in Ioffe-time convolution form.

	In lattice calculations, the equal-time correlator is typically normalized by its zero-momentum value, defining the reduced pseudo-distribution in the ratio scheme~\cite{Radyushkin:2017cyf,Orginos:2017kos}, $\mathcal{M}(\nu, z^2)/\mathcal{M}(0, z^2)$. This ratio nonperturbatively eliminates the ultraviolet divergences of the Wilson line and sets the correlator to unity at $\nu = 0$, normalizing the target distribution by its lowest moment: the valence quark number for quarks and the momentum fraction $\langle x \rangle_g$ for gluons. In perturbation theory, where ultraviolet divergences are subtracted in \MSBAR{}, the ratio-scheme matching kernel is the \MSBAR{} kernel divided by the $\nu=0$ Wilson coefficient, and at one loop the division becomes a finite subtraction (\cref{sec:matrixelems}).

	This formulation avoids the need to Fourier transform the pseudo Ioffe-time distribution directly. Lattice calculations provide a discrete sampling of the pseudo Ioffe-time distribution, with statistical and systematic uncertainties, over a limited range of Ioffe times. Reconstructing the Fourier transform of these noisy, discrete data is an ill-posed inverse problem, which can be circumvented by parametrizing the gluon \PDF{} and analyzing the lattice results in the spirit of the global fitting paradigm~\cite{Ma:2014jla,Ma:2017pxb,Cichy:2019ebf,Bringewatt:2020ixn,DelDebbio:2020cbz,DelDebbio:2020rgv,JeffersonLabAngularMomentumJAM:2022aix,Karpie:2023nyg}, or through other reconstruction approaches~\cite{Karpie:2019eiq,Liang:2019frk,Alexandrou:2020tqq,Karpie:2021pap,Khan:2022vot,Chowdhury:2024ymm,Candido:2024hjt,Dutrieux:2024rem,Dutrieux:2025jed}.

\subsection{Gluon correlator scalar form factors}\label{sec:formfactors}
To work with scalar expressions we use the tensor decomposition into spin-averaged form factors
	of refs.~\cite{Radyushkin:2017cyf} and~\cite{Balitsky:2019krf}. These decompositions are built from the four-vectors $p^\mu$ and $z^\mu$ and, for the gluon case, from the
	metric tensor $g^{\mu \nu}$. For quarks, the decomposition is given in \cref{eq:quarkFF}. For gluons, the tensor decomposition incorporates the antisymmetry of $G^{\mu\nu}$ with
	respect to its indices and can be written as:
	\begin{align}
		\mathcal{G}^{\mu\nu\rho\sigma,\text{bare}}(z,p) & {} = \left( g^{\mu\rho} p^\nu p^\sigma - g^{\mu\sigma} p^\nu p^\rho -
		g^{\nu\rho} 
		p^\mu p^\sigma + g^{\nu\sigma} p^\mu p^\rho \right) 
		\mathcal{M}_{pp}^{g,\text{bare}}(\nu,z^2)  \nonumber \\
		& \enspace + \left( g^{\mu\rho} z^\nu z^\sigma - g^{\mu\sigma} z^\nu z^\rho - 
		g^{\nu\rho} 
		z^\mu z^\sigma + g^{\nu\sigma} z^\mu z^\rho \right) \mathcal{M}_{zz}^{g,\text{bare}}(\nu,z^2)
		\nonumber 
		\\
		& \enspace + \left( g^{\mu\rho} z^\nu p^\sigma - g^{\mu\sigma} z^\nu p^\rho - 
		g^{\nu\rho} z^\mu p^\sigma + g^{\nu\sigma} z^\mu p^\rho \right) 
		\mathcal{M}_{zp}^{g,\text{bare}}(\nu,z^2)
		\nonumber \\
		& \enspace + \left( g^{\mu\rho} p^\nu z^\sigma - g^{\mu\sigma} p^\nu z^\rho - 
		g^{\nu\rho} p^\mu z^\sigma + g^{\nu\sigma} p^\mu z^\rho \right) \mathcal{M}_{pz}^{g,\text{bare}}(\nu,z^2)
		\nonumber \\
		& \enspace + \left( p^\mu z^\nu  - p^\nu z^\mu\right) \left(  p^\rho z^\sigma -  
		p^\sigma 
		z^\rho\right) \mathcal{M}_{ppzz}^{g,\text{bare}}(\nu,z^2)
		\nonumber \\ 
		& \enspace + \left(g^{\mu\rho} g^{\nu\sigma} -g^{\mu\sigma} g^{\nu\rho} 
		\right)\mathcal{M}_{gg}^{g,\text{bare}}(\nu,z^2)    \,.
		\label{eq:Manb}
	\end{align}
	
	The scalar form factors ${\cal M}^g$ are functions of $z^2$ and the Ioffe time.
    By contracting this decomposition with its tensor coefficients we construct a set of projection
	operators onto the form factors ${\cal M}^g$. Here, and throughout this work, we use ``twist'' to refer to operators and form factors and we reserve the term ``power'' for the expansion in $1/P^z$ (see, for example, \cref{sec:matching}). In this sense $\mathcal{M}_{pp}$ is the leading-twist form factor, the only one with a twist-2 contribution at tree level.

\paragraph{Tilde form factors.}
	The six scalar form factors have different mass dimensions due to the kinematic prefactors $p^\mu$, $z^\mu$, $g^{\mu\nu}$ in the tensor structures. A specific index combination of $\mathcal{G}^{\mu\nu\rho\sigma}$ therefore produces a linear combination of form factors with dimensionful kinematic prefactors. %
	To present results in a basis-independent way, we define the ``tilde form factors'':
	\begin{equation}\label{eq:tildeformfactors}
		\begin{aligned}
		\mathcal{\widetilde M}_{pp}^{g,\text{bare}}(\nu,z^2) &= (p^z)^2 \mathcal{M}_{pp}^{g,\text{bare}}(\nu,z^2)\,, &
		\mathcal{\widetilde M}_{zz}^{g,\text{bare}}(\nu,z^2) &= \zt^2 \mathcal{M}_{zz}^{g,\text{bare}}(\nu,z^2)\,, \\
		\mathcal{\widetilde M}_{zp}^{g,\text{bare}}(\nu,z^2) & = \nu \mathcal{M}_{zp}^{g,\text{bare}}(\nu,z^2)\,, &
		\mathcal{\widetilde M}_{pz}^{g,\text{bare}}(\nu,z^2) &= \nu \mathcal{M}_{pz}^{g,\text{bare}}(\nu,z^2)\,, \\
		\mathcal{\widetilde M}_{ppzz}^{g,\text{bare}}(\nu,z^2) &= \nu^2 \mathcal{M}_{ppzz}^{g,\text{bare}}(\nu,z^2)\,, &
		\mathcal{\widetilde M}_{gg}^{g,\text{bare}}(\nu,z^2) &= \mathcal{M}_{gg}^{g,\text{bare}}(\nu,z^2)\,.
		\end{aligned}
	\end{equation}
	With these definitions, any index combination of $\mathcal{G}^{\mu\nu\rho\sigma}$ is a sum of tilde form factors with numerical coefficients, independent of kinematics. The tilde form factors are also the natural basis for \UV{} renormalization, because the mixing matrix in this basis is dimensionless (see \cref{sec:gluon_operator_renorm}).

	The advantage of this tensor decomposition is that a single form factor, $-2\mathcal{M}_{pp}$ (or $-2\mathcal{\widetilde M}_{pp}$), carries the tree-level contribution for the matching onto light-cone distributions. The remaining five form factors, $\mathcal{M}_{zz}$, $\mathcal{M}_{zp}$, $\mathcal{M}_{pz}$, $\mathcal{M}_{ppzz}$, and $\mathcal{M}_{gg}$, do not enter the definition of the light-cone gluon \PDF{}, and we refer to them throughout this work as the \emph{auxiliary form factors}. They have no tree-level matrix element in partonic states (\cref{sec:tree}), but at one loop they contribute to the matching kernel of any operator that does not isolate $\mathcal{M}_{pp}$, and are needed to reconstruct arbitrary operator choices $\mathcal{G}^{\mu\nu\rho\sigma}$ on the lattice. They mix with the leading-twist form factor under ultraviolet renormalization. On the lattice the leading-twist contribution is therefore obtained by considering a combination of operators~\cite{Balitsky:2019krf}
	\begin{gather}
		\mathcal{G}_{\text{ET}}^{0ii0,\text{bare}}(\zt,p^z) + \mathcal{G}_{\text{ET}}^{jiij,\text{bare}}(\zt,p^z) = 2(p^0)^2 \mathcal{M}_{pp}^{g,\text{bare}}(\nu,z^2)\,,
		\label{eq:Gcombination_Mpp}
	\end{gather}
	where the indices $i$ and $j$ satisfy $i,j\in\{1,2\}$. A second choice, with a minimal contamination~\cite{Balitsky:2021qsr}, is
	\begin{gather}
		\mathcal{G}_{\text{ET}}^{3ii3,\text{bare}}(\zt,p^z) + 2\mathcal{G}_{\text{ET}}^{3003,\text{bare}}(\zt,p^z) = 2
		(p^0)^2\mathcal{M}_{pp}^{g,\text{bare}}(\nu,z^2) - 2
		(p^0)^2\zt^2\,\mathcal{M}_{ppzz}^{g,\text{bare}}(\nu,z^2)\,.
		\label{eq:Gcombination_ziiz}
	\end{gather}
	With the transverse sums over the two lattice directions, both relations hold exactly in $d$ dimensions. If the sums are instead continued to the $d-2$ transverse directions of dimensional regularization, the coefficient $2$ of $\mathcal{M}_{pp}^{g,\text{bare}}$ in \cref{eq:Gcombination_Mpp} becomes $d-2$, and both combinations acquire $\mathcal{O}(\epsilon)$ admixtures of auxiliary form factors, for example $(d-2)(4-d)\,\mathcal{M}_{gg}^{g,\text{bare}}$ in \cref{eq:Gcombination_Mpp}. We show in \cref{sec:tree} that this choice does not change the gluon-in-gluon matching.
	These relations follow from the tensor decomposition \eqref{eq:Manb} for general external states with $P=(P^0,0,0,P^z)$: for a hadron of mass $M$ one has $(p^0)^2 = (p^z)^2 + M^2$, while for the massless on-shell partonic states of our calculation $p^0 = p^z$. Here and throughout, the contractions use the mostly-minus convention fixed at the start of this section.
	Both of these combinations are multiplicatively renormalizable
	\cite{Balitsky:2021qsr,Balitsky:2019krf,Zhang:2018diq}.
	In the tilde form factor basis, and for on-shell massless states ($p^0 = p^z$, so that $(p^0)^2\zt^2 = \nu^2$), these combinations read
	\begin{align}
		\mathcal{G}_{\text{ET}}^{0ii0}+\mathcal{G}_{\text{ET}}^{jiij} &= 2\,\mathcal{\widetilde M}_{pp}^{g,\text{bare}}\,,
		\label{eq:Gcombination_Mpp_tilde} \\
		\mathcal{G}_{\text{ET}}^{3ii3}+2\,\mathcal{G}_{\text{ET}}^{3003} &= 2\,\mathcal{\widetilde M}_{pp}^{g,\text{bare}} - 2\,\mathcal{\widetilde M}_{ppzz}^{g,\text{bare}}\,.
		\label{eq:Gcombination_ziiz_tilde}
	\end{align}

	\paragraph{General pseudo-\PDF{} from form factor combinations.}
	For a general operator combination
	\begin{equation}
	\label{eq:general_G}
	\mathcal{G}_{g,\text{ET}}^{\text{bare}} = \sum_k c_k \mathcal{\widetilde M}_k^{g,\text{bare}}\,,
	\end{equation}
with purely numerical coefficients $c_k$, the generalized pseudo-\PDF{} of \cref{eq:fgpseudogen} takes the form
	\begin{equation}\label{eq:pseudo_tilde_expansion}
		x \tilde{f}^\text{pseudo,bare}_{g/H}(x,z^2) = -\sum_k c_k \int_{-\infty}^{\infty}\frac{\mathrm{d}\nu}{2\pi (P^z)^2}\, e^{ix\nu}\, \mathcal{\widetilde M}_k^{g,\text{bare}}(\nu,z^2) \equiv - \sum_k c_k \, x \tilde{f}^\text{pseudo,bare}_{g/H,k}(x,z^2)\,.
	\end{equation}
	Here, we explicitly define the individual partial distributions $\tilde{f}_{g/H,k}(x,z^2)$ via their Fourier transforms, which cleanly separates the form factor integrals from the tensor projection coefficients $c_k$. Exactly analogous linear combinations apply for the quasi-\PDF{} components $\tilde{f}^\text{quasi,bare}_{g/H,k}(x,P^z)$.
	Since $\mathcal{\widetilde M}^{g,\text{bare}}_k = \kappa_k \mathcal{M}^{g,\text{bare}}_k$ (with prefactors $\kappa_k$ from \cref{eq:tildeformfactors}), the effective coefficient multiplying each form factor $\mathcal{M}^{g,\text{bare}}_k$ in the Fourier integrand is $\kappa_k/(P^z)^2$. Using $P^z = \nu/\zt$ at fixed $z^2$, the ratio $\kappa_k/(P^z)^2$ is either a constant
	or cancels kinematic $\nu$-factors present in $\mathcal{M}_k^{g,\text{bare}}$ from the tensor decomposition.
	This ensures that the pseudo-\PDF{} Fourier transform is well-defined for all six form factors, and one can Fourier-transform the individual contributions at fixed $z^2$.

	To obtain the result for a specific choice of operator Lorentz indices 
	$(\mu,\nu,\rho,\sigma)$, one can insert the indices 
	directly and derive Feynman rules for a particular choice. For example ref.~\cite{Wang:2017qyg} 
	used the combination $\mu,\nu,\rho,\sigma=z,i,i,z$, where $i$ is 
	summed over transverse 
	components. The method of projectors provides a different approach, with the
	advantage that once all tensor form factors
	have been obtained, any particular choice of operator Lorentz indices can be derived.
	For example, we find that the combination $\mu,\nu,\rho,\sigma=z,i,i,z$ translates into 
	the 
	following sum of form factors:
	\begin{align}
	\mathcal{G}_{\text{ET}}^{ziiz,\text{bare}}{} & (\zt,p^z) =  %
	{-}2 \mathcal{\widetilde M}_{gg}^{g,\text{bare}}(\nu,z^2) \nonumber \\
	+ {} & 2 \left[
	\mathcal{\widetilde M}_{pz}^{g,\text{bare}}(\nu,z^2)+ \mathcal{\widetilde M}_{zp}^{g,\text{bare}}(\nu,z^2) +
	\mathcal{\widetilde M}_{pp}^{g,\text{bare}}(\nu,z^2) +
	\mathcal{\widetilde M}_{zz}^{g,\text{bare}}(\nu,z^2) \right].\label{eq:Mziiz}
    \end{align}
	Up to an overall sign (\cref{fn:wang_sign}), this is the operator of ref.~\cite{Wang:2017qyg} expressed in our basis, and we use it below to
	place that operator in the eigenvalue classification of \cref{sec:gluon_operator_renorm} and to
	compare matrix elements in \cref{sec:matrixelems}.

	In all index combinations considered here, the leading-twist contribution $\mathcal{M}_{pp}^{g,\text{bare}}$
	appears with 
	a positive sign, while the matching to the \PDF{} is performed with a negative sign, 
	corresponding to the tree-level matrix element $\mathcal{M}_{pp}^{g,\text{bare}} = -\frac{1}{2}(e^{i pz}+e^{-i pz})$, which can be seen in
	\cref{eq:fgpseudo}. This means
	that in the matching to \PDF{}s the negative of the given index combinations have to be used, as evident in our definition of the quasi and pseudo \PDF{}s.\footnote{\label{fn:wang_sign}%
	Note that ref.~\cite{Wang:2017qyg}, with which we compare in \cref{sec:matrixelems}, includes this minus sign in the
	operator, while we include it in the quasi-\PDF{} definition, \cref{eq:fgquasi}. The operator of
	ref.~\cite{Wang:2017qyg} is $\sum_{i=1,2}G^{z}{}_{i}(z)\,W(z,0)\,G^{iz}(0)$, with the transverse index pair contracted
	covariantly. Since $G^{z}{}_{i}=g_{ii}\,G^{zi}=-G^{zi}$ for each transverse direction $i$, its matrix element is
	$-\mathcal{G}^{ziiz}$, whose leading-twist contribution already has the sign that reproduces the \DGLAP{} kernels.
	The two conventions therefore give the same quasi-\PDF{}.}

\section{Renormalization\label{sec:renormalization}}

	The factorization theorems of \cref{eq:quasimatching,eq:pseudomatching} can be applied to distributions renormalized in any scheme. In our one-loop calculation we identify infrared (\IR{}) and ultraviolet (\UV{}) poles in dimensional regularization, $\epsilon_\IR{}=\epsilon_\UV{}\equiv\epsilon$ (\cref{sec:preliminaries}), so the \UV{} pole of the operator and the collinear pole removed by the matching appear in a single $1/\epsilon$. We therefore fix the renormalization constants independently. We extract the renormalization constants for the gluon operator from a two-loop vacuum calculation, in which only \UV{} divergences occur; this calculation reproduces ref.~\cite{Braun:2020ymy} and is described in \cref{app:vacuum}. For the quark operator, we determine the renormalization constant from the known result quoted in ref.~\cite{Braun:2020ymy}. Subtracting these counterterms from the one-loop matrix elements must leave, for $\lvert x\rvert\leq1$, only the collinear poles that cancel in the matching, which we verify explicitly.

	\paragraph{Distinguishing renormalization and mixing.}
	We distinguish three notions of renormalization and mixing in this work, and we note as we go which of the statements below are scheme specific.

	First, there is the \UV{} renormalization of the space-like bilocal operators themselves, which we discuss in this section. In the \MSBAR{} scheme this renormalization is multiplicative and flavor diagonal to all orders, because, at fixed separation $\zt\neq0$, the \UV{} divergences are local to the operator endpoints and to the Wilson line~\cite{Ji:2017oey,Zhang:2018diq,Balitsky:2019krf}. Thus, any putative counterterm that could mix the quark bilocal into the gluon bilocal (or vice versa) would have to change the field content at a single endpoint, which is forbidden by fermion number conservation. In addition, the two operators carry different mass dimensions, so a mixing counterterm would need a dimensionful coefficient. Such a counterterm is prohibited in the \MSBAR{} scheme, for which renormalization constants are pure numbers that are independent of the field separation. There is therefore no quark--gluon mixing under operator renormalization. For the gluon operator, the renormalization operator of \cref{eq:Zhat} is diagonal in the transverse/parallel projector basis with two vacuum renormalization constants, one for each of the two eigenvalue sectors of $\hat{Z}$. The six form factors of \cref{eq:Manb} are not eigenvectors of that operator, so they mix among themselves under renormalization, but this mixing is a change of basis among the form factors, independent of flavor, and fixed entirely by the two vacuum constants (see \cref{sec:gluon_operator_renorm}). Note that flavor diagonality is an \MSBAR{} statement: in momentum-subtraction schemes, such as RI/MOM~\cite{Stewart:2017tvs}, off-diagonal quark--gluon mixing is generically induced and must be subtracted (see ref.~\cite{Ji:2020ect} for a review of scheme choices).

	Second, there is the \UV{} renormalization of the light-cone distributions, which we denote $Z_\text{LC}$.
	In contrast to the space-like operators, the light-like operators do mix in the flavor-singlet sector. Both arguments above assume that the separation is space-like, so that the two operator endpoints are at a finite invariant distance and the renormalization constants are numbers that do not depend on $z$. On the light cone the invariant distance vanishes, and a light-ray operator is instead renormalized by an integral operator along the ray~\cite{Balitsky:1987bk}. Such a kernel is not attached to the endpoints, and it is not restricted by the dimension count either, so that the quark and gluon light-ray operators can mix. In momentum-fraction space this kernel is the \DGLAP{} kernel, so the mixing is \DGLAP{} evolution and all quark--gluon mixing under renormalization enters through $Z_\text{LC}$.

	Third, there is the matching. The matching kernel matrix $C_{ik}$ has flavor off-diagonal entries, and in our partonic quasi- and pseudo-distributions the \DGLAP{} mixing structure appears through collinear \IR{} poles $\propto\mathcal{P}_{ik}$ (the one-loop \DGLAP{} kernels, defined below), which the matching removes. Mixing is possible at this level because a matching coefficient can depend on the field separation $z$. In particular, the off-diagonal kernel $C_{gq}$ contains the power of the separation that compensates the mass-dimension mismatch between the gluon and quark operators, and is thus singular at vanishing $z$ (see the discussion of ref.~\cite{Ji:2025cbb} in \cref{sec:intro}). Quark--gluon mixing therefore enters the gluon distribution beyond tree level through $Z_\text{LC}$ and $C_{ik}$, while the operator renormalization remains flavor diagonal (in the \MSBAR{} scheme).

	The arguments above hold at fixed $\zt\neq0$. The quasi-\PDF{}, however, is a Fourier transform over all separations, so it contains the local limit, where quark and gluon operators do mix. Our $z^\mu=(0,0,0,\zt)$ is purely spacelike, so that $z^\mu\to0$ and $z^2\to0$ are the same limit, which we write as $\zt\to0$. At $\zt=0$ the bilocal operators reduce to local currents, and within the vicinity of this limit, the bilocal operator can be Taylor expanded in $\zt$. This generates towers of local operators, and these do mix under renormalization. However, this mixing does not spoil flavor diagonality at any finite momentum fraction. The Fourier transform maps the short-distance region to large momentum fractions, and as we discuss in \cref{sec:boundary}, this region is mapped to boundary terms supported at $x=\pm\infty$, which probe the light-cone \PDF{}s only at vanishing momentum fraction. Local-current mixing can therefore enter the operator renormalization only through off-diagonal boundary terms at infinity, which drop out of the matching (\cref{sec:boundary}).

	\paragraph{Renormalization preliminaries.}
	We calculate the quasi- and pseudo-distributions with partonic targets $k \in\{q,g,S\}$: non-singlet quarks ($q$) and, in the singlet sector, singlet quarks ($S$) and gluons ($g$). We write the renormalization of the quasi- and pseudo-distributions in the singlet sector schematically as
	\begin{equation}\label{eq:Zmatrix}
	\begin{pmatrix}
		\tilde{f}_{S/H}^{\text{ren}} \\
		\tilde{f}_{g/H}^{\text{ren}}
	\end{pmatrix} =
	\begin{pmatrix}
		Z_{SS} & Z_{Sg} \\
		Z_{gS} & Z_{gg}
	\end{pmatrix}
	\begin{pmatrix}
		\tilde{f}_{S/H}^{\text{bare}} \\
		\tilde{f}_{g/H}^{\text{bare}}
	\end{pmatrix}\,.
	\end{equation}
	For the space-like operators in the \MSBAR{} scheme the off-diagonal entries vanish to all orders, $Z_{Sg} = Z_{gS} = 0$, by the endpoint-locality and dimensional arguments above. The diagonal entry $Z_{gg}$ should be understood as a schematic representation that incorporates the corresponding tensor structure. For example, for a multiplicatively renormalizable combination this component is the corresponding eigenvalue of the form-factor mixing matrix of \cref{sec:gluon_operator_renorm}, while for a general index combination this component is matrix-valued (in the form-factor basis). Our explicit one-loop results in \cref{sec:matrixelems} are consistent with these expectations.
	We emphasize that this one-loop verification does not by itself establish the all-order statement, which rests on the structural arguments given earlier. The matrix form of \cref{eq:Zmatrix} is nevertheless useful, because the structure becomes non-trivial for the light-cone renormalization $Z_\text{LC}$ of \cref{sec:matching} and in schemes other than \MSBAR{}.

	We express perturbative expansions in the \MSBAR{}-renormalized coupling $\alphas$, using
	\begin{equation}\label{eq:g2expansion}g_{s,0}^2 = \frac{4\pi\alphas\, \mu^{2\epsilon}\, e^{\epsilon\gamma_E}}{(4\pi)^\epsilon}\,.
	\end{equation}
	At the one-loop level, for bare and renormalized distributions alike, we write
	\begin{equation}\label{eq:fexpansion}
	f_{j/H} = f^{(0)}_{j/H} + f^{(1)}_{j/H} + \mathcal{O}(\alphas^2)\,,
	\end{equation}
	where $f^{(1)}_{j/H}$ denotes the complete one-loop term, proportional to $\alphas$, which we display with an explicit prefactor $\alphas/(2\pi)$.

	With massless on-shell states the loop integrals of the partonic light-cone \PDF{}s are scaleless in dimensional regularization, so \UV{} and \IR{} poles arise with equal and opposite coefficients. In the \MSBAR{} scheme, after renormalization of the coupling and fields, the partonic light-cone \PDF{} is
	\begin{equation}
	f_{i/j}^{\text{bare}}(x) = \delta(1-x) \delta_{ij} +  \frac{\alphas}{2\pi} \left( \frac{1}{\epsilon_\UV{}} - \frac{1}{\epsilon_\IR{}}\right) \mathcal{P}_{ij}(x)+ \mathcal{O}(\alphas^2)\,,
	\end{equation}
	where the $\mathcal{P}_{ij}$ are the one-loop \DGLAP{} kernels with their color factors included, e.g.~$\mathcal{P}_{gg} = C_A P_{gg}$, and the color-stripped kernels $P_{ij}$ are given in \cref{eq:DGLAP}.
	For $\epsilon_\UV{}=\epsilon_\IR{}\equiv\epsilon$ this expression vanishes beyond tree level, i.e.~the light-cone \PDF{}s receive no perturbative corrections.

	We choose $\Gamma=\gamma^0$ for all quark distributions, $\tilde{f}_{S/i}$ with $i\in\{S,g\}$, leading to matching kernels specific to this choice. The renormalization constants $Z_{Si}$ are independent of the $\Gamma$ choice.
	We decompose the gluon distributions, $\tilde{f}_{g/i}$, into six scalar form factors in \cref{eq:Manb}, so that any field-strength tensor-index combination can be reconstructed directly.

	\subsection{Gluon operator renormalization}\label{sec:gluon_operator_renorm}

	The \UV{} renormalization of the gluon correlator is complicated by operator mixing between form factors.

	\paragraph{Vacuum eigenbasis.} In vacuum, the bilocal gluon operator admits two independent structures~\cite{Braun:2020ymy}:
	\begin{equation}
	g^{\mu\nu}_\perp = g^{\mu\nu} - \frac{n^\mu n^\nu}{n^2}\,,\qquad g^{\mu\nu}_\parallel = \frac{n^\mu n^\nu}{n^2}\,.
	\end{equation}
	The projectors onto antisymmetric rank-2 tensors with definite transverse/parallel content are
	\begin{align}
	\hat{\Pi}^{\perp\perp,\mu\nu}{}_{\alpha\beta} &= g_\perp^{\mu}{}_{[\alpha}\, g_\perp^{\nu}{}_{\beta]}\,,\qquad
	\hat{\Pi}^{\parallel\perp,\mu\nu}{}_{\alpha\beta} = g_\parallel^{\mu}{}_{[\alpha}\, g_\perp^{\nu}{}_{\beta]} + g_\perp^{\mu}{}_{[\alpha}\, g_\parallel^{\nu}{}_{\beta]}\,,
	\end{align}
	where $[\alpha\beta]$ denotes antisymmetrization. %
	These projectors are complete on the index space of $F^{\mu\nu}$,
	\begin{equation}
	\hat{\Pi}^{\perp\perp,\mu\nu}{}_{\alpha\beta} + \hat{\Pi}^{\parallel\perp,\mu\nu}{}_{\alpha\beta} = g^{\mu}{}_{[\alpha}\, g^{\nu}{}_{\beta]}\,,
	\end{equation}
	where the purely parallel component drops out because $n^\mu n_{[\alpha} n^\nu n_{\beta]}$ vanishes by antisymmetry, and each of the two components, which we call the $\perp\perp$ and $\parallel\perp$ sectors, renormalizes multiplicatively. We find it helpful to introduce a renormalization operator that is diagonal in this basis,
	\begin{equation}\label{eq:Zhat}
	\hat{Z}^{\mu\nu}{}_{\alpha\beta} = Z_{\perp\perp}\, \hat{\Pi}^{\perp\perp,\mu\nu}{}_{\alpha\beta} + Z_{\parallel\perp}\, \hat{\Pi}^{\parallel\perp,\mu\nu}{}_{\alpha\beta}\,.
	\end{equation}
	The bilocal operator renormalizes with one factor of $\hat{Z}$ per field strength. Defining
	\begin{equation}
	Z = 1 + \frac{\alphas}{4\pi\epsilon}Z^{(1)}
	\end{equation}
	for every renormalization constant, we determine the two parameters from a two-loop vacuum calculation, in which only \UV{} divergences occur, and which we present in \cref{app:vacuum}. The bare vacuum correlator renormalizes with the square of the parameter of its sector, and its pole in each sector, \cref{eq:ZPi}, gives at one loop
	\begin{equation}\label{eq:sector_values}
	Z_{\perp\perp}^{(1)} = -\frac{5C_A-4T_Rn_f}{6} = C_A - \frac{\beta_0}{2}\,,\qquad
	Z_{\parallel\perp}^{(1)} = -\frac{11C_A-4T_Rn_f}{6} = -\frac{\beta_0}{2}\,,
	\end{equation}
	where $\beta_0 = 11C_A/3 - 4T_Rn_f/3$ is the one-loop coefficient of the $\beta$-function and $n_f$ is the number of massless quark flavors; the $n_f$ terms are the quark loop of the vacuum diagrams. The corresponding anomalous dimensions are $\gamma^{(1)} = -Z^{(1)}$.
	Ref.~\cite{Braun:2020ymy} defines the bilocal operator with an explicit factor $g^2$ and obtains its renormalization parameters to two-loop accuracy from vacuum expectation values computed to three loops. The two definitions differ by the renormalization of that coupling factor, and removing it from \cref{eq:sector_values} reproduces their one-loop values, $C_A$ and $0$, exactly; we give the translation in \cref{app:vacuum}.
	We use the parameters of \cref{eq:sector_values} to renormalize our one-loop non-vacuum matrix elements below, where the cancellation of the \UV{} poles is a strong cross-check of our computational framework. We give the details after introducing the form-factor mixing matrix.

\paragraph{External state eigenbasis.}

To express the renormalization in the form factor basis of \cref{sec:formfactors}, we compute the $6\times 6$ mixing matrix relating renormalized and bare form factors,
\begin{equation}\label{eq:Mij_ren}
	\mathcal{M}_{j}^{g,\text{ren}} = \sum_k (Z_\mathcal{M})_{jk}\, \mathcal{M}_{k}^{g,\text{bare}}\,,\qquad
	\mathcal{\widetilde M}_{j}^{g,\text{ren}} = \sum_k (\widetilde{Z}_\mathcal{M})_{jk}\, \mathcal{\widetilde M}_k^{g,\text{bare}}\,,
\end{equation}
by applying $\hat{Z}\otimes\hat{Z}$, which acts on the left ($\mu\nu$) and right ($\rho\sigma$) index pairs independently, to each tensor structure $\mathcal{O}_k$ from \cref{eq:Manb} and decomposing the result.

In the tilde form factor basis, the mixing matrix takes the form:
\begin{equation}\label{eq:Ztilde_general}
	\begin{pmatrix}
		\mathcal{\widetilde M}_{pp}^{g,\text{ren}}\\ \mathcal{\widetilde M}_{zz}^{g,\text{ren}}\\ \mathcal{\widetilde M}_{zp}^{g,\text{ren}}\\ \mathcal{\widetilde M}_{pz}^{g,\text{ren}}\\ \mathcal{\widetilde M}_{ppzz}^{g,\text{ren}}\\ \mathcal{\widetilde M}_{gg}^{g,\text{ren}}
	\end{pmatrix} =
	\begin{pmatrix}
		Z_{\perp\perp}^2 & 0 & 0 & 0 & 0 & 0 \\
		(\Delta Z)^2 & Z_{\parallel\perp}^2 & Z_{\parallel\perp}\Delta Z & Z_{\parallel\perp}\Delta Z & 0 & Z_{\perp\perp}^2 - Z_{\parallel\perp}^2 \\
		\Delta Z\, Z_{\perp\perp} & 0 & Z_{\parallel\perp}Z_{\perp\perp} & 0 & 0 & 0 \\
		\Delta Z\, Z_{\perp\perp} & 0 & 0 & Z_{\parallel\perp}Z_{\perp\perp} & 0 & 0 \\
		Z_{\perp\perp}^2 - Z_{\parallel\perp}^2 & 0 & 0 & 0 & Z_{\parallel\perp}^2 & 0 \\
		0 & 0 & 0 & 0 & 0 & Z_{\perp\perp}^2
	\end{pmatrix}
	\begin{pmatrix}
		\mathcal{\widetilde M}_{pp}^{g,\text{bare}}\\ \mathcal{\widetilde M}_{zz}^{g,\text{bare}}\\ \mathcal{\widetilde M}_{zp}^{g,\text{bare}}\\ \mathcal{\widetilde M}_{pz}^{g,\text{bare}}\\ \mathcal{\widetilde M}_{ppzz}^{g,\text{bare}}\\ \mathcal{\widetilde M}_{gg}^{g,\text{bare}}
	\end{pmatrix}
	\,,
\end{equation}
where %
$\Delta Z = Z_{\parallel\perp} - Z_{\perp\perp}$.
The eigenvalues of this mixing matrix correspond to the three renormalization parameters $Z_{\perp\perp}^2$, $Z_{\perp\perp}\,Z_{\parallel\perp}$, and $Z_{\parallel\perp}^2$, and we refer to the corresponding eigenspaces as the three eigenvalue sectors.

Each of the three eigenvalues has a two-dimensional left eigenspace, and together the three eigenspaces span the six-dimensional form factor space, as expected from the two-dimensional vacuum decomposition. A linear combination $\sum_k c_k\,\mathcal{\widetilde M}_k$ renormalizes multiplicatively when its coefficient vector $(c_k)$, defined as in \cref{eq:general_G}, is a left eigenvector of $\widetilde{Z}_\mathcal{M}$, and its renormalization constant is then the corresponding eigenvalue. We explicitly verify the classification that follows in \cref{app:vacuum}.
\Cref{eq:Ztilde_general} is written in terms of $Z_{\perp\perp}$ and $Z_{\parallel\perp}$ themselves, so its eigenvectors, and with them this classification, do not depend on the order to which those two constants are computed.

The left eigenspace of $Z_{\perp\perp}^2$ is spanned by $\mathcal{\widetilde M}_{pp}^{g,\text{bare}}$ and $\mathcal{\widetilde M}_{gg}^{g,\text{bare}}$, which therefore renormalize multiplicatively on their own. %
This subspace contains the leading-twist combination %
of \cref{eq:Gcombination_Mpp_tilde}. The eigenspace of $Z_{\perp\perp}Z_{\parallel\perp}$ is spanned by
\begin{equation}\label{eq:eigenspace_perppar}
\mathcal{\widetilde M}_{pp}^{g,\text{bare}}+\mathcal{\widetilde M}_{zp}^{g,\text{bare}}
\qquad \mathrm{and} \qquad \mathcal{\widetilde M}_{pp}^{g,\text{bare}}+\mathcal{\widetilde M}_{pz}^{g,\text{bare}}\,,
\end{equation}
corresponding to $c_{pp} = c_{zp}+c_{pz}$ with $c_{zz} = c_{ppzz} = c_{gg} = 0$. The eigenspace of $Z_{\parallel\perp}^2$ is spanned by
\begin{equation}\label{eq:eigenspace_parpar}
\mathcal{\widetilde M}_{pp}^{g,\text{bare}} - \mathcal{\widetilde M}_{ppzz}^{g,\text{bare}} \qquad \mathrm{and} \qquad \mathcal{\widetilde M}_{zz}^{g,\text{bare}} + \mathcal{\widetilde M}_{zp}^{g,\text{bare}} + \mathcal{\widetilde M}_{pz}^{g,\text{bare}} + \mathcal{\widetilde M}_{ppzz}^{g,\text{bare}} - \mathcal{\widetilde M}_{gg}^{g,\text{bare}}\,,
\end{equation}
 corresponding to $c_{zp} = c_{pz} = c_{zz}$, $c_{gg} = -c_{zz}$ and $c_{pp}+c_{ppzz} = c_{zz}$. This eigenspace contains both the minimal contamination combination of \cref{eq:Gcombination_ziiz_tilde} and the combination of \cref{eq:Mziiz},
 the negative of the operator used in ref.~\cite{Wang:2017qyg}.

The multiplicative renormalizability of the lattice combinations in \cref{eq:Gcombination_Mpp,eq:Gcombination_ziiz}, which was established in refs.~\cite{Balitsky:2021qsr,Balitsky:2019krf,Zhang:2018diq}, follows directly from this eigenvalue analysis. Our analysis provides the complete classification of all multiplicatively-renormalizable combinations, expressed in terms of the form factors $\mathcal{\widetilde M}_k^{g,\text{bare}}$ rather than the operator-index combinations of $G^{\mu\nu\rho\sigma}$, and their associated renormalization parameters.\footnote{We verify this classification through the (left) eigenvector computation of $\widetilde{Z}_\mathcal{M}$.}
In \cref{app:vacuum} we compare our classification with the operator combinations quoted in refs.~\cite{Balitsky:2019krf,Balitsky:2021qsr,Zhang:2018diq,Wang:2017qyg}: every multiplicatively renormalizable combination quoted there lies in one of the three eigenvalue sectors above. We also check a combination that ref.~\cite{Zhang:2018diq} states does not renormalize multiplicatively, and find that it is not a left eigenvector of $\widetilde{Z}_\mathcal{M}$.

The matrix in \cref{eq:Ztilde_general} is the result of applying the renormalization operator, \cref{eq:Zhat}, to the bare form factor matrix and expressed in terms of the vacuum eigenvalues. Inserting the values of \cref{eq:sector_values}, we write the one-loop mixing matrix as
\begin{equation}\label{eq:Z1_tilde}
	(\widetilde{Z}_\mathcal{M}^{(1)})_{ij} =
	\begin{pmatrix}
		2C_A{-}\beta_0 & 0 & 0 & 0 & 0 & 0 \\
		0 & {-}\beta_0 & {-}C_A & {-}C_A & 0 & 2C_A \\
		{-}C_A & 0 & C_A{-}\beta_0 & 0 & 0 & 0 \\
		{-}C_A & 0 & 0 & C_A{-}\beta_0 & 0 & 0 \\
		2C_A & 0 & 0 & 0 & {-}\beta_0 & 0 \\
		0 & 0 & 0 & 0 & 0 & 2C_A{-}\beta_0
	\end{pmatrix}\,.
\end{equation}

Only the diagonal entries depend on $n_f$. The off-diagonal entries involve the difference $Z_{\parallel\perp}^{(1)} - Z_{\perp\perp}^{(1)} = -C_A$, from which the quark loop drops out.

Expanding the renormalization equation, \cref{eq:Mij_ren}, to one-loop order, and noting that only $\mathcal{\widetilde M}_{pp}^g$ is nonzero at tree level, the \UV{} counterterm to be added to each bare one-loop form factor, $\mathcal{\widetilde M}_j^{g,\text{ren}} = \mathcal{\widetilde M}_j^{g,\text{bare}} + \delta\mathcal{\widetilde M}_j^g$ at this order, is determined by the first column of $(\widetilde{Z}_\mathcal{M}^{(1)})_{ij}$:
\begin{equation}\label{eq:counterterm}
	\delta \mathcal{\widetilde M}_j^g = \frac{\alphas}{4\pi\epsilon}\, (\widetilde{Z}_\mathcal{M}^{(1)})_{j,pp}\,\mathcal{\widetilde M}_{pp}^{g,(0)}\,.
\end{equation}
The one-loop counterterms for $\mathcal{\widetilde M}_{zz}$ and $\mathcal{\widetilde M}_{gg}$ vanish, %
while $\mathcal{\widetilde M}_{zp}$, $\mathcal{\widetilde M}_{pz}$, and $\mathcal{\widetilde M}_{ppzz}$ receive nonzero counterterms from the off-diagonal entries.

In the one-loop gluon distributions of \cref{sec:matrixelems} the coefficient of the logarithm of any tree-normalized, multiplicatively renormalizable combination of form factors is
\begin{equation}\label{eq:log_coefficient}
-\frac{\alphas C_A}{2\pi}\left[\frac{\lambda}{2C_A}\,\delta(1-x) + x P_{gg}(x)\right]\,,
\end{equation}
with $\lambda$ the corresponding eigenvalue listed in \cref{eq:Z1_tilde} and $P_{gg}$ the \DGLAP{} kernel given in \cref{eq:DGLAP}. Its integral is the coefficient $\gamma_\mathcal{O}$ that fixes the large-$\lvert x\rvert$ tail of the quasi-\PDF{}. We evaluate it sector by sector in \cref{sec:symmetries}, where we also relate it to the anomalous dimension $-\frac{\alphas}{2\pi}\frac{\lambda}{2}$ of the operator.

We verify the renormalization matrix by adding these counterterms explicitly to the bare one-loop Fourier-transformed form factors. We carry out this cross-check for both the pseudo- and the quasi-\PDF{} cases, without expanding in $\epsilon$. We find that, in the region $|x|\leq 1$, which we denote the ``interior region'', the $1/\epsilon$ contact poles of the auxiliary form factors cancel identically, leaving only the collinear pole of the leading-twist component $\mathcal{\widetilde M}_{pp}$. In the rest of this work, we refer to the region $\lvert x\rvert>1$ as the ``exterior'' region, and reserve ``boundary'' for the endpoints $x=\pm\infty$ themselves.

For the quasi-\PDF{}, whose support extends beyond $|x|=1$, the counterterm, which is nonvanishing at the endpoint contact terms $\delta(1\mp x)$, vanishes at the boundary at $x=\pm\infty$. We discuss these ultraviolet boundary terms, together with the quasi-\PDF{} results, in \cref{sec:matrixelems}.

\paragraph{Renormalization-basis cross-check.}
As a cross-check of the calculation from the projection onward, we repeat the calculation using a second decomposition of the same correlator. We build an orthogonalized basis, $B_i$, from the renormalization projectors of \cref{eq:Zhat} instead of the tensor structures of \cref{sec:formfactors}. The first two elements of this basis, $B_1$ and $B_2$, are the $\perp\perp$ and $\parallel\perp$ structures themselves, and Gram--Schmidt orthogonalization supplies the remaining four, $B_3,\dots,B_6$. We project the corresponding form factors $\mathcal{\widetilde M}_i^{B}$ from the diagrams independently and carry them through to the renormalized pseudo- and quasi-\PDF{} distributions, so that the projection, the reduction of a different set of loop integrals to the master integrals, the Fourier transform and the counterterms of \cref{eq:counterterm} all run a second time, and the agreement of the two calculations tests each of these stages. The form factors $\mathcal{\widetilde M}_i^{B}$ are related to the original form factors $\mathcal{\widetilde M}_j^{g}$ by a transformation matrix $\widetilde{U}_{ij}$ that is a rational function of $d$ alone and therefore commutes with the Fourier transform. We check this relation in both directions, before and after the transform.

In this new basis, the counterterms are related to the original matrix elements via
\begin{equation}
 \delta \mathcal{\widetilde M}_i^{B} = \frac{\alphas}{4\pi\epsilon}\,\big[\sum_j \widetilde{U}_{ij}\,(\widetilde{Z}_\mathcal{M}^{(1)})_{j,pp}\big]\,\mathcal{\widetilde M}_{pp}^{g,(0)}\,.
\end{equation}
These counterterms are no longer defined in the \MSBAR{} scheme, because the $d$ dependence of $\widetilde{U}$ generates additional finite $\delta(1\mp x)$ terms. These finite pieces are the ones required by the closure relation below, so that the renormalization also tests the exact-in-$\epsilon$ ordering of the calculation. We confirm that our renormalization procedure commutes with this change of basis,
\begin{equation}\label{eq:MB_closure}
	\mathcal{\widetilde M}_i^{B,\text{ren}} = \sum_j \widetilde{U}_{ij}\Big|_{d=4-2\epsilon}\, \mathcal{\widetilde M}_j^{g,\text{ren}}\,,
\end{equation}
for all regions of momentum fraction and without expanding in $\epsilon$. In particular, the two combinations without $\mathcal{\widetilde M}_{pp}$ admixture, $\mathcal{\widetilde M}_{5}^{B}$ and $\mathcal{\widetilde M}_{6}^{B}$, are entirely pole-free for the pseudo-\PDF{}, while the remaining four combinations retain the residual collinear pole of the renormalized $\mathcal{\widetilde M}_{pp}$. For the quasi-\PDF{}, the boundary poles at $x=\pm\infty$ that survive renormalization (cf.\ \cref{sec:matrixelems}) satisfy the closure relation \cref{eq:MB_closure} as well, and $\mathcal{\widetilde M}_{5}^{B}$ and $\mathcal{\widetilde M}_{6}^{B}$ carry no other pole.
\section{Matching}
\label{sec:matching}

With this discussion of renormalization in mind, the matching with renormalized light-cone distributions can be written schematically as:
\begin{equation}
	\tilde{f}^\text{bare} = Z^{-1} \otimes C \otimes Z_\text{LC} \otimes f^\text{bare}\,,
\end{equation}
where $Z$ is the renormalization matrix defined in \cref{eq:Zmatrix}, $C$ the matrix with matching coefficients, and $Z_\text{LC}$ the \UV{}-renormalization matrix for the light-cone distributions. Note that any renormalization related to the \QCD{} Lagrangian enters for the quasi-, pseudo- and light-cone \PDF{}s alike and drops out in this matching relation.

Expanding the relation between bare and renormalized distributions to one-loop order, using vanishing tree-level off-diagonal elements in the renormalization and distributions, we find:
\begin{equation}
	\tilde{f}^{(1),\text{bare}}_{i/k} - f^{(1),\text{bare}}_{i/k} = \left[ C_{ik}^{(1)} + Z_{\text{LC},ik}^{(1)} - Z^{(1)}_{ik} \right] \otimes f^{(0),\text{bare}}_{k/k}\,.\label{eq:matching1loop}
\end{equation}
The renormalization constants $Z_{\text{LC},ik}$ are the \MSBAR{} \UV{} poles of the bare light-cone \PDF{}s and are therefore determined by the one-loop \DGLAP{} kernels, $Z_{\text{LC},ik}^{(1)} = - \mathcal{P}_{ik}/\epsilon_{\text{UV}}$ (up to coupling prefactors). We verify by explicit calculation that the off-diagonal elements of the operator renormalization vanish at one loop, $Z_{Sg}^{(1)} = Z_{gS}^{(1)} = 0$, as expected from the flavor-diagonal \MSBAR{} renormalization of the space-like operators (\cref{sec:renormalization}). The left-hand side of \cref{eq:matching1loop} is free of collinear \IR{} divergences: with massless partonic states the collinear \IR{} behavior of the bare light-cone \PDF{}s, proportional to $-\mathcal{P}_{ik}/\epsilon_\IR{}$, coincides with that of the twist-2 component of the quasi- and pseudo-distributions and cancels in the difference. Soft divergences cancel between real and virtual diagrams for each distribution separately, so the matching coefficients $C_{ik}$ are infrared finite. Beyond one loop, the diagonal elements of $Z$ are known through three loops for the quark and two loops for the gluon bilocal~\cite{Braun:2020ymy}, and the matching coefficients of the quark quasi distribution have been computed to \NNLO{}~\cite{Li:2020xml,Chen:2020ody} and \NNNLO{}~\cite{Cheng:2024wyu}.

The gluon tree-level partonic light-cone \PDF{}, $f^{(0)}_{g/g}(y) = \delta(1{-}y) - \delta(1{+}y)$, is odd, so the matching convolution projects onto the odd part of the kernel, $C_{gg}(x) - C_{gg}(-x)$.

\subsection{The region $\lvert x\rvert>1$}
\label{sec:symmetries}

	We perform all Fourier transforms in $d$ dimensions before expanding in $\epsilon$ (\cref{app:FT}), since the $\epsilon$-expansion does not commute with the short-distance limit $\zt\to0$ (\cref{sec:boundary}).

	\paragraph{Short-distance origin of the region $\lvert x\rvert>1$.}
	The quasi-\PDF{} has support at $\lvert x\rvert>1$, where the light-cone \PDF{} vanishes. The origin of this exterior region can be understood through the Paley--Wiener theorem~\cite{Paley:1934,Hormander:1990}: a function of $\nu$ transforms into a distribution with compact support on $\lvert x\rvert\leq1$ if and only if it is analytic in the whole complex $\nu$ plane and grows no faster than $e^{\lvert\operatorname{Im}\nu\rvert}$, up to powers of $\nu$.\footnote{For a different, but related, application of the Paley--Wiener theorem to the extraction of continuous functions from noisy lattice data, see ref.~\cite{Jay:2026qoh}.} The non-zero exterior arises because the equal-time correlator carries a branch point at $\nu=0$ from the short-distance logarithm.

	We spell the mechanism out in a toy example first, and then apply it to the quasi-\PDF{}.
	The mechanism is visible in the simplest function supported on $[0,1]$, the box $w(u)=1$, whose transform
	\begin{equation}
	h(\nu) = \int_0^1\mathrm{d}u\,e^{-iu\nu} = \frac{1-e^{-i\nu}}{i\nu}
	\end{equation}
	is analytic in the whole $\nu$ plane %
	and grows like $e^{\lvert\operatorname{Im}\nu\rvert}$, where the exponent is the length of the support.
	To invert the transform for $x>1$, we shift the integration contour of $\int\mathrm{d}\nu\,e^{i\nu x}h(\nu)$ upward by $iY$: the exponential contributes $e^{-Yx}$, $h$ contributes at most $e^{Y}$, and the product $e^{-Y(x-1)}$ vanishes as $Y\to\infty$ for every $x>1$. Because $h(\nu)$ is an entire function, this contour shift crosses no singularities, so the transform vanishes for $x>1$ (and, by shifting downward, for $x<0$).

	In contrast, consider instead $(\nu^2)^\epsilon h(\nu)$, which is $\lvert\nu\rvert^{2\epsilon}$ on the real axis but has a branch point at $\nu=0$. Since $h(-\nu)=\overline{h(\nu)}$, the transform becomes
	\begin{equation}
	\frac{1}{\pi}\operatorname{Re}\int_0^\infty\mathrm{d}\nu\,e^{i\nu x}(\nu^2)^\epsilon h(\nu).
	\end{equation}
	Rotating the contour onto the positive imaginary axis, $\nu=it$, we obtain $e^{i\nu x}\to e^{-tx}$, $h(it)\to (e^{t}-1)/t$, and $(\nu^2)^\epsilon \to e^{i\pi\epsilon}t^{2\epsilon}$; for $x>1$ this yields
	\begin{equation}\label{eq:toy_exterior}
	\int\frac{\mathrm{d}\nu}{2\pi}\,e^{i\nu x}\,(\nu^2)^\epsilon\,h(\nu) = -\frac{\sin(\pi\epsilon)}{\pi}\,\Gamma(2\epsilon)\left[(x-1)^{-2\epsilon}-x^{-2\epsilon}\right].
	\end{equation}
	Here the prefactor is $\operatorname{Re}(ie^{i\pi\epsilon}) = -\sin(\pi\epsilon)$; in the analytic case ($\epsilon=0$) it reduces to $\operatorname{Re}(i)=0$, which recovers the vanishing of the transform for $x>1$ found above.

	In this toy model, the exterior region is therefore given by the discontinuity of the non-analytic factor at $\nu=0$, weighted by the analytic part continued to the imaginary axis. The non-analytic factor is of order $\epsilon$, so if the analytic part carries a $1/\epsilon$ pole, the exterior stays finite. Dividing \cref{eq:toy_exterior} by $\epsilon$ and taking $\epsilon\to0$, we find
	\begin{equation}
	 -\log\left[\frac{x}{x-1}\right] = -\int_0^1\frac{\mathrm{d}u}{x-u}\,,
	\end{equation}
	which is the Cauchy transform of the box, defined here as
	\begin{equation}
	 -\int\mathrm{d}u\,\frac{w(u)}{x-u}.
	\end{equation}
	The tail of this transform, $-1/x$, has coefficient $\int_0^1\mathrm{d}u\,w(u)=1$.

	For the quasi-\PDF{}, the correlator has this structure at every order in $\alphas$. At fixed $z^2$ it is the transform of the pseudo-distribution, which has support $\lvert x\rvert\leq1$ diagram by diagram (\cref{sec:definitions}), and for massless partons its $z^2$ dependence at order $\alphas^n$ enters only through factors $(\zt^2\mu^2)^{k\epsilon}$ with $k\leq n$, that is, through powers of $\log(\zt^2\mu^2)$. The exterior is therefore generated by these logarithms alone.

	In particular, at one loop there is a single such factor, $(\zt^2\mu^2)^\epsilon$, which generates the exterior alone~\cite{Balitsky:2019krf,Yao:2022vtp}. We write $\mathcal{L}(u)$ for the coefficient of the $1/\epsilon$ pole multiplying this factor. Once we expand the factor, $\mathcal{L}(u)$ is also the coefficient of the short-distance scale logarithm $\log(\zt^2\mu^2)$. Since $\mathcal{L}(u)$ vanishes outside the support of the light-cone distribution ($[0,1]$ for quarks and $[-1,1]$ for gluons, where it is even), the one-loop quasi-\PDF{} in the exterior region is
	\begin{equation}\label{eq:exterior_constraint}
		\tilde{f}^{(1)}(x)\Big|_{\lvert x\rvert>1} = -\int\!\mathrm{d}u\, \frac{\mathcal{L}(u)}{\lvert x-u\rvert}\,.
	\end{equation}
	For $x>1$ this is the Cauchy transform of $\mathcal{L}$, and for $x<-1$ it is the same expression with the opposite overall sign, as stated in ref.~\cite{Balitsky:2019krf}. This fixes the exterior pointwise from the scale logarithm alone, as we have verified for each of our exterior region results in \cref{sec:matrixelems}. Individual terms and diagrams can also carry poles $1/\nu^n$ at $\nu=0$, which transform into tails that approach a constant ($n=1$) or grow as $\lvert x\rvert^{n-1}$ ($n\geq2$) (\cref{app:FT}). These poles cancel in the complete correlator (\cref{sec:quarkinquark}), so that at one loop the tail of a complete distribution falls off at least as fast as $1/\lvert x\rvert$. If we expand $(\nu^2)^\epsilon = 1+\epsilon\log\nu^2+\dots$ before Fourier transforming the logarithm, we obtain the same exterior pointwise, since $\lvert x-u\rvert$ never vanishes for $\lvert x\rvert>1$; the two orders of operations differ only at $x\to\pm\infty$ under an integral, which is the boundary-term question of \cref{sec:boundary}.

	If we expand \cref{eq:exterior_constraint} at large $\lvert x\rvert$, the interval $0\leq u\leq1$ contributes $-\gamma_\mathcal{O}/\lvert x\rvert$, where
	\begin{equation}\label{eq:gammaO}
		\gamma_\mathcal{O} = \int_0^1\!\mathrm{d}u\,\mathcal{L}(u)\,.
	\end{equation}
	We refer to $\gamma_\mathcal{O}$ as the tail coefficient. In both quark and gluon channels, the scale logarithm $\mathcal{L}(u)$ decomposes into the ultraviolet counterterm of the equal-time operator and the \DGLAP{} splitting kernel (\cref{eq:log_coefficient}). For quarks, the non-singlet splitting kernel $\mathcal{P}_{qq}(u)$ integrates to zero by quark-number conservation ($\int_0^1\mathrm{d}u\,\mathcal{P}_{qq}(u) = 0$), so $\gamma_\mathcal{O}$ equals the operator anomalous dimension, $\gamma_{\text{op},q} = \alphas C_F/(2\pi)\cdot 3/2$. For gluons, the momentum-weighted distribution carries $u\mathcal{P}_{gg}(u)$, whose momentum integral does not vanish because gluons mix with quarks under evolution: by the momentum sum rule,
	\begin{equation}
	 \int_0^1\mathrm{d}u\,u\,\mathcal{P}_{gg}(u) = -\frac{2}{3}n_f T_R = -\frac{11}{6}C_A + \frac{\beta_0}{2}\,.
	\end{equation}
	Combining this with the operator counterterm $-\frac{\lambda}{2}\delta(1-u)$ from \cref{eq:log_coefficient} yields
	\begin{equation}
	\gamma_\mathcal{O} = \frac{\alphas}{2\pi}\left[\frac{11C_A}{6} - \frac{\lambda+\beta_0}{2}\right]\,,
	\end{equation}
	where $\lambda$ is the eigenvalue of the mixing matrix for the sector (\cref{sec:gluon_operator_renorm}). Because $\lambda$ contains the $-\beta_0$ of coupling renormalization, the combination $(\lambda+\beta_0)/2$ takes the values $C_A$, $C_A/2$ and $0$ in the three sectors, so that $\mathcal{L}(u)$, and with it $\gamma_\mathcal{O}$ and the one-loop exterior, are strictly independent of the number of fermions. For gluons, $\gamma_\mathcal{O}$ therefore differs from the anomalous dimension $\gamma_{\text{op},g} = -\alphas/(2\pi)\cdot\lambda/2$ of the operator by the momentum integral,
	\begin{equation}
	\gamma_\mathcal{O} - \gamma_{\text{op},g} = -\frac{\alphas}{2\pi}\int_0^1\mathrm{d}u\,u\,\mathcal{P}_{gg}(u) = \frac{\alphas}{2\pi}\frac{2}{3}n_fT_R\,,
	\end{equation}
	which carries the entire $n_f$ dependence of $\gamma_{\text{op},g}$.

	Evaluating \cref{eq:gammaO} for the quark bilocal and for representative gluon operator choices yields
	\begin{equation}\label{eq:gammaO_values}
	\gamma_\mathcal{O} = \frac{\alphas }{2\pi}\left\{
	\begin{array}{cc}
	3C_F/2 & \qquad \mathrm{quark\,bilocal}\\
	5C_A/6 & \qquad \mathcal{M}_{pp}\\
	11C_A/6 & \qquad \mathcal{G}^{ziiz}\\
	\end{array}
	\right.\,.
	\end{equation}
	At large $\lvert x\rvert$, \cref{eq:exterior_constraint} therefore fixes the asymptotic tail of the quasi-distribution. For a partonic quark target, this produces $-\gamma_\mathcal{O}/\lvert x\rvert = -\alphas C_F/(2\pi)\cdot3/(2\lvert x\rvert)$. For gluons, crossing symmetry doubles the contribution because both endpoints $y=\pm1$ contribute, yielding $-2\gamma_\mathcal{O}/\lvert x\rvert$, which evaluates to $-5/(3\lvert x\rvert)$ and $-11/(3\lvert x\rvert)$ in units of $\alphas C_A/(2\pi)$. In both channels, these reproduce the large-$\lvert x\rvert$ asymptotics of the explicit one-loop matrix elements presented in \cref{sec:matrixelems} (and classified by eigenvalue sector in \cref{eq:tail_classification}).

	\paragraph{Hadronic exterior.}
	For a physical hadron $H$, convolving the partonic relation \cref{eq:exterior_constraint} with the light-cone distributions $f_{j/H}(y,\mu)$ gives the exterior quasi-\PDF{} at one loop,
	\begin{align}\label{eq:exterior_hadron}
		\tilde{f}_{i/H}(x,\mu)\Big|_{\lvert x\rvert>1} {} & = -\int_{-1}^{1}\!\mathrm{d}w\,\frac{\sum_j(\mathcal{L}_{ij} \otimes f_{j/H})(w)}{\lvert x-w\rvert} \\
		{} & =\int_{-1}^{1}\!\mathrm{d}w\,\frac{1}{\lvert x-w\rvert}\left[\frac{\mathrm{d}f_{i/H}(w,\mu)}{\mathrm{d}\log\mu^2} - \gamma_{\text{op},i}\,f_{i/H}(w,\mu)\right]\,,
	\end{align}
	where $\mathcal{L}_{ij}$ is the scale-logarithm coefficient of the partonic channel $i/j$ and
	\begin{equation}
	 (\mathcal{L}_{ij}\otimes f_{j/H})(w) = \int\mathrm{d}y\,\mathcal{L}_{ij}\left(\frac{w}{y}\right)\,\frac{f_{j/H}(y)}{\lvert y\rvert}\,,
	\end{equation}
	as in \cref{eq:quasimatching}. The second form follows from the renormalization group: in the matching relation the scale logarithm has to cancel the difference between the \DGLAP{} evolution of the light-cone distribution,
	\begin{equation}
	\frac{\mathrm{d}f_{i/H}}{\mathrm{d}\log\mu^2} = \frac{\alphas}{2\pi}\sum_j\mathcal{P}_{ij}\otimes f_{j/H}\,,
	\end{equation}
	and the running of the equal-time operator,
	\begin{equation}
	\frac{\mathrm{d}\tilde{f}_{i/H}}{\mathrm{d}\log\mu^2} = \gamma_{\text{op},i}\,\tilde{f}_{i/H}\,,
	\end{equation}
	with the anomalous dimensions $\gamma_{\text{op},i}$ given below \cref{eq:gammaO}. In the flavor-singlet sector the \DGLAP{} term contains the quark--gluon mixing, while $\gamma_{\text{op},i}$ is flavor diagonal in the \MSBAR{} scheme (\cref{sec:renormalization}), so that the off-diagonal kernels $\mathcal{L}_{ij}$ receive no contribution from operator running and are given entirely by the splitting functions $\mathcal{P}_{ij}$. For gluons, \cref{eq:exterior_hadron,eq:exterior_tower} apply to the momentum-weighted distributions $x\tilde{f}_{g/H}$ and $w f_{j/H}$.

	Expanding $1/\lvert x-w\rvert$ at large $\lvert x\rvert$ yields the asymptotic tower
	\begin{equation}\label{eq:exterior_tower}
		\tilde{f}_{i/H}(x,\mu)\Big|_{\lvert x\rvert>1} = -\frac{1}{\lvert x\rvert}\sum_{n\geq0}\frac{1}{x^{n}}\left[\gamma_{\text{op},i}\,\mathcal{A}_{n,i/H} - \frac{\mathrm{d}\mathcal{A}_{n,i/H}}{\mathrm{d}\log\mu^2}\right]\,,
	\end{equation}
	where
	\begin{equation}
	\mathcal{A}_{n,q/H} = \int_{-1}^{1}\mathrm{d}y\,y^n f_{q/H}(y,\mu)\,,\qquad
	\mathcal{A}_{n,g/H} = \int_{-1}^{1}\mathrm{d}y\,y^{n}\,y f_{g/H}(y,\mu)\,.
	\end{equation}
	The gluon moments are those of the momentum-weighted distribution $y f_{g/H}(y)$, which is even in $y$ by crossing symmetry. The odd-$n$ gluon moments therefore vanish, and the even ones are $\mathcal{A}_{n,g/H} = 2\langle x^{n+1}\rangle_g$ in terms of the conventional moments $\langle x^m\rangle_g = \int_0^1\mathrm{d}x\,x^m f_{g/H}(x,\mu)$, of which $\langle x\rangle_g$ is the gluon momentum fraction.
	This is the momentum-space form of the short-distance operator product expansion. Around $z=0$ the bilocal operator expands into local twist-2 operators of increasing spin, with the $n$-th operator contributing to the asymptotic tower at order $1/\lvert x\rvert^{n+1}$. In the non-singlet sector the $n$-th coefficient is $(\gamma_{\text{op},i}-\gamma_n)\mathcal{A}_{n,i/H}$, where $\gamma_n$, the $n$-th Mellin moment of the \DGLAP{} kernel, is the anomalous dimension of the $n$-th local operator. At higher orders the scale logarithm appears with higher powers, and the exterior is no longer given by the kernel $1/\lvert x-w\rvert$ alone; however, the coefficients of all these logarithms are fixed by anomalous dimensions and lower-order matching coefficients. At leading power, the exterior therefore contains only the renormalization-group evolution of the correlator, with no independent non-perturbative parameters appearing outside $\lvert x\rvert\leq1$.

	The leading $1/\lvert x\rvert$ tail is the $n=0$ term. For non-singlet quarks, the anomalous dimension $\gamma_0$ vanishes by vector current conservation. For gluons, the terms with odd $n$ vanish, and $\gamma_0$ is the momentum integral of $\mathcal{P}_{gg}$. On partonic targets, the resulting coefficients $\gamma_{\text{op},i}-\gamma_0$ recover the respective tail coefficients $\gamma_\mathcal{O}$ of \cref{eq:gammaO}. In a hadron, the scale derivative of $\mathcal{A}_{0,g/H}$ also contains the quark momentum fraction through $\mathcal{P}_{gq}$, and on a quark target this is the entire $1/\lvert x\rvert$ tail of the gluon quasi-distribution (\cref{eq:fquasi_gq_pp_bare}).

	In the ratio scheme, dividing the equal-time correlator by its zero-momentum value $\mathcal{M}(0, z^2)$ removes both the $1/\lvert x\rvert$ tail and the boundary terms at $x=\pm\infty$. At $P^z=0$ ($\nu=0$), only the lowest local operators ($n=0$) survive in the operator product expansion, so that $\mathcal{M}(0, z^2)$ isolates their forward matrix elements multiplied by the short-distance Wilson coefficients. For the quark correlator this is the vector current alone, and for the gluon correlator these are the gluon and the quark momentum operators. For the gluon correlator in a hadron, $\mathcal{M}(0, z^2)$ also contains the quark momentum fraction, so that at one loop the reduced gluon distribution receives a term proportional to the ratio of the quark and gluon momentum fractions~\cite{Balitsky:2021qsr}.
	Dividing by $\mathcal{M}(0, z^2)$ therefore subtracts this $n=0$ contribution from the asymptotic tower of \cref{eq:exterior_tower}, removing the operator running $\gamma_{\text{op},i}$ along with the $1/\lvert x\rvert$ tail. In the pseudo-distribution, the $\nu=0$ value is the zeroth moment, which the ratio normalizes to unity. In the quasi-distribution, the Fourier transform maps this value to the boundaries: at each of $x\to\pm\infty$ the weight is $-1/2$ times the one-loop zeroth moment of the corresponding bare pseudo-distribution, after replacing $L_\mu^\mathcal{P}$ by $L_\mu^\mathcal{Q}$ (\cref{sec:matrixelems}).

	With the $n=0$ term absent, the exterior fall-off is accelerated. For non-singlet quarks, the scale derivative of $\mathcal{A}_{0,q/H}$ vanishes by vector current conservation, and the exterior begins at $n=1$: the distribution falls off as $\gamma_1\mathcal{A}_{1,q/H}/(x\lvert x\rvert)$, with $\gamma_1 = -\alphas C_F/(2\pi)\cdot 4/3$ the anomalous dimension of the momentum fraction and $\mathcal{A}_{1,q/H}$ normalized to the valence number. For gluons %
	the momentum-weighted distribution $x\tilde{f}_g$, which is even in $x$, begins at $n=2$ and falls off as $\lvert x\rvert^{-3}$.

	\paragraph{Reconstruction from lattice data.}
	These results have direct consequences for $x$-space quasi-distributions obtained from lattice data~\cite{Chen:2025cxr,Dutrieux:2025jed,Dutrieux:2025axb,Xiong:2025obq}. At leading power, the exterior region $\lvert x\rvert>1$ of an $x$-space quasi-distribution is nonvanishing but contains no information beyond the light-cone distribution on $[-1,1]$, from which the exterior is determined through perturbative matching (\cref{eq:quasimatching}), at one loop through \cref{eq:exterior_hadron}. The exterior also enters the inverse relation of \cref{sec:definitions}: at one loop the inverse kernel is the tree-level kernel minus the one-loop kernel, and therefore contains the one-loop exterior of \cref{eq:exterior_constraint} with the opposite sign.

	Since the exterior is fixed at leading power by the light-cone distribution, any reconstruction of the quasi-distribution from lattice data can be tested on the exterior. In a coordinate-space extrapolation~\cite{Chen:2025cxr}, the exterior follows from the transform of the extrapolated correlator. In an $x$-space reconstruction, such as the Gaussian-process reconstruction of ref.~\cite{Dutrieux:2025jed}, the exterior is part of the reconstructed function. In both cases the exterior has to agree with the matching relation up to power corrections, which grow as $x\to1$ (\cref{eq:quasimatching}); at one loop the matching relation is \cref{eq:exterior_hadron}, once converted to the renormalization scheme of the analysis (\cref{sec:matrixelems}). The difference between the reconstructed exterior and the matching relation therefore measures the power corrections together with the systematics of the Fourier transform, in a region where the leading-power quasi-distribution vanishes at tree level and starts at order $\alphas$.

\subsection{The boundary terms at $x=\pm\infty$}
\label{sec:boundary}

	The one-loop quasi-distribution also has support at the points $x=\pm\infty$ themselves. These boundary terms are forced by a constraint on its integral. Integrating the quasi-\PDF{} over $x$ sets $\zt=0$ in its Fourier representation, since $\int\mathrm{d}x\,e^{i\zt xP^z} = 2\pi\delta(\zt)/P^z$. The integrals $\int\mathrm{d}x\,\tilde{f}_q(x)$ and $\int\mathrm{d}x\,x\tilde{f}_g(x)$ are therefore proportional to the correlator at $\zt=0$, the forward matrix element of the local operator: the vector current $\overline{\psi}\gamma^0\psi$, whose matrix element is the quark number, and for gluons the product of the two field strengths at a single point. In dimensional regularization this local matrix element receives no one-loop correction. At short distances the one-loop correlator behaves as $(\zt^2\mu^2)^\epsilon/\epsilon$, which vanishes at $\zt=0$, so the bare one-loop quasi-distribution integrates to zero. Its interior, however, carries poles whose integral does not vanish. The compensating contribution must therefore sit outside every finite interval, at $x=\pm\infty$.

	The mechanism is the same short-distance factor. For $\zt\neq0$ it is a pole and a logarithm, and the logarithm transforms into the $1/\lvert x\rvert$ tail of \cref{sec:symmetries}. In $d$ dimensions the tail falls off as $\lvert x\rvert^{-1-2\epsilon}$ and is integrable, and its integral is a $1/\epsilon$ pole that cancels the integral of the interior poles. When we expand in $\epsilon$, we write the tail as a plus distribution at infinity, which integrates to zero, and the pole moves onto boundary terms $\delta^\pm(1/x)/x^2$ at $x=\pm\infty$ (\cref{eq:plusInfinity}). The boundary poles are therefore fixed by the tail coefficient of \cref{eq:gammaO}: they add up to $-\gamma_\mathcal{O}/\epsilon$ for quarks and to $-2\gamma_\mathcal{O}/\epsilon$ for gluons. This is where the $\epsilon$-expansion fails to commute with the limit $\zt\to0$. The two boundaries carry equal weight, since at short distances the correlator depends on $\nu$ only through $(\zt^2\mu^2)^\epsilon=(\nu^2\mu^2/(P^z)^2)^\epsilon$.\footnote{Individual contributions can produce one-sided boundary terms, such as the transforms of \cref{app:FTadvanced}; their asymmetric parts cancel in the complete correlator.}

	Expanding in $\epsilon$ first instead leaves $\log(\zt^2)$ terms whose transforms are ill-defined at $x\to\pm\infty$. Ref.~\cite{Izubuchi:2018srq} defined them by writing $\log(\zt^2)$ as the derivative of a power, which ref.~\cite{Chou:2022drv} characterized as undoing the $\epsilon$-expansion. As ref.~\cite{Chou:2022drv} showed, this leaves an arbitrary scale in the boundary terms, which they proposed to fix by requiring formal quark-number conservation. In $d$ dimensions the boundary terms are determined without such a condition.

	Since the renormalization constant of the bilocal operator does not depend on $\zt$, its counterterm sits at $\delta(1\mp x)$ and leaves the boundary poles behind (\cref{sec:quarkinquark}). Renormalization therefore breaks the balance between the interior and the boundaries, and the \MSBAR{} quasi-distribution no longer integrates to zero; this is the statement of ref.~\cite{Izubuchi:2018srq} that the moments of the \MSBAR{} quasi-\PDF{} do not exist. The boundary terms are unrelated to the linear divergence of the Wilson-line self-energy, which vanishes identically in \MSBAR{}~\cite{Izubuchi:2018srq}.\footnote{In dimensional regularization the Wilson-line self-energy power divergence does not appear as a pole in $d=4$ but as a pole in $d=3$, and therefore leaves no trace in the \MSBAR{} subtraction. On the lattice it reappears as the linear divergence canceled by the counterterm $\delta m(a)\sim 1/a$~\cite{Chen:2016fxx}.}

	In the matching convolution \cref{eq:quasimatching} the argument $x/y$ of the kernel diverges as $y\to0$, so the boundary terms evaluate the light-cone \PDF{} at vanishing momentum fraction, as $\lim_{\beta\to0}\beta f(\beta x)$~\cite{Izubuchi:2018srq}. At such small $y$ the parton momentum $\lvert y\rvert P^z$ is no longer a hard scale, so this region lies outside the reach of the perturbative matching. The boundary terms do carry poles, however, and since we expand in $\epsilon$ only after the convolution, it suffices to show that each pole multiplies a vanishing limit. We make this argument for the convolution with the light-cone distribution in \cref{eq:quasimatching}. In the inverse relation (\cref{sec:definitions}), the boundary terms of the kernel instead evaluate the quasi-distribution at vanishing momentum fraction, which is finite at fixed $P^z$ because the correlator is expected to decay at large $\zt$~\cite{Chen:2025cxr}; in the ratio scheme these terms are absent (\cref{sec:symmetries}).

	\paragraph{Quark channel.}
	Ref.~\cite{Izubuchi:2018srq} showed for the isovector distribution that the boundary terms drop out of the convolution, because $\lim_{\beta\to0}\beta f(\beta x)=0$ for $f(y)\sim y^{-1+a}$ with $a>0$. For a single quark flavor this is not sufficient, since global fits contain infinitely many sea quarks and antiquarks~\cite{Chou:2022drv}, with $a_S<0$~\cite{Ball:2016spl}. However, the two boundaries enter with equal weight, so the convolution only requires $\lim_{\beta\to0}\beta\,[f_q(\beta x)+f_q(-\beta x)]=0$. With crossing symmetry, $f_q(-y)=-f_{\overline{q}}(y)$~\cite{Collins:2011zzd}, this is the valence combination $\lim_{\beta\to0}\beta\,[f_q(\beta x)-f_{\overline{q}}(\beta x)]$, from which the sea cancels. The valence distribution integrates to the finite net quark number, so the limit vanishes. Ref.~\cite{Chou:2022drv} arrived at the same combination in the non-singlet sector, from the symmetric boundary terms of the $\log(\zt^2)$ prescription; in our calculation the symmetry holds for the complete correlator in $d$ dimensions.

	\paragraph{Gluon channel.}
	For gluons this argument does not work. The gluon is its own antiparticle, $f_g(-y)=-f_g(y)$~\cite{Collins:2011zzd}, so there is no net gluon number to isolate, and at $\mu^2\approx10\,\text{GeV}^2$ the gluon \PDF{} grows at small $y$ as $f_g(y)\sim y^{-1+a_g}$ with $a_g\simeq-0.2$~\cite{Ball:2016spl}, so that $\lim_{\beta\to0}\beta f_g(\beta x)$ diverges. The gluon boundary terms are nevertheless harmless. The gluon correlator measures the momentum density $y f_g(y)$ and not the number density~\cite{Balitsky:2019krf}, and accordingly the $1/\lvert x\rvert$ tail sits on $x\tilde{f}_g$ (\cref{eq:fgquasi}), whereas the factorization theorem \cref{eq:quasimatching} is formulated for $\tilde{f}_g$. The gluon matching kernel therefore falls off one power faster than the quark kernel,
	\begin{equation}\label{eq:tail_normalization}
		C_{qq}(x)\;\xrightarrow{\;\lvert x\rvert\to\infty\;}\;-\frac{3}{2\lvert x\rvert}\,,\qquad
		C_{gg}(x)\;\xrightarrow{\;\lvert x\rvert\to\infty\;}\;-\frac{5}{6\,x\lvert x\rvert}\,,
	\end{equation}
	in units of $\alphas C_F/(2\pi)$ and $\alphas C_A/(2\pi)$, where the gluon coefficient is quoted for $\mathcal{\widetilde M}_{pp}$ (the other eigenvalue sectors are classified in \cref{eq:tail_classification}). With the measure of \cref{eq:quasimatching}, the extra power of $x$ becomes an extra power of $y$, and for $x>0$ the small-$y$ regions of the quark and gluon convolutions are
	\begin{equation}\label{eq:boundary_moments}
		-\frac{3}{2x}\int \mathrm{d}y\,f_q(y) \qquad \mathrm{and}
		\qquad
		-\frac{5}{6x^2}\int \mathrm{d}y\,y\,f_g(y)
	\end{equation}
	per boundary. The gluon boundary terms therefore probe $y f_g(y)$, whose integral, the gluon momentum fraction, is bounded by the momentum sum rule; hence $1+a_g>0$ and $\lim_{\beta\to0}\beta\,[\beta x\,f_g(\beta x)]\sim\beta^{1+a_g}\to0$. Unlike the quark case, this holds at each boundary separately. In a hadron, the gluon quasi-distribution also receives boundary terms from the gluon-in-quark channel, where the local gluon operator mixes into the quark momentum operator. By the same argument these probe $y f_q(y)$, whose integral, the quark momentum fraction, is bounded by the momentum sum rule as well. The quark-in-gluon channel has no boundary terms at one loop (\cref{eq:fquasi_qg_Assembled_Rules}), since at $\zt=0$ the quark operator reduces to the conserved vector current, which does not mix with gluon operators.
	\section{General elements of the calculation\label{sec:preliminaries}}

	We calculate the quark matrix element \cref{eq:Qmatelem} with $\Gamma = \gamma^0$ and the gluon matrix element \cref{eq:matelem} for arbitrary index structure, the latter through the form-factor decomposition of \cref{sec:formfactors}. All contributions are computed in both a general covariant gauge and an axial gauge, which provides a very strong check of all results and shows which combinations of diagrams are gauge invariant. We collect the conventions of the calculation in \cref{sec:conventions}, describe the projection of the diagrams onto scalar quantities and their reduction to three master integrals in \cref{sec:reduction}, and give the master integrals in closed form in \cref{sec:oneloopints}.

	\subsection{Conventions}
	\label{sec:conventions}

	Our perturbative calculation includes only partonic states, so we replace $P^z$ in all formulae with $p^z$ and we assume massless onshell external states that satisfy $p^\mu = (p^z,0,0,p^z)$.

	For the calculation of \cref{eq:Qmatelem,eq:matelem} one has to pick a scheme for the regularization of \IR{} and \UV{} divergences and a renormalization scheme. We use dimensional regularization in $d=4-2\epsilon$ dimensions and do not distinguish between $\epsilon_\IR$ and $\epsilon_\UV$ because the reduction to a set of master integrals mixes $\epsilon_\IR$ and $\epsilon_\UV$. We subtract poles in $\epsilon$, i.e.~implement the \MSBAR{} scheme. The matrix elements start at $\mathcal{O}(\alphas^0)$, so it is sufficient to replace the bare \QCD{} coupling $\alpha_0$, where we define $g_s^2=4\pi\alpha_0$, with the \MSBAR{} renormalized coupling at leading order $\alpha_0 = \alphas e^{\epsilon \gamma_\mathrm{E}} / (4\pi)^\epsilon$.

	An alternative approach, used e.g.~in ref.~\cite{Izubuchi:2018srq}, is to distinguish $\epsilon_\IR\neq\epsilon_\UV$ in dimensional regularization, which leads to relatively simple matching expressions because the one-loop corrections to the light-cone distributions are given by the well-known \DGLAP{} splitting kernels multiplied by $1/\epsilon_\UV - 1/\epsilon_\IR$. In schemes with a gluon mass or off-shell momenta to regulate the \IR{} divergences one additionally needs to calculate the non-vanishing corrections to light-cone distributions, see e.g.~ref.~\cite{Wang:2017qyg} and \cref{sec:matrixelems}.

	A further technical choice for the calculation is the gauge. In an axial gauge, contributions with Wilson lines can be eliminated at the cost of introducing significantly more terms through larger propagator expressions. Axial gauge can also eliminate the need to include potential ghost contributions. In Feynman or in $R_\xi$ gauge both Wilson line and ghost contributions are non-vanishing. We perform the calculation in both axial and $R_\xi$ gauge, with gauge parameter $\xi_G$, as a cross-check.

	\paragraph{The axial gauge and its ambiguities.}
	Our axial gauge choice, defined by the condition $r^\mu A_\mu = 0$, removes the ghost fields and the Wilson line diagrams, but it does not fix the gauge completely. A residual freedom remains under any gauge transformation $A^\mu \rightarrow A^\mu + \partial^\mu \alpha$ with $\alpha(x)$ satisfying $r \cdot \partial \alpha = 0$, and it leaves the pole $1/(r \cdot p)$ of the axial-gauge propagator (\cref{app:feynman}) undetermined, so that an additional prescription is required.

	The choice of prescription for this pole corresponds to specifying the physical boundary conditions for the gauge field at infinity \cite{Belitsky:2002sm}, see also ref.~\cite{Belitsky:2005qn} appendix G.\footnote{Note that a more rigorous approach to the axial gauge is the Mandelstam--Leibbrandt gauge \cite{Mandelstam:1982cb,Leibbrandt:1987qv}, which we do not consider in this paper. This method resolves the ambiguity (in a light-cone gauge) from the start by using a second, conjugate light-like vector $\tilde{r}$ to define an unambiguous pole prescription. While more complicated to handle in practice, it provides a consistent and Lorentz-covariant way to define the $r\cdot p$ pole in the propagator.}

	The physical context of the process then dictates the choice \cite{Belitsky:2002sm,Belitsky:2005qn}. The retarded prescription $1/(r\cdot p + i\eta)$ corresponds to a future-pointing Wilson line with $A(-\infty)=0$, and is required for quantities sensitive to final-state interactions, such as transverse-momentum-dependent distributions in semi-inclusive deep inelastic scattering. The advanced prescription $1/(r\cdot p - i\eta)$ corresponds to a past-pointing line with $A(+\infty)=0$, and is required for quantities sensitive to initial-state interactions, as in the Drell--Yan process \cite{Belitsky:2002sm}. For time-reversal even quantities such as the unpolarized \PDF{} the appropriate choice is the principal value, $\frac{1}{2} ( 1/(r \cdot p + i\eta) + 1/(r \cdot p - i\eta) )$, which corresponds to $A(\infty) = -A(-\infty)$, and this is the prescription we use in axial gauge.

	\subsection{Projection and reduction}
	\label{sec:reduction}

	For the quark operators we do not include the anti-quark diagrams as the results for anti-quarks can be easily obtained via crossing symmetry. To handle the Dirac structure of the higher-order quark-diagram contributions, we take the Dirac trace. This corresponds to a projection onto the tree-level Dirac structure $\frac{1}{2}\sum_s \bar{u}_s(p)\gamma^\mu u_s(p) = \frac{1}{2}\mathrm{tr}\left(\slashed{p}\,\gamma^\mu\right) = \frac{d}{2}\, p^\mu $, where we consider the case $\mu=0$, for which $p^0 = p^z$, and choose external projectors in $d$ dimensions.\footnote{The other case $\mu=z$ has been documented in ref.~\cite{Izubuchi:2018srq}.} The projection introduces a factor $\operatorname{Tr}\mathbf{1}/2 = d/2$, which cancels from the matching relation against the quark tree-level normalization whenever the target parton is a quark. In the quark-in-gluon channel, where the target is a gluon, it does not cancel (\cref{sec:matrixelems}). After Fourier transformation, the tree-level diagram contributes $(d/4)\,\delta(1-x)$, the distribution \cref{eq:ftree_qq}.

	For the gluon case, we extract the scalar form factors of \cref{eq:Manb} by constructing projection operators in $d$ dimensions. We apply these projectors to the one-loop matrix element \cref{eq:matelem} in $R_\xi$ and axial gauges. The projection operators are lengthy, with 48 terms, but straightforward to derive. This generalizes previous calculations in Feynman gauge that employed fixed index combinations. We verify in general dimension, through an independent index-contraction routine, that each projector is dual to the corresponding tensor structure as displayed in \cref{eq:Manb}, and that the renormalization basis of \cref{sec:gluon_operator_renorm} is orthogonal.

	In addition to lengthy projection operators in the gluon case, working in both $R_\xi$ and axial gauges leads to a proliferation of terms. We generate the Feynman diagrams with \qgraf{}~\cite{Nogueira:1991ex} and apply the Feynman rules, the algebraic reductions and the identification of integrals in Mathematica. For the integrals we construct integration-by-parts (\IBP{}) reduction rules, generate a system of equations with sufficient seeds, and solve this linear system via the custom-system reduction feature of \texttt{Kira} \cite{Maierhofer:2017gsa,Klappert:2020nbg}. The \IBP{} identities are a consequence of the translational invariance of dimensionally regularized integrals~\cite{Chetyrkin:1981qh} and hold here.\footnote{While we only make use of the IBP reduction of the integrals before Fourier transform, they in fact also hold when including the Fourier transform, and this has been used in the literature \cite{Chen:2020iqi}.}

	The \IBP{} reduction reduces all expressions to a set of just three master integrals that cover both quark and gluon cases. These master integrals arise from two families of integrals. The first of these families, which we denote $I_z$, arises from the real corrections, which include an exponential factor $\exp(ik\cdot z)$ that stems from connecting the two field strength tensors in \cref{eq:matelem}. The second family, which we denote $I$, is generated through the virtual corrections. The propagators in both families are otherwise the same, and the two families are defined as
\begin{gather}
	I_z^\pm(a_1,a_2,a_3,a_4) =  \int_k  \frac{e^{i k \cdot z}}{D_1^{a_1} D_2^{a_2}  D_3^{a_3} 
		D_4^{a_4}}
\end{gather}
and
\begin{gather}
	I^\pm(a_1,a_2,a_3,a_4) =  \int_k  \frac{1}{D_1^{a_1} D_2^{a_2}  D_3^{a_3} D_4^{a_4}}\,,
\end{gather}
with $\int_k = \mu^{4-d}\int \mathrm{d}^dk/(2\pi)^d$, $D_1=k^2+i\epsilon$, $D_2=(k-p)^2+i\epsilon$, $D_3=k\cdot z\pm i\eta$, and $D_4 = (k-p)\cdot z\pm i\eta$. The superscript \enquote{$\pm$} denotes our choice of sign for the $i\eta$ term in the linear propagators. For integrals without linear propagators, we drop this superscript. We assume $\epsilon > 0$ and $\eta > 0$.

	\subsection{Master integrals}
	\label{sec:oneloopints}

The three master integrals are $I_z(1,0,0,0)$, $I_z^\pm(1,0,0,1)$ and $I^\pm(1,0,0,1)$. The first is a function of $z^2={-}\zt^2$ alone, while $I_z^\pm(1,0,0,1)$ and $I^\pm(1,0,0,1)$ are functions of both $z^2$ and $p\cdot z=-\zt p^z$. Further integrals that appear in intermediate steps are related to these by shifts of the loop momentum. The integral $I_z(0,1,0,0)$ is related to $I_z(1,0,0,0)$ via $I_z(0,1,0,0) = e^{i p\cdot z} I_z(1,0,0,0)$. With the loop-momentum shift $k\to -k+p$ one sees that
\begin{equation}
I^\pm(0,1,1,0) = - I^\mp(1,0,0,1)\,.
\end{equation}
The integrals $I_z^\pm(0,1,1,0)$ can be mapped to $I_z^\pm(1,0,0,1)$ with the shift $k\to k+p$, which gives
\begin{align}
	I_z^\pm(0,1,1,0)(p)  & = e^{i p\cdot z}I_z^\pm(1,0,0,1)(-p)\,, \\
	I_z^\pm(1,0,0,1)(p) &=  e^{i p\cdot z}  I_z^\pm(0,1,1,0)(-p)  \,,
\end{align}
where the argument denotes the external momentum at which the integral is evaluated. We give the master integrals with the dimensional regularization scale set to unity, $\mu^2=1$.

\subsubsection*{Master integral $I_z(1,0,0,0)$}

The master integral $I_z(1,0,0,0)$ is calculated by applying Schwinger parameters to the propagator to obtain
\begin{align} \label{eq:Iz1000}
	I_z(1,0,0,0) & =  \int_k \frac{e^{i k \cdot z}}{k^2 + i\epsilon} 
	\nonumber \\ 
	 & = -\frac{1}{4} i \pi^{-d/2} (\zt^2)^{1-d/2}\Gamma(d/2-1) \,.
\end{align}

\subsubsection*{Master integral $I(1,0,0,1)$}

Similarly, we solve this integral through Schwinger parameters to find
\begin{align}
	 I^\pm(1,0,0,1) & = 	\int_k \frac{1}{ (k^2+i\epsilon) \, ((k-p)\cdot z \pm i\eta)}  \nonumber
	 \\
	& {} = \, \mp \frac{1}{4} \pi^{-d/2} (\mp i p^z \zt)^{d-3}
	(\zt^2)^{1-d/2} 	\Gamma(3-d)\Gamma(d/2-1)\,.
\end{align}
We check the consistency of this result with the relation
\begin{align}
	\label{eq:deltarelation}
	2\pi i \,\delta((k-p)\cdot z) = \frac{1}{(k-p)\cdot z-i\eta} - \frac{1}{(k-p)\cdot z+i\eta}\,,
\end{align}
and compare with the numerical evaluation in eq.~(26) of ref.~\cite{Liu:2022tji} and to 
numerical 
results obtained with \texttt{FIESTA}~\cite{Smirnov:2008py,Smirnov:2013eza,Smirnov:2015mct} and 
\texttt{pySecDec} \cite{Borowka:2015mxa,Borowka:2017idc} to higher powers in $\epsilon$ and find 
full agreement.

\subsubsection*{Master integral $I_z(1,0,0,1)$}
We now consider the integral
\begin{gather}
	\label{eq:masterIz1001}
	I_z^\pm(1,0,0,1)= \int_k \frac{e^{i k \cdot z}}{(k^2 + i\epsilon )\, ((k-p)\cdot z \pm
	i\eta)}\,.
\end{gather}
Applying Schwinger parameters, we find
\begin{align}
 I_z^+(1,0,0,1) & = \int_k \frac{e^{i k\cdot z}}{
		(k^2+i\epsilon) ((k-p)\cdot z +i\eta)} \nonumber \\
		& = - \frac{1}{4}  \pi^{-d/2} (-i p^z \zt)^{d-3} (\zt^2)^{1-d/2} e^{-i p^z \zt}
		\Gamma(d/2-1) \Gamma(3-d,-i p^z \zt)\,.
\end{align}
In refs.~\cite{Izubuchi:2018srq,Chou:2022drv} the incomplete gamma function appears as $\Gamma(4-d,-i p^z \zt)$. Using the recurrence relation $\Gamma(s+1,x)=s\Gamma(s,x) + x^s e^{-x}$, one can write $\Gamma(3-d,-i p^z \zt)$ in terms of $\Gamma(4-d,-i p^z \zt)$, so that the two results can be compared directly. For the opposite prescription we find
\begin{align}
 I_z^-(1,0,0,1) & = \int_k \frac{e^{i k\cdot z}}{
	(k^2+i\epsilon) ((k-p)\cdot z -i\eta)} \nonumber \\
	& = I_z^+(1,0,0,1) + \frac{1}{4} \pi^{-d/2}  e^{-i p^z \zt} (\zt^2)^{1-d/2} \Gamma(3-d)\Gamma(d/2-1) \nonumber \\
	 & \qquad \qquad \qquad \qquad \cdot \left( (-ip^z \zt)^{d-3} + (+i p^z \zt)^{d-3}  \right)\,,
\end{align}
and the consistency of the two results can again be checked via \cref{eq:deltarelation}.

\subsubsection*{Prescription-independent combinations}

The real and virtual corrections are individually dependent on the regularization of the linear propagators, essentially relying on the principal value prescription, but their sum is independent of any prescription to regulate the linear propagators. For every integral $I_z(1,0,0,1)$ in the real corrections there is a virtual counterpart $-e^{i p\cdot z}I(1,0,0,1)$ that eliminates the linear propagator singularity, and a similar cancelation occurs for the shifted integrals $I_z(0,1,1,0)$. We expect that this property holds at higher loop orders. It will therefore prove convenient to write the results in terms of the combination
\begin{align}
	e^{i p\cdot z}I^\pm(1,0,0,1) - I_z^\pm(1,0,0,1) = {} & \frac{1}{4}  \pi^{-d/2} (-i p^z \zt)^{d-3}
	(\zt^2)^{1-d/2} e^{-i p^z \zt} \nonumber\\
	{} & \quad \cdot \Gamma(d/2-1) \Big(- \Gamma(3-d) + \Gamma(3-d,-i p^z \zt) \Big)\nonumber\\
	=: {} & \diffmaster{}\,,
	\label{eq:diffmaster}
\end{align}
which is independent of the $i\eta$ prescription. This can be seen directly from the explicit results or at the momentum-integral level, as the linear propagator singularity cancels in the numerator. Similarly, we define the regularization-independent combination
\begin{align}
	    I^\pm(0,1,1,0) - I_z^\pm(0,1,1,0) = {} & \frac{1}{4}  \pi^{-d/2} (+i p^z \zt)^{d-3}
	    	(\zt^2)^{1-d/2} \nonumber \\
	    & \qquad \cdot \Gamma(d/2-1) \Big(-\Gamma(3-d) + \Gamma(3-d,+i p^z \zt)
	    \Big) \nonumber \\
	    =: {} & \diffmastertwo{}\,,
    \label{eq:diffmastercross}
\end{align}
and have the relation $I^\eta(0,1,1,0)(p) = e^{i p\cdot z} I^\eta(1,0,0,1)(-p)$, in the same notation as above.
\section{Results: Matrix elements and Fourier transforms \label{sec:matrixelems}}

The one-loop results in this section are given in the \MSBAR{} scheme. The ratio scheme divides the correlator by its value at zero hadron momentum, $P^z=0$, that is, at $\nu=0$ for fixed $\zt$, and the hybrid scheme applies the same normalization at short distances and switches to a Wilson-line mass subtraction beyond a fixed coordinate-space distance $z_s$~\cite{Ji:2020brr,Ji:2020ect}. Both are finite renormalizations of the same correlator and follow from the results below without further loop calculations. The ratio-scheme kernel is the \MSBAR{} kernel divided by the $\nu\to0$ limit of the coordinate-space matching coefficient, a finite function of $\log(\mu^2\zt^2)$~\cite{Yao:2022vtp}, and at one loop the division becomes a subtraction of its one-loop term times the tree-level kernel. We discuss the consequences for the region $\lvert x\rvert>1$ and for the gluon normalization in a hadron in \cref{sec:symmetries}.

\subsection{Tree-level results}\label{sec:tree}

To set the overall normalization of the quark and gluon matrix elements, we first list our tree-level results. The tree-level quark-in-quark matrix element, averaged over the $n_s=2$ quark spin states and over color with a factor of $1/N_c$, is given by
\begin{equation}\label{eq:Qtree_quark}
	\mathcal{Q}_{\gamma^0,\text{ET}}^{(0)}(\zt,p^z) = \frac{d}{2} p^0
	\mathcal{M}_{pp}^{q,(0)}
= \frac{d}{2} p^z
e^{i p\cdot z}\,,
\end{equation}
so that $\mathcal{M}_{pp}^{q,(0)}=e^{i p\cdot z}$, in line with the decomposition of $\mathcal{Q}_{\gamma^\alpha,\text{ET}}^\text{bare}$ into structures $P^\alpha$ and $z^\alpha$ in \cref{eq:quarkFF}.

The prefactor $\eta_{pp} = d/2$ of \cref{eq:quarkFF} originates from the spin sum $\sum_s \bar{u}_s(p)\gamma^0 u_s(p) = \operatorname{Tr}\mathbf{1}\,p^0$ continued as $\operatorname{Tr}\mathbf{1} = d$, averaged over the $n_s = 2$ spin states of a four-dimensional quark. With the conventional $\operatorname{Tr}\mathbf{1} = 4$, the prefactor is $2p^z$.

For the gluon-in-gluon case we have, averaging over
colors with a factor $1/(N_c^2-1)$ and over the $n_\text{pol}$ polarizations of the external gluon, here taken at the four-dimensional value $n_\text{pol}=2$,
\begin{equation}
	\mathcal{M}_{pp}^{g,(0)}(\nu,z^2)  = -\frac{1}{2}\left( e^{i p\cdot z} + e^{-i
		p\cdot z}\right) = -\frac{1}{2}\left( e^{-i \nu} + e^{i
		\nu}\right)\,. \label{eq:Mpp_g_tree}
\end{equation}
All other gluonic scalar form factors vanish at tree level. The polarization sum contracts into the $\mathcal{M}_{pp}$ tensor structure of \cref{eq:Manb} with a $d$-independent coefficient, so the number of polarizations enters the gluon form factors only through the averaging divisor $n_\text{pol}$. The tree-level distribution also carries the tensor coefficient $c_{pp}$ of \cref{eq:general_G}, which multiplies $\mathcal{\widetilde M}_{pp}$ in the operator combination. For the leading-twist combination \cref{eq:Gcombination_Mpp_tilde} it takes the value $c_{pp}=2$, corresponding to the factor of $2$ in \cref{eq:fgpseudo}.

These matrix elements yield the following quasi- and pseudo-\PDF{}s:
\begin{gather}
	\tilde{f}_{q/q}^\text{quasi,tree}(x) = \tilde{f}_{q/q}^\text{pseudo,tree}(x) = \frac{\operatorname{Tr}\mathbf{1}}{2n_s}\,\delta(x-1) = \frac{d}{4}\,\delta(x-1)\,,\label{eq:ftree_qq}\\
	x \tilde{f}_{g/g}^\text{quasi,tree}(x) = x \tilde{f}_{g/g}^\text{pseudo,tree}(x) = \frac{c_{pp}}{n_\text{pol}}\left[\delta(x-1) + \delta(1+x)\right] = \delta(x-1) + \delta(1+x)\,.\label{eq:ftree_gg}
\end{gather}
For the gluon distributions the symmetry $f_g(x)=-f_g(-x)$ originates from the factor $x$. The anti-quark distributions can be obtained through the crossing relation $f_{\bar{q}}(x)=-f_q(-x)$.

\paragraph{Trace and polarization, \MSBAR{} conventions.}
In the matching relation \cref{eq:matching1loop}, the matching coefficient $C_{ik}$ is convolved with the target parton's tree-level distribution $f_{k/k}^{(0)}$. Because $f_{k/k}^{(0)}$ is a combination of contact terms, any overall $\epsilon$-dependent factor shared between the one-loop amplitude and the target tree distribution cancels out of $C_{ik}$. In the diagonal channels, these factors cancel directly:
for quark-in-quark, both the one-loop diagrams and the tree distribution \cref{eq:ftree_qq} carry the open quark Dirac trace $\operatorname{Tr}\mathbf{1} = d$, which factors out as $d/4 = 1-\epsilon/2$;
for gluon-in-gluon, the polarization divisor $n_\text{pol}$ enters both the one-loop amplitude and the tree distribution \cref{eq:ftree_gg} as an overall factor and cancels in $C_{gg}$.
In the off-diagonal channels ($q/g$ and $g/q$), however, the target parton differs from the operator, so no such cancellation occurs.

There, the bilocal operator has no ultraviolet divergence at one loop at fixed $\zt\neq0$ ($Z_{Sg}^{(1)} = Z_{gS}^{(1)} = 0$), so in the interior the uncancelled factors multiply the collinear $1/\epsilon$ pole alone. Because the light-cone \MSBAR{} counterterm $Z_{\text{LC},ik}^{(1)}$ is a pure pole with a four-dimensional residue, any difference between this factor and its four-dimensional limit produces a finite $\mathcal{O}(\epsilon^0)$ shift in the matching coefficient, proportional to the \DGLAP{} splitting function. To match onto standard light-cone \PDF{}s, we therefore align our conventions with those of the standard splitting functions:
in the quark-in-gluon channel, we average over the $n_\text{pol} = d-2$ physical polarizations of the external gluon, while with the four-dimensional value $n_\text{pol} = 2$ an evanescent factor $(d-2)/2$ would remain on the collinear pole and shift the matching kernel by a finite term proportional to $P_{qg}$;
for the Dirac trace, we recover the conventional $\operatorname{Tr}\mathbf{1} = 4$ normalization by setting the overall factor $d/4$ to $1$ instead of expanding it against the collinear pole;
and in the gluon-in-quark channel, we continue the transverse index sum of the gluon operator, $i\in\{1,2\}$ on the lattice, to all $d-2$ transverse directions in the kernel.
The last choice follows from the light-cone definition \cref{eq:gluonPDF}, whose covariant contraction $g^{\nu\sigma}$ runs over the $d-2$ directions transverse to the light cone. Inserting the decomposition \cref{eq:Manb} with a light-like separation, every auxiliary structure vanishes, since $z^+=0$, $z^2=0$ and $g^{++}=0$, and the contraction is exactly $d-2$ times the contribution of a single transverse direction, proportional to $(p^+)^2\mathcal{M}_{pp}$. The operator that defines $f_g^{\MSBAR}$ therefore carries the factor $d-2$ on its leading-twist form factor, and in the gluon-in-quark channel the collinear poles on the two sides of \cref{eq:matching1loop} cancel with the correct finite remainder only if the coefficient $c_{pp}$ of $\mathcal{\widetilde M}_{pp}$ in the equal-time operator is continued in the same way. For $\mathcal{G}^{ziiz}$ and the two combinations of \cref{eq:Gcombination_Mpp_tilde,eq:Gcombination_ziiz_tilde}, $c_{pp}$ equals the transverse count, so that the condition amounts to summing $i$ over $d-2$ directions. In the gluon-in-quark quasi-distribution the factor $d-2$ also multiplies the boundary poles at $x=\pm\infty$ (\cref{eq:fquasi_gq_pp_bare}), and its $\mathcal{O}(\epsilon)$ part leaves a finite boundary term there, which drops out of the matching together with the poles (\cref{sec:boundary}).
The transverse count of the gluon operator plays the same role as the Dirac trace of the quark operator. Both are normalizations of the operator, and both survive only in the off-diagonal channel in which the operator is evaluated in the other parton: the Dirac trace in the quark-in-gluon channel and the transverse count in the gluon-in-quark channel. These choices do not change the operator measured on the lattice, only the light-cone distribution onto which the kernel matches. A different convention would define a different light-cone distribution, related to the standard \MSBAR{} \PDF{} by a finite renormalization proportional to the splitting function.

In its own diagonal channel, the Dirac trace is an overall factor. The transverse count $n_t$ is an overall factor only for combinations in which every term carries the same number of transverse sums, such as $\mathcal{G}^{ziiz}$ and $\mathcal{G}^{0ii3}+\mathcal{G}^{3ii0}$. In the leading-twist combination \cref{eq:Gcombination_Mpp_tilde}, however, $\mathcal{G}^{0ii0}$ carries one transverse sum and $\mathcal{G}^{jiij}$ two, so that with $n_t$ transverse directions the combination is $n_t\mathcal{\widetilde M}_{pp}+n_t(2-n_t)\mathcal{\widetilde M}_{gg}$. After tree normalization, continuing to $n_t=d-2$ therefore adds $-(d-4)\,\mathcal{\widetilde M}_{gg}$, and the minimal-contamination combination \cref{eq:Gcombination_ziiz_tilde} acquires $(d-4)/(d-2)$ times the second combination of \cref{eq:eigenspace_parpar}. Neither admixture changes $C_{gg}$. The renormalization operator \cref{eq:Zhat} distinguishes only the direction of $z$, so that every component $\mathcal{G}^{\mu\nu\rho\sigma}$ with fixed indices renormalizes multiplicatively, with an eigenvalue set by the number of indices along $z$, and a transverse sum cannot leave the eigenvalue sector of \cref{sec:gluon_operator_renorm}. Since tree normalization fixes the coefficient of $\mathcal{\widetilde M}_{pp}$, the admixture is a left eigenvector without an $\mathcal{\widetilde M}_{pp}$ component. Its counterterm \cref{eq:counterterm} is proportional to that component and vanishes, and since the collinear pole resides in $\mathcal{\widetilde M}_{pp}$ alone (\cref{sec:gluon_operator_renorm}), its one-loop matrix element is finite, as the gluon-in-gluon displays below confirm. An $\mathcal{O}(\epsilon)$ coefficient times a finite matrix element vanishes at $d=4$. This is why $c_{pp}=2$ can be used unchanged for gluon-in-gluon in every multiplicatively renormalizable combination. The index contracted inside the loop is different: as we discuss around \cref{eq:bmr_residual}, contracting it in two instead of $d-2$ directions drops an evanescent operator on the collinear pole and leaves the finite polynomial difference in the gluon matching kernel of refs.~\cite{Balitsky:2019krf,Balitsky:2021qsr}.

\subsection{One-loop results}

Beyond leading order we write every result as in \cref{eq:fexpansion}, for bare and renormalized quantities alike: the superscript $(0)$ denotes the tree-level term and $(1)$ the complete one-loop term, which we display with an explicit prefactor $\alphas/(2\pi)$.

\subsubsection{Conventions and definitions}
We use the single-subtraction notation
\begin{equation}
\int_a^b \mathrm{d}x\, \left[g(x)\right]_{(+c)}^{[a,b]} f(x) = \int_a^b\mathrm{d}x\,g(x)\left(f(x) - f(c)\right)\,,\label{eq:plusdef}
\end{equation}
that is, the plus distribution subtracts the test-function value at $x=c$ on the region $[a,b]$. We give the full definition, including the subtractions at $\pm\infty$ needed for the quasi-\PDF{}, in \cref{app:FT}.

In writing our one-loop results we make use of the regularized spin-averaged \DGLAP{} splitting functions \cite{Dokshitzer:1977sg,Gribov:1972ri,Altarelli:1977zs} without overall color factors:
\begin{align} \label{eq:DGLAP}
	C_F P_{qq}(x) &= C_F \left[ \frac{1+x^2}{[1-x]_{(+1)}^{[0,1]}} + \frac{3}{2}\delta(1-x) \right] =
		C_F \left[ \frac{1+x^2}{1-x} \right]^{[0,1]}_{(+1)}\,, \\
	C_A P_{gg}(x) &= 2 C_A \left[ \frac{x}{[1-x]_{(+1)}^{[0,1]}} + \frac{1-x}{x} + x(1-x) \right] + \delta(1-x)\frac{11
	C_A - 4 n_f T_R}{6}\,,\label{eq:DGLAP_gg}\\
	T_R P_{qg}(x) &= T_R \left[ x^2 + (1-x)^2 \right]\,, \label{eq:DGLAP_qg} \\
	C_F P_{gq}(x) &= C_F \left[ \frac{1+(1-x)^2}{x} \right]\,. \label{eq:DGLAP_gq}
\end{align}
We use these expressions as a cross-check of the pole contributions in our results.

For quasi-\PDF{}s we define the logarithm containing the renormalization scale as
\begin{equation}
	L_\mu^\mathcal{Q} = \log(\mu^2 / (4 (p^z)^2))\,,
	\label{eq:LmuQ}
\end{equation}
and for pseudo \PDF{}s we define the corresponding logarithm as
\begin{equation}
	L_\mu^\mathcal{P} = \log(e^{2\gamma_\text{E}}\tilde{z}^2 \mu^2 / 4)\,.
	\label{eq:LmuP}
\end{equation}

Results for the gluonic operator $\mathcal{G}^{\mu\nu\rho\sigma}$ are given in the tilde form factor basis of \cref{eq:tildeformfactors}, from which any index combination follows as in \cref{eq:Gcombination_Mpp_tilde,eq:Gcombination_ziiz_tilde}.

\subsubsection{Quark in quark}\label{sec:quarkinquark}

As a cross-check of our framework and of the results in the literature, we reproduce the existing one-loop results in dimensional regularization \cite{Izubuchi:2018srq,Chou:2022drv}. The calculations of refs.~\cite{Izubuchi:2018srq,Chou:2022drv} were carried out in Feynman gauge and separate the \IR{} and \UV{} poles, while we identify them as $\epsilon_\IR{}=\epsilon_\UV{}\equiv\epsilon$ (\cref{sec:conventions}).

For $\epsilon_\IR{}=\epsilon_\UV{}$, the light-cone \PDF{} vanishes perturbatively in dimensional regularization (\cref{sec:renormalization}), so the poles that remain in the interior region after subtracting the independently known \UV{} counterterm must be collinear and proportional to the \DGLAP{} kernel, as they are in \cref{eq:fquasiAssembled,eq:fpseudo_qq_bare}.

\begin{figure}[t]
	\centering
	\begin{subfigure}[b]{0.19\textwidth}
		\centering\includegraphics[width=\linewidth]{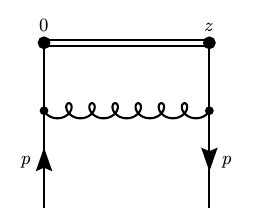}
		\caption{vertex}\label{fig:qq-vertex}
	\end{subfigure}
	\hfill
	\begin{subfigure}[b]{0.19\textwidth}
		\centering\includegraphics[width=\linewidth]{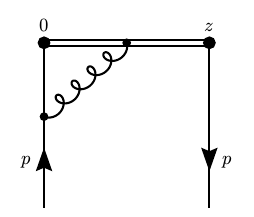}
		\caption{sail}\label{fig:qq-sail1}
	\end{subfigure}
	\hfill
	\begin{subfigure}[b]{0.19\textwidth}
		\centering\includegraphics[width=\linewidth]{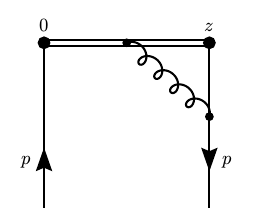}
		\caption{sail}\label{fig:qq-sail2}
	\end{subfigure}
	\hfill
	\begin{subfigure}[b]{0.19\textwidth}
		\centering\includegraphics[width=\linewidth]{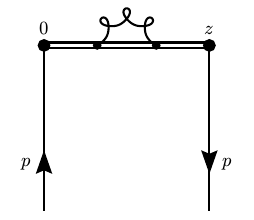}
		\caption{Wilson-line self-energy}\label{fig:qq-tadpole}
	\end{subfigure}
	\hfill
	\begin{subfigure}[b]{0.19\textwidth}
		\centering\includegraphics[width=\linewidth]{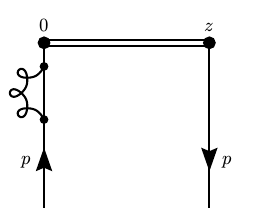}
		\caption{self-energy}\label{fig:qq-selfenergy}
	\end{subfigure}
	\caption{The one-loop diagrams contributing to the quark-in-quark matrix element. The
	incoming and outgoing quarks carry momentum $p$, and the double line denotes the gauge link
	$W(z,0)$ joining the field insertions at $0$ and $z$. The gluon connects (a) the two quark legs
	(vertex), (b,c)~a quark leg and the gauge link (sail), (d)~the gauge link with itself (Wilson-line
	self-energy), and (e)~a quark leg with itself (quark self-energy).}
	\label{fig:qq-diagrams}
\end{figure}

Only five diagrams contribute (\cref{fig:qq-diagrams}): the vertex (\subref{fig:qq-vertex}), sails (\subref{fig:qq-sail1},~\subref{fig:qq-sail2}), Wilson-line self-energy (\subref{fig:qq-tadpole}), and quark self-energy (\subref{fig:qq-selfenergy}).

At one loop we have, after color and spin averaging:
\begin{align}
\mathcal{Q}_{\gamma^0,\text{ET}}^{\text{bare}}(\zt,p^z) = {} & \left[\frac{d}{2}p^z\right] e^{i p\cdot z} + \left[\frac{d}{2}p^z\right]   i g_s^2 C_F
	\frac{d-2}{d-4} \frac{\zt^2}{p\cdot z} \mu^{4-d} \mathcal{\widetilde{Q}}^{(1)}_{\gamma^0,\text{ET}}(\zt,p^z) + \mathcal{O}(g_s^4) \,, \\
	\mathcal{\widetilde Q}^{(1)}_{\gamma^0,\text{ET}}(\zt,p^z)  = {} & {-}\left( e^{i p\cdot z} I^\pm(1,0,0,1) - I_z^\pm(1,0,0,1) \right) + \left( \frac{i}{2}+\frac{1}{2 p\cdot z} \right)  I_z(1,0,0,0)
	\nonumber \\
	& +
	\left( \frac{i}{d-3} - \frac{1}{2 p\cdot z} + \frac{p\cdot z}{(d-4)(d-3)} \right)  I_z(0,1,0,0)	\,.
	\end{align}
The term proportional to $I(1,0,0,1)$ constitutes the \enquote{virtual} corrections, while the rest are the \enquote{real} corrections. The diagrams that make up each set depend on the gauge. In $R_\xi$ gauge the real corrections are the vertex, the Wilson-line self-energy and the sail contributions in which the gluon connects one half of the Wilson line, split at infinity, to the quark on the other side, and the virtual corrections are the sail contributions within one side; the quark self-energy vanishes there for $\epsilon_\IR{}=\epsilon_\UV{}$. The axial gauge with auxiliary vector in the $\hat{z}$ direction eliminates every Wilson-line diagram, so that only the vertex survives in the real corrections and only the quark self-energy, which does not vanish in this gauge, in the virtual corrections.

The integrals that depend on $\pm i\eta$ enter only through the combination $e^{i p\cdot z} I^\pm(1,0,0,1) - I_z^\pm(1,0,0,1) \equiv \diffmaster{}$ of \cref{sec:oneloopints}, so the matrix element is independent of the $i\eta$ regularization of the linear propagator, although the real and virtual corrections separately are not. This cancellation of the linear-propagator (eikonal) singularity between the real and virtual families is the equal-time, spacelike-Wilson-line counterpart of the real/virtual cancellation of the light-like rapidity singularity, discussed in Sec.~13.14 of ref.~\cite{Collins:2011zzd} (see also ref.~\cite{Collins:2008ht}).

We test this property by comparing the $R_\xi$ and axial gauge results, which carry different prescriptions for the linear propagator. In $R_\xi$ gauge the prescription is uniquely fixed, since the $\pm i0$ is set by the direction of momentum flow along the Wilson line, while in axial gauge the spurious $1/(r\cdot p)$ pole is not fixed by the gauge condition and we use the principal value (\cref{sec:conventions,app:feynman}). A quantity can therefore agree between the two gauges only if it is independent of that prescription. The sum of real and virtual corrections does agree, while the real and virtual corrections separately are not gauge invariant.

After inserting the master integral solutions and converting to the \MSBAR{} coupling with \cref{eq:g2expansion}, the result reads
\begin{align}\label{eq:Q1quarkinquark}
\mathcal{Q}_{\gamma^0,\text{ET}}^{\text{bare}}& (\zt,p^z) =
\left[\frac{d}{2}p^z\right] e^{i p\cdot z} + \frac{\alphas}{2\pi} \left[\frac{d}{2}p^z\right] \cdot
\frac{1}{2}C_F 2^{d-2} e^{\epsilon \gamma_\text{E}}(d-2)(\zt^2)^{2-d/2}\Gamma(d/2-1) \mu^{2\epsilon} \nonumber \\
		\times & \bigg[
			\frac{1+i p\cdot z}{2 (d-4) (p\cdot z)^2}+e^{i p\cdot z} \bigg(-\frac{(i p\cdot z)^d (\Gamma (3-d)-\Gamma (3-d,i p\cdot z))}{(d-4) (p\cdot z)^4} \nonumber \\
			\hfill & \qquad\qquad -\frac{1}{2 (d-4) (p\cdot z)^2}+\frac{i}{(d-4) (d-3) p\cdot z}+\frac{1}{(d-4)^2 (d-3)}\bigg)
		\bigg]\,.
\end{align}
We reproduce the result in eq.~(44) of ref.~\cite{Izubuchi:2018srq} and eq.~(22) of ref.~\cite{Chou:2022drv}.\footnote{We compare with arXiv version v2 of ref.~\cite{Izubuchi:2018srq}, whose equation numbers we quote throughout, and with version v4 of ref.~\cite{Chou:2022drv}, where there is a typo in the incomplete Gamma function which should read $\Gamma(0,-i p^z z)$ instead of $\Gamma(0,i p^z z)$.}

Although the individual rational terms and the incomplete gamma function in \cref{eq:Q1quarkinquark} diverge as $1/\nu^2$ and $1/\nu$ at $\nu=0$, these kinematic poles cancel in the sum. At fixed $\zt^2$, the matrix element is therefore regular at $\nu=0$, admitting a Taylor expansion in powers of $\nu$. This regularity reflects the short-distance operator product expansion, where the coefficient of $\nu^n$ is the forward matrix element of a local twist-2 operator, a Mellin moment of the light-cone \PDF{}, multiplied by a perturbative Wilson coefficient~\cite{Izubuchi:2018srq,Karpie:2018zaz}. Because these local twist-2 operators are finite after renormalization, no negative powers of $\nu$ can appear. In particular, the leading term at $\nu=0$ isolates the local vector current $\overline{\psi}\gamma^0\psi$, whose forward matrix element is the net quark number protected by current conservation. In the ratio scheme this $\nu=0$ value is divided out, so that the reduced correlator is unity at $\nu=0$ and the matching kernel preserves the net quark number (\cref{sec:symmetries}).

\paragraph{Quasi-\PDF{}.}

The quasi-\PDF{} is obtained by Fourier transforming the bare matrix element, \cref{eq:Q1quarkinquark}, using the transforms of \cref{app:FT}. In contrast to the pseudo-\PDF{}, the support of the quasi-\PDF{} extends over the entire real axis and carries the boundary delta functions $\delta^\pm(1/x)/x^2$ at $x=\pm\infty$. We discuss these boundary terms before assembling the full distribution.

We find a net, symmetric ultraviolet pole $-3/(4\epsilon)$ on each of the boundary deltas $\delta^\pm(1/x)/x^2$, which reproduces the boundary pole of the $x$-space \MSBAR{} quasi-\PDF{} counterterm of ref.~\cite{Izubuchi:2018srq}. Note that this counterterm subtracts the boundary pole together with the contact pole, whereas we subtract only the contact pole and let the boundary pole drop out of the matching convolution (\cref{sec:boundary}). The boundary pole is ultraviolet because $x\to\pm\infty$ at fixed $p^z$ is the limit $\zt\to0$, and it survives renormalization (\cref{sec:boundary}). The gluon boundary poles of \cref{eq:quasi_gg_towers} arise in the same way.

The individual transforms of \cref{app:FT} also generate derivatives of the boundary deltas, because a $1/\nu^n$ kernel transforms into a tail that grows as $\lvert x\rvert^{n-1}$. However, the $1/\nu^n$ terms cancel in \cref{eq:Q1quarkinquark}, and the derivative deltas therefore cancel identically in the assembled quasi-\PDF{}. The value deltas survive because they come not from the $\nu$ dependence at fixed $\zt^2$ but from the factor $(\zt^2\mu^2)^\epsilon$ (\cref{sec:boundary}).

Assembling the regions into a single distribution on $x\in\mathbb{R}$, the bare quasi-\PDF{} coefficient reads
\begin{align}
	\label{eq:fquasiAssembled}
	\tilde{f}_{q/q}^{(1),\text{quasi,bare}}(x) = {} & \frac{\alphas}{2\pi}\,C_F\,\frac{d}{4}\Bigg[
	\frac{1}{\epsilon}\left(\frac{3}{2}\delta(1-x) - P_{qq}(x)\right)
	+ \left(\frac{5}{2} + \frac{3}{2}L_\mu^\mathcal{Q}\right)\delta(1-x) \nonumber \\
	& + \left( -\frac{3}{4\epsilon} - \frac{5}{4} - \frac{3}{4}L_\mu^\mathcal{Q} \right) \left[ \frac{\delta^+(1/x)}{x^2} + \frac{\delta^-(1/x)}{x^2} \right] \nonumber \\
	& + \begin{dcases}
	\left( 1 + \frac{1+x^2}{1-x}\log\frac{x}{x-1} + \frac{3}{2x} \right)_{(+1)}^{[1,\infty)} - \left( \frac{3}{2x} \right)_{(+\infty)}^{[1,\infty)} & x>1 \\[1ex]
	\left( 1 + \frac{1+x^2}{1-x}\big(\log(1-x)+\log(x)-1-L_\mu^\mathcal{Q}\big) \right)_{(+1)}^{[0,1]} & 0<x<1 \\[1ex]
	\left( -1 + \frac{3}{2(1-x)} - \frac{1+x^2}{1-x}\log\frac{-x}{1-x} \right)_{(+1)}^{(-\infty,0]} - \left( \frac{3}{2(1-x)} \right)_{(+(-\infty))}^{(-\infty,0]} & x<0
	\end{dcases}
	\Bigg]\,,
\end{align}
where $d/4 = \operatorname{Tr}\mathbf{1}/4 = 1-\epsilon/2$ is the Dirac-trace factor from \cref{eq:ftree_qq}. We keep this factor factored out and unexpanded. In the matching relation it cancels against the target quark tree-level distribution $f_{q/q}^{(0)}$, and setting $d/4\to 1$ recovers the standard four-dimensional \MSBAR{} distribution.

The other $1/\epsilon$ poles in the bare distribution have support entirely in the interior region. The $-P_{qq}(x)/\epsilon$ term is the collinear singularity removed by matching to the light-cone \PDF{} and the $\tfrac{3}{2}\delta(1-x)/\epsilon$ part is the \UV{} pole removed by the  multiplicative renormalization of the bilocal operator. We exactly reproduce eq.~(56) of ref.~\cite{Izubuchi:2018srq}, which displays the \MSBAR{} counterterm $\delta\tilde q^{(1)}$ together with the renormalized quasi-\PDF{}. Our bare result \cref{eq:fquasiAssembled} corresponds to their sum.

\paragraph{Pseudo-\PDF{}.}

Integrating the matrix element in \cref{eq:Q1quarkinquark} with respect to the Ioffe time gives the bare pseudo \PDF{}
\begin{align}
	\tilde{f}_{q/q}^{(1),\text{pseudo,bare}}(x,z^2) = {} &  \frac{\alphas}{2\pi} C_F \frac{d}{4} \Bigg[
	\frac{1}{\epsilon} \left( -P_{qq}(x) \, \theta(0\leq x \leq 1) + \frac{3}{2}\delta(1-x)  \right) \nonumber \\
	& + \left( 2-2x -4\left[\frac{\log(1-x)}{1-x}\right]^{[0,1]}_{(+1)} - (1+\LmuP{}) P_{qq}(x) \right) \theta(0\leq x \leq 1) \nonumber \\
	& + \left(\frac{3}{2} + \frac{3}{2}\LmuP{} \right)\delta(1-x)
	\Bigg] \,. \label{eq:fpseudo_qq_bare}
\end{align}
For $\epsilon_\IR{}=\epsilon_\UV=\epsilon$ this expression agrees with ref.~\cite{Izubuchi:2018srq}. The matching with the light-cone \PDF{} removes the $\frac{1}{\epsilon} P_{qq}(x)$ term, while the remaining $\frac{1}{\epsilon} \frac{3}{2} \delta(1-x)$ is removed by an additional composite operator $\UV{}$ renormalization.

\subsubsection{Gluon in quark}

\begin{figure}[t]
	\centering
	\begin{subfigure}[b]{0.19\textwidth}
		\centering\includegraphics[width=\linewidth]{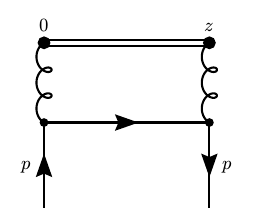}
		\caption{uncrossed}\label{fig:gq-direct}
	\end{subfigure}
	\hspace{0.06\textwidth}
	\begin{subfigure}[b]{0.19\textwidth}
		\centering\includegraphics[width=\linewidth]{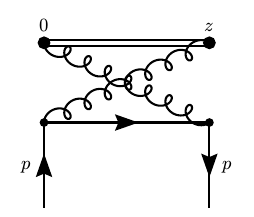}
		\caption{crossed}\label{fig:gq-crossed}
	\end{subfigure}
	\caption{The two one-loop diagrams contributing to the gluon-in-quark matrix element. The external
	quark carries momentum $p$, and the gluon bilocal operator inserts two field strengths at $0$ and
	$z$ (joined by the gauge link $W(z,0)$, double line), each emitting a gluon onto the internal
	quark line. In the crossed diagram (\subref{fig:gq-crossed}) the two operator gluons attach to
	the opposite ends of the internal quark line. Both diagrams must be summed for gauge invariance.}
	\label{fig:gq-diagrams}
\end{figure}

For the gluon-in-quark matrix element, two one-loop diagrams contribute (\cref{fig:gq-diagrams}), the uncrossed (\subref{fig:gq-direct}) and crossed diagrams (\subref{fig:gq-crossed}). %

We sum over colors and spins with averaging factors $1/N_c$ and $1/2$, include both diagrams, express the tensor components in terms of the master integral $I_z(1,0,0,0)$, and verify gauge invariance. Factoring out a common prefactor from the tilde form factors,
\begin{equation}
\mathcal{\widetilde M}_k^g = \frac{1}{2} g_s^2 C_F I_z(1,0,0,0) \mathcal{\widetilde{M}}_k^*\,,
\label{eq:Mqgstar}
\end{equation}
we obtain
\begin{align}
	\mathcal{\widetilde M}_{pp}^*(\nu,z^2) & =
	\frac{d }{(d-4)}\left[\frac{\left(\nu^2-2\right)}{4 \nu }\left( e^{-i \nu}-e^{i\nu} \right)-i\right]\,, \label{eq:Mtilde_gq_pp_reduced} \\
	\mathcal{\widetilde M}_{gg}^*(\nu,z^2) & =
	\frac{d }{2 \nu}\left(e^{i \nu}-e^{-i \nu}\right)-i d\,, \label{eq:Mtilde_gq_gg_reduced} \\
	\mathcal{\widetilde M}_{zz}^*(\nu,z^2) & =
	\frac{(d-2) d }{2}\left[\frac{1}{2 \nu}\left(e^{i \nu}-e^{-i \nu}\right)-i \right]\,, \label{eq:Mtilde_gq_zz_reduced} \\
	\mathcal{\widetilde M}_{pz}^*(\nu,z^2) & = \mathcal{\widetilde M}_{zp}^*(\nu,z^2) = \frac{d}{4 \nu} \left[ e^{i \nu} \left(1-i \nu\right) - i e^{-i \nu} \left(\nu-i\right)\right]\,, \label{eq:Mtilde_gq_zp_reduced} \\
	\mathcal{\widetilde M}_{ppzz}^*(\nu,z^2) & = \frac{i d}{2} \left[e^{i \nu}+e^{-i\nu}-2\right]\,. \label{eq:Mtilde_gq_ppzz_reduced}
\end{align}

\paragraph{Quasi-\PDF{}.}

Fourier-transforming each tilde form factor with respect to the spatial separation $\zt$ at fixed $p^z$ gives the bare quasi-\PDF{} components $x\tilde{f}^{(1),\text{quasi,bare}}_{g/q,k}$, analogous to \cref{eq:pseudo_tilde_expansion} for the pseudo case. As for the gluon in a gluon, we quote the momentum-weighted distribution $x\tilde{f}$, which is the transform of the correlator itself (\cref{eq:fgquasi}), and we keep its boundary terms at $x=\pm\infty$. As in the quark-in-quark channel, the open external quark line contributes a Dirac trace $\operatorname{Tr}\mathbf{1} = d$, which we factor out as $d/4$; this factor cancels against the target quark tree distribution in the matching relation. Every component is even in $x$. We quote the leading-twist component for $x>0$, and the result for $x<0$ follows from $x\to-x$. For the leading-twist form factor we find
\begin{align}
	x\tilde{f}^{(1),\text{quasi,bare}}_{g/q,\mathcal{M}_{pp}}(x,p^z) = {} & \frac{\alphas C_F}{2\pi}\frac{d}{4} \Bigg\{ -\frac{2}{3}\left(\frac{1}{\epsilon}+L_\mu^\mathcal{Q}\right)\frac{\delta^+(1/x)}{x^2} - \frac{4}{3}\left[\frac{1}{x}\right]^{[1,\infty)}_{(+\infty)} \nonumber \\
	& + \begin{cases}
		x+\frac{4}{3x}+\frac{1}{2}\, x P_{gq}(-x) \log \left(\frac{x+1}{x}\right)+\frac{1}{2}\, x P_{gq}(x) \log \left(\frac{x-1}{x}\right) & x>1 \\
		 1-\frac{3x}{2}+\frac{5x^2}{4} +\frac{1}{2}\, x P_{gq}(-x) \log \left(\frac{x+1}{x}\right) \\
		 \quad+\frac{1}{2}\left[\frac{1}{\epsilon}+L_\mu^\mathcal{Q}+\frac{1}{2}-\log(1-x)-\log(x)\right] x P_{gq}(x)& 0<x<1
	\end{cases} \Bigg\}\,,
	\label{eq:fquasi_gq_pp_bare}
\end{align}
where $x P_{gq}(x) = 1+(1-x)^2$ and $x P_{gq}(-x) = -1-(1+x)^2$. The transforms $x\tilde{f}_{g/q,k}$ are defined before the contraction of \cref{eq:pseudo_tilde_expansion}, that is, before the tensor coefficients $c_k$ and the matching minus sign enter, which is why the collinear pole of \cref{eq:fquasi_gq_pp_bare} carries a coefficient $\frac{1}{2}$ and no overall minus sign. For the leading-twist index combination \cref{eq:Gcombination_Mpp_tilde} the coefficient is $c_{pp}=2$ for $i\in\{1,2\}$ and $c_{pp}=d-2$ once the transverse sum over $i$ is continued to the $d-2$ transverse directions (\cref{sec:tree}), so that the physical distribution $x\tilde{f}^\text{full} = -\sum_k c_k\, x\tilde{f}_k$ carries the collinear pole $-x P_{gq}(x)/\epsilon$ through its leading-twist term $-(d-2)\,x\tilde{f}_{g/q,\mathcal{M}_{pp}}$. This pole is removed by the matching.

At large $\lvert x\rvert$ the exterior approaches $-4/(3\lvert x\rvert)$. This tail is the transform of the short-distance logarithm, which is fixed at leading power by the splitting kernel $P_{gq}$ alone (\cref{sec:symmetries}), and its coefficient is minus the integral $\int_{-1}^{1}\mathrm{d}x\,\frac{1}{2}\lvert x\rvert P_{gq}(\lvert x\rvert) = \frac{4}{3}$ of the interior collinear pole. In $d$ dimensions the tail falls off as $\lvert x\rvert^{-1-2\epsilon}$, and its integral places the boundary poles of \cref{eq:fquasi_gq_pp_bare} at $x=\pm\infty$, which cancel the interior pole in $\int\mathrm{d}x\,x\tilde{f}_{g/q}$, since the local gluon operator receives no one-loop correction in dimensional regularization (\cref{sec:boundary}). These boundary terms are the one-loop mixing of the local gluon operator into the quark momentum operator. In the matching convolution they probe the quark momentum fraction, which the momentum sum rule bounds, so they drop out of the matching in the same way as the gluon-in-gluon boundary terms (\cref{sec:boundary}).

	The auxiliary form factors give the remaining bare quasi-\PDF{} components, which we quote on the whole real line,
	\begin{align}
		x\tilde{f}^{(1),\text{quasi,bare}}_{g/q,\mathcal{M}_{pz}}(x,p^z) = x\tilde{f}^{(1),\text{quasi,bare}}_{g/q,\mathcal{M}_{zp}}(x,p^z) = {} & \frac{\alphas C_F}{2\pi}\frac{d}{4}\Bigg\{\frac{1-x^2}{2}\, \theta(1-x^2) \nonumber \\
		& \qquad - \frac{1}{3}\left[\frac{\delta^+(1/x)}{x^2}+\frac{\delta^-(1/x)}{x^2}\right]\Bigg\}
		\label{eq:fquasi_gq_zp_bare} \\
		x\tilde{f}^{(1),\text{quasi,bare}}_{g/q,\mathcal{M}_{zz}}(x,p^z) = x\tilde{f}^{(1),\text{quasi,bare}}_{g/q,\mathcal{M}_{gg}}(x,p^z) = {} & \frac{\alphas C_F}{2\pi}\frac{d}{4}\Bigg\{-(1-\lvert x \rvert)^2\, \theta(1-x^2) \nonumber \\
		& \qquad + \frac{1}{3}\left[\frac{\delta^+(1/x)}{x^2}+\frac{\delta^-(1/x)}{x^2}\right]\Bigg\}
		\label{eq:fquasi_gq_zz_bare} \\
		x\tilde{f}^{(1),\text{quasi,bare}}_{g/q,\mathcal{M}_{ppzz}}(x,p^z) = {} & \frac{\alphas C_F}{2\pi}\frac{d}{4}\Bigg\{2(\lvert x \rvert-1)\, \theta(1-x^2) \nonumber \\
		& \qquad + \frac{\delta^+(1/x)}{x^2}+\frac{\delta^-(1/x)}{x^2}\Bigg\}\,.
	\label{eq:fquasi_gq_ppzz_bare}
	\end{align}
	On $[-1,1]$ they coincide with the pseudo-\PDF{} result \cref{eq:fpseudo_gq_aux}, and they differ from it only by finite boundary terms at $x=\pm\infty$, whose coefficient is minus one half of the integral of the interior, as for the auxiliary gluon-in-gluon components $\mathcal{\widetilde M}_{zz}$ and $\mathcal{\widetilde M}_{gg}$. Unlike $x\tilde{f}_{\mathcal{M}_{pp}}$, the auxiliary components carry neither an ultraviolet nor an infrared pole, and hence no scale logarithm (\cref{eq:fpseudo_gq_aux,eq:fpseudo_gq_zz_bare,eq:fpseudo_gq_ppzz_bare}), so by \cref{eq:exterior_constraint} they vanish at every finite $|x|>1$, and outside $[-1,1]$ they contribute only the boundary terms at $x=\pm\infty$ above. At finite $|x|>1$, the gluon-in-quark quasi-distribution is therefore the same for every gluon operator up to the coefficient $c_{pp}$, unlike in the gluon-in-gluon channel, where the auxiliary form factors carry sector-dependent tails (\cref{eq:tail_classification}).

\paragraph{Comparison with ref.~\cite{Ji:2025cbb}.}

Ref.~\cite{Ji:2025cbb} gives the forward unpolarized quasi kernel of this channel for the transverse-trace operator $g_\perp^{\mu\nu}F_{z\mu}F_{\nu z}$ and for the singlet quark distribution, so that in a quark state the kernel enters, in their notation, as $\frac{1}{2}\left[C_{gq}(x,1)-C_{gq}(x,-1)\right]$. Since the gluon operator has mass dimension four and the quark operator three, the coordinate-space kernel of this channel carries a factor $1/\zt$ (\cref{sec:renormalization}), and its transform to momentum space is ambiguous unless the singularity at $\zt=0$ is regulated. Ref.~\cite{Ji:2025cbb} regulates it by keeping the $d$-dimensional factor $(\zt^2)^{\epsilon}/\zt$ unexpanded in the transform and expanding in $\epsilon$ only afterwards. Because our transforms are exact in $d$ throughout, we implement this prescription by construction.

In our basis, the transverse-trace operator is the combination \cref{eq:Mziiz} with the transverse sum taken over the $d-2$ transverse directions of dimensional regularization, which multiplies the combination by $(d-2)/2$.
We assemble the $x$-weighted combination $x\tilde{f}_{g/q} = -\sum_k c_k\, x\tilde{f}_{g/q,k}$ with these coefficients, set $d/4\to1$ and subtract the collinear pole, and find that our result agrees with the matching kernel of ref.~\cite{Ji:2025cbb} exactly in all four regions. Here the transverse sum has to run over all $d-2$ directions, as required in \cref{sec:tree} for the gluon-in-quark channel. With two directions, the evanescent factor $(d-2)/2$ would remain on the collinear pole and leave an extra finite term $-\lvert x\rvert P_{gq}(\lvert x\rvert)$ for $\lvert x\rvert<1$, in units of $\alphas C_F/(2\pi)$.

\paragraph{Comparison with refs.~\cite{Wang:2017qyg,Wang:2019tgg}.}

For the uncrossed diagram alone, the auxiliary form factors retain spurious exterior support at $|x|>1$, which cancels once the crossed diagram is included to form the gauge-invariant matrix element. This explains why $x\tilde{f}$ from eq.~(55) of ref.~\cite{Wang:2017qyg}, which evaluated only the uncrossed diagram for the operator $-\mathcal{G}^{ziiz}$ (\cref{fn:wang_sign}) with a gluon-mass regulator, approaches a constant at large $|x|$ instead of falling off.
We can nevertheless compare this uncrossed contribution in the regions $x>1$ and $x<-1$, which are \IR{} and \UV{} finite and therefore regulator independent. When we calculate \cref{eq:Mziiz} for the uncrossed diagram and Fourier transform it, we reproduce eq.~(55) of ref.~\cite{Wang:2017qyg} exactly, up to the overall sign explained in \cref{fn:wang_sign}. We have also checked \cref{eq:Mziiz} against an explicit calculation of $\sum_{i=1,2}\mathcal{G}^{ziiz}_{\text{ET}}$ with the indices inserted first. The agreement is nontrivial at the level of the individual form factors: $2\,\mathcal{\widetilde M}_{pp}$ alone reproduces the logarithms and the linear term but leaves a constant $-\tfrac{3}{2}$ instead of $-\tfrac{5}{2}$ for $x>1$, and the remainder comes from the four auxiliary form factors $\mathcal{\widetilde M}_{zz}$, $\mathcal{\widetilde M}_{gg}$, $\mathcal{\widetilde M}_{zp}$ and $\mathcal{\widetilde M}_{pz}$.

The power-law tail of $x\tilde{f}_{g/q}$ in the exterior region, carried entirely by $\mathcal{M}_{pp}$, is a property of the \MSBAR{} kernel: the gluon matching of ref.~\cite{Balitsky:2019krf} and the forward limit of ref.~\cite{Yao:2022vtp} both retain $|x|>1$ support, and for $|x|>1$ our result agrees pointwise with the kernel of ref.~\cite{Ji:2025cbb}. The one-loop quasi-distributions of ref.~\cite{Wang:2019tgg} instead approach constants at large $|x|$, $\pm\alphas C_F/(4\pi)$ in $x\tilde{f}_{g/q}$ as $x\to\pm\infty$, and ref.~\cite{Balitsky:2019krf}, checking the gluon-in-gluon result of ref.~\cite{Wang:2019tgg} against its own exterior constraint, found a constant difference of $\mp\frac{2}{3}\frac{\alphas C_A}{2\pi}$. Both constants come from the uncrossed diagrams alone, which is all that ref.~\cite{Wang:2019tgg} displays, leaving the crossed diagrams to the rule $\tilde{f}(x)=-\tilde{f}(-x)$. The constants are odd in $x$ and cancel once the crossed diagrams are added (\cref{sec:symmetries}), in the same way as the spurious exterior support of the uncrossed diagram above. After crossing, the exterior of ref.~\cite{Wang:2019tgg} agrees with ours: exactly for the gluon in a quark, and for the gluon in a gluon pointwise with our result for the same eigenvalue sector, $Z_{\perp\perp}Z_{\parallel\perp}$ (\cref{eq:tail_classification}).

\paragraph{Pseudo-\PDF{}.}

Inserting the master integral and Fourier transforming with respect to the Ioffe time, we obtain the momentum-weighted components $x\tilde{f}^{(1),\text{pseudo,bare}}_{g/q,\mathcal{M}_k}$. As for the gluon in a gluon, the support is confined to $x\in[-1,1]$ exactly in $\epsilon$ and every component is even in $x$, so we quote the region $0<x<1$, and the results for $-1<x<0$ follow from $x\to-x$. We find
\begin{align}
	x\tilde{f}^{(1),\text{pseudo,bare}}_{g/q,\mathcal{M}_{pp}}(x,z^2) = {} & \frac{\alphas C_F}{2\pi}\frac{d}{4}\,
		 \frac{1}{2}\, x P_{gq}(x) \left[ \frac{1}{\epsilon} + \LmuP{} \right]\,.
	\label{eq:fpseudo_gq_pp_bare}
\end{align}
The $1/\epsilon$ pole here is of pure \IR{} origin and is removed entirely by the matching to the light-cone
\PDF{}. With $c_{pp}=2$, the pole coefficient is the gluon-quark splitting kernel
$\mathcal{B}_{gq}(x) = x\,P_{gq}(x) = 1+(1-x)^2$ obtained from the nucleon matching relation of
refs.~\cite{Balitsky:2019krf,Balitsky:2021qsr}, whose Ioffe-time distributions are momentum weighted as ours.

The remaining tilde form factors lead to
\begin{align}\label{eq:fpseudo_gq_aux}
	x\tilde{f}^{(1),\text{pseudo,bare}}_{g/q,\mathcal{M}_{pz}}(x,z^2) = x\tilde{f}^{(1),\text{pseudo,bare}}_{g/q,\mathcal{M}_{zp}}(x,z^2) = {} & \frac{\alphas C_F}{2\pi}\frac{d}{4}\,\frac{1-x^2}{2}
	\\
	x\tilde{f}^{(1),\text{pseudo,bare}}_{g/q,\mathcal{M}_{zz}}(x,z^2) = x\tilde{f}^{(1),\text{pseudo,bare}}_{g/q,\mathcal{M}_{gg}}(x,z^2) = {} & -\frac{\alphas C_F}{2\pi}\frac{d}{4}\,(1-x)^2
	\label{eq:fpseudo_gq_zz_bare} \\
	x\tilde{f}^{(1),\text{pseudo,bare}}_{g/q,\mathcal{M}_{ppzz}}(x,z^2) = {} & \frac{\alphas C_F}{2\pi}\frac{d}{4}\,2(x-1)
	\label{eq:fpseudo_gq_ppzz_bare}
\end{align}
for $0<x<1$. As in the quasi case, these are the transforms of the individual form factors, and the physical distribution carries the pole $-x P_{gq}(x)/\epsilon$ through its leading-twist term $-(d-2)\,x\tilde{f}_{g/q,\mathcal{M}_{pp}}$.

\subsubsection{Gluon in gluon}

\begin{figure}[t]
	\centering
	\begin{subfigure}[b]{0.19\textwidth}
		\centering\includegraphics[width=\linewidth]{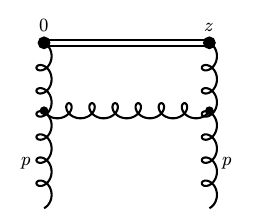}
		\caption{vertex}\label{fig:gg-vertex}
	\end{subfigure}
	\hfill
	\begin{subfigure}[b]{0.19\textwidth}
		\centering\includegraphics[width=\linewidth]{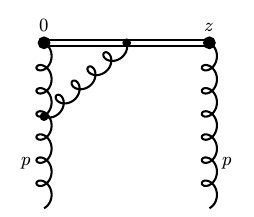}
		\caption{sail}\label{fig:gg-sail}
	\end{subfigure}
	\hfill
	\begin{subfigure}[b]{0.19\textwidth}
		\centering\includegraphics[width=\linewidth]{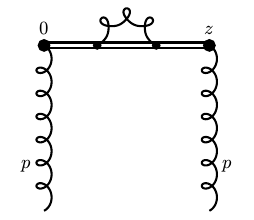}
		\caption{Wilson-line self-energy}\label{fig:gg-tadpole}
	\end{subfigure}
	\hfill
	\begin{subfigure}[b]{0.19\textwidth}
		\centering\includegraphics[width=\linewidth]{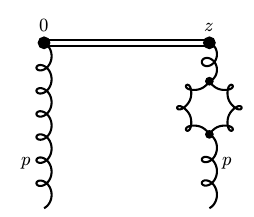}
		\caption{self-energy}\label{fig:gg-selfenergy}
	\end{subfigure}
	\hfill
	\begin{subfigure}[b]{0.19\textwidth}
		\centering\includegraphics[width=\linewidth]{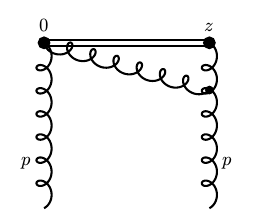}
		\caption{operator vertex}\label{fig:gg-opvertex}
	\end{subfigure}
	\caption{A representative sample of the one-loop diagrams contributing to the gluon-in-gluon
	matrix element. The incoming and outgoing gluons carry momentum $p$, and the double line denotes
	the gauge link $W(z,0)$ joining the field-strength insertions at $0$ and $z$. We show
	(\subref{fig:gg-vertex})~a triple-gluon vertex correction spanning the two external legs,
	(\subref{fig:gg-sail})~a sail joining a leg to the gauge link, (\subref{fig:gg-tadpole})~a
	Wilson-line self-energy, (\subref{fig:gg-selfenergy})~a gluon self-energy on a leg, and
	(\subref{fig:gg-opvertex})~the non-abelian two-gluon operator vertex. The gluon self-energy
	also receives tadpole, ghost- and quark-loop contributions, which are not shown.}
	\label{fig:gg-diagrams}
\end{figure}

The gluon-in-gluon matrix element receives contributions from $68$ diagrams in total, organized by the number of gluons emitted from the Wilson line into $36$, $20$, and $12$ diagrams with zero, one, and two emissions, respectively. We show a representative sample of distinct one-loop topologies in \cref{fig:gg-diagrams}. Many of these diagrams vanish for massless on-shell external states, or when projected onto the form-factor basis; our count represents the diagrams generated by \qgraf{} before on-shell conditions and projections are applied. In $R_\xi$ gauge, the external-leg self-energy corrections are scaleless on shell and integrate to zero in dimensional regularization ($\epsilon_\IR=\epsilon_\UV$). Quark loops enter the gluon-in-gluon matrix element at one loop only through external-leg corrections, so the results for the bare matrix elements below are proportional to $C_A$, with no $n_f$ contributions. The $n_f$ dependence enters exclusively through the $\beta_0$ term in the renormalization counterterm \cref{eq:counterterm}, which is proportional to the tree-level form factor and therefore contributes only to the endpoint contact terms $\delta(1\mp x)$, leaving the exterior tails in \cref{eq:tail_classification} independent of $n_f$.

In axial gauge with auxiliary vector $r$ along $\hat{z}$, the non-vanishing scale $(r\cdot p)^2$ prevents the external-leg self-energy from vanishing on shell. As an independent check of our setup, we calculated this self-energy by projecting onto its two transverse tensor structures~\cite{Leibbrandt:1987qv}, reproduced the coefficients of that reference, and verified that its inclusion in the complete axial-gauge matrix element agrees with the $R_\xi$ result.

\paragraph{Matrix elements.}

After summing all diagrams and projecting onto the scalar form factors, the complete bare one-loop gluon-in-gluon matrix element reduces to the three master integrals of \cref{sec:oneloopints}. The virtual corrections vanish identically for $\mathcal{M}_{zz}$ and $\mathcal{M}_{gg}$, and the dependence on the gauge parameter $\xi_G$ cancels in the sum of diagrams. We have also computed these matrix elements independently in an axial gauge with auxiliary vector in the $\hat{z}$ direction, and find agreement with the $R_\xi$ result. We quote the tilde form factors of \cref{eq:tildeformfactors}, averaged over color with a factor $1/(N_c^2-1)$ and over the $n_\text{pol}=2$ polarizations as at tree level. We write
\begin{equation}\label{eq:Mtilde_gg_oneloop_def}
	\mathcal{\widetilde M}_k^{g,\text{bare}}(\nu,z^2) = \mathcal{\widetilde M}_k^{g,(0)}(\nu,z^2) + \frac{1}{2}\, g_s^2\, C_A\, \mu^{4-d}\, \mathcal{\widetilde M}_k^{g,*}(\nu,z^2) + \mathcal{O}(g_s^4)\,,
\end{equation}
where $\mathcal{\widetilde M}_{pp}^{g,(0)} = (p^z)^2\mathcal{M}_{pp}^{g,(0)}$ (recalling the general definitions in \cref{eq:tildeformfactors}, with $\mathcal{M}_{pp}^{g,(0)}$ given in \cref{eq:Mpp_g_tree}) is the only non-vanishing tree-level tilde form factor, and the one-loop coefficients $\mathcal{\widetilde M}_k^{g,*}$ receive contributions from all three master integrals of \cref{sec:oneloopints}.
We find
\begin{align}\label{eq:Mtilde_gg_ioffe}
	\mathcal{\widetilde M}_{pp}^{g,*} = {} & -\frac{2 i \nu}{d-4} \left( \diffmaster{} - e^{i\nu}\,\diffmastertwo{} \right) \nonumber \\
	& + \frac{I_z(1,0,0,0)}{(d-4)^2(d-3)\,\nu^2} \Big[ -4i(d-1)(d-3)(d-4)\left(e^{i\nu}+e^{-i\nu}-2\right) \nonumber \\
	& \qquad - 2d(d-3)(d-4)\,\nu\left(e^{i\nu}-e^{-i\nu}\right) \nonumber \\
	& \qquad + i(d-3)(d-4)\,\nu^2\left((d+2)\left(e^{i\nu}+e^{-i\nu}\right)-2d\right) \nonumber \\
	& \qquad + 2(d-2)(d-4)\,\nu^3\left(e^{i\nu}-e^{-i\nu}\right)
	- i(d-2)\,\nu^4\left(e^{i\nu}+e^{-i\nu}\right) \Big] \,,
\end{align}
\begin{align}
	\mathcal{\widetilde M}_{zz}^{g,*} = {} & -i(d-2)^2 \left[ \frac{e^{i\nu}+e^{-i\nu}-2}{\nu^2} + 1 \right] I_z(1,0,0,0) \,, \label{eq:Mtilde_gg_zz_ioffe} \\
	\mathcal{\widetilde M}_{zp}^{g,*} = {} & \mathcal{\widetilde M}_{pz}^{g,*} \nonumber \\
	 = {} & -\frac{i (d-3) \nu}{d-2} \left( \diffmaster{} - e^{i\nu}\,\diffmastertwo{} \right) \nonumber \\
	& + \frac{I_z(1,0,0,0)}{(d-2)\,\nu^2} \Big[ -2i(d-1)(d-2)\left(e^{i\nu}+e^{-i\nu}-2\right) \nonumber \\
	& \qquad - d(d-2)\,\nu\left(e^{i\nu}-e^{-i\nu}\right) \nonumber \\
	& \qquad + i\,\nu^2\left((2d-5)\left(e^{i\nu}+e^{-i\nu}\right)-2(d-3)\right) \Big] \,, \label{eq:Mtilde_gg_zp_ioffe} \\
	\mathcal{\widetilde M}_{ppzz}^{g,*} = {} & \frac{2 i (d-3) \nu}{d-2} \left( \diffmaster{} - e^{i\nu}\,\diffmastertwo{} \right) \nonumber \\
	& + \frac{i \left(d^2-6d+10\right)}{d-2} \left(e^{i\nu}+e^{-i\nu}-2\right) I_z(1,0,0,0) \,, \label{eq:Mtilde_gg_ppzz_ioffe} \\
	\mathcal{\widetilde M}_{gg}^{g,*} = {} & \frac{2\, I_z(1,0,0,0)}{\nu^2} \Big[ -id\left(e^{i\nu}+e^{-i\nu}-2\right) - 2\nu\left(e^{i\nu}-e^{-i\nu}\right) \nonumber \\
	& + i\,\nu^2\left(e^{i\nu}+e^{-i\nu}-(d-2)\right) \Big] \,. \label{eq:Mtilde_gg_gg_ioffe}
\end{align}
All six tilde form factors are even under $\nu\to-\nu$ (as required by the symmetry of the correlator).

\paragraph{Pseudo-\PDF{}.}

We Fourier transform the tilde form factors with respect to the Ioffe time at fixed $z^2$ (\cref{eq:pseudo_tilde_expansion}) to yield the bare pseudo-\PDF{} components $x\tilde{f}^{(1),\text{pseudo,bare}}_{g/g,\mathcal{M}_k}$. In contrast to the quasi-\PDF{} below, the support is confined to $x\in[-1,1]$ exactly in $\epsilon$, and every component is even in $x$, as required by the $\nu$-evenness of the correlator.
We therefore quote the region $0<x<1$, and the results for $-1<x<0$ follow from $x\to-x$.
Using the notation of \cref{eq:DGLAP}, we obtain
\begin{align}
	x\tilde{f}^{(1),\text{pseudo,bare}}_{g/g,\mathcal{M}_{pp}} = {} & \frac{\alphas C_A}{2\pi} \Bigg\{ \left(\frac{1}{\epsilon} + \LmuP{}\right) \times \nonumber \\
	& \qquad \left( x\left(-2+x-x^2\right) + \frac{1}{2}\,\delta(1-x) + \left[\frac{1}{1-x}\right]^{[0,1]}_{(+1)} \right) \nonumber \\
	& \qquad - \frac{8}{3} + x - x^2 + \frac{2x^3}{3} + \delta(1-x) \nonumber \\
	& \qquad + 2\left[\frac{1}{1-x}\right]^{[0,1]}_{(+1)} + 2\left[\frac{\log(1-x)}{1-x}\right]^{[0,1]}_{(+1)} \Bigg\} \,, \label{eq:fpseudo_gg_pp_bare} \\
	x\tilde{f}^{(1),\text{pseudo,bare}}_{g/g,\mathcal{M}_{zp}} = {} & x\tilde{f}^{(1),\text{pseudo,bare}}_{g/g,\mathcal{M}_{pz}} \nonumber \\
	 = {} & \frac{\alphas C_A}{2\pi} \Bigg\{ -\frac{1}{4}\,\delta(1-x) \left(\frac{1}{\epsilon} + \LmuP{}\right) \nonumber \\
	& \qquad + \frac{-1+x-2x^2+2x^3}{2} - \frac{1}{4}\,\delta(1-x) - \frac{1}{2}\left[\frac{1}{1-x}\right]^{[0,1]}_{(+1)} \Bigg\} \,, \label{eq:fpseudo_gg_zp_bare} \\
	x\tilde{f}^{(1),\text{pseudo,bare}}_{g/g,\mathcal{M}_{zz}} = {} & \frac{\alphas C_A}{2\pi}\, \frac{2\left(x-1\right)^3}{3} \,, \label{eq:fpseudo_gg_zz_bare} \\
	x\tilde{f}^{(1),\text{pseudo,bare}}_{g/g,\mathcal{M}_{ppzz}} = {} & \frac{\alphas C_A}{2\pi} \Bigg\{ \frac{1}{2}\,\delta(1-x) \left(\frac{1}{\epsilon} + \LmuP{}\right) \nonumber \\
	& \qquad - 1 + x + \frac{1}{2}\,\delta(1-x) + \left[\frac{1}{1-x}\right]^{[0,1]}_{(+1)} \Bigg\} \,, \label{eq:fpseudo_gg_ppzz_bare} \\
	x\tilde{f}^{(1),\text{pseudo,bare}}_{g/g,\mathcal{M}_{gg}} = {} & \frac{\alphas C_A}{2\pi}\, \frac{2\left(2x^3-3x^2+3x-2\right)}{3} \,, \label{eq:fpseudo_gg_gg_bare}
\end{align}
for $0<x<1$. The mixing counterterms of \cref{eq:counterterm} cancel the $\delta(1-x)/\epsilon$ contact poles of the auxiliary form factors identically, so the renormalized auxiliary components follow from the displays above by dropping the $1/\epsilon$ terms, with the $\LmuP\,\delta(1-x)$ terms surviving as the scale dependence required by the renormalization group.

For the leading-twist combination \cref{eq:Gcombination_Mpp_tilde}, only $\mathcal{M}_{pp}$ contributes: multiplying its renormalized component by $c_{pp}=2$ and the overall matching minus sign of \cref{eq:pseudo_tilde_expansion} yields the physical renormalized distribution
\begin{align}\label{eq:fpseudo_gg_ren}
	x\tilde{f}^{(1),\text{pseudo,ren}}_{g/g}(x,z^2) = {} & -\frac{\alphas}{2\pi}\, \frac{1}{\epsilon}\, x\, C_A P_{gg}(x) \nonumber \\
	& + \frac{\alphas C_A}{2\pi} \Bigg\{ \frac{16}{3} - 2x + 2x^2 - \frac{4x^3}{3} - 2\,\delta(1-x) \nonumber \\
	& - 4\left[\frac{1}{1-x}\right]^{[0,1]}_{(+1)} - 4\left[\frac{\log(1-x)}{1-x}\right]^{[0,1]}_{(+1)} \nonumber \\
	& + \LmuP{} \left( 2x\left(2-x+x^2\right) - \delta(1-x) - 2\left[\frac{1}{1-x}\right]^{[0,1]}_{(+1)} \right) \Bigg\} \,,
\end{align}
which we again quote for $0<x<1$. The remaining $1/\epsilon$ pole is of pure collinear origin. Its minimal (pure-pole) subtraction is performed by the matching onto the light-cone gluon \PDF{}, which leaves the brace as the complete finite one-loop matching contribution. The $\delta$-pole coefficient at each endpoint is $-\beta_0/2 = -(11 C_A - 4 T_R n_f)/6$, summing to $-\beta_0$ over the two mirrored endpoints, and the pole density satisfies the momentum sum rule
\begin{equation}
\int_0^1 \mathrm{d}x\, x \left[ C_A P_{gg}(x) + 2 n_f\, T_R P_{qg}(x) \right] = 0\,,
\end{equation}
together with the quark-in-gluon result below.

\paragraph{Comparison with refs.~\cite{Balitsky:2019krf,Balitsky:2021qsr}.}

Refs.~\cite{Balitsky:2019krf,Balitsky:2021qsr} compute the same one-loop correction in coordinate space with the background-field method, and present it as a convolution of a kernel, which we denote $K(u)$, with the tree-level correlator, $\int_0^1\mathrm{d}u\,K(u)\,\mathcal{M}(u\nu,z^2)$, for three index combinations that all contain $\mathcal{M}_{pp}$: the leading-twist combination \cref{eq:Gcombination_Mpp}, the minimal-contamination combination \cref{eq:Gcombination_ziiz}, and $\mathcal{G}^{0ii3}+\mathcal{G}^{3ii0}$ (their $M_{03}^+$). Once we insert the tree-level partonic distribution and take the Fourier transform, their kernel $K(u)$ is our $x\tilde{f}^{(1),\text{pseudo,ren}}_{g/g}(x)$ at $x=u$, in units of $\alphas C_A/(2\pi)$ and with the collinear pole minimally subtracted, so that the two calculations can be compared term by term.

Compared in this way, every distributional structure agrees. The coefficient of $\LmuP{}$ is in each case $\gamma_\mathcal{O}\,\delta(1-x) - \mathcal{B}_{gg}(x)$, where $\mathcal{B}_{gg}$ is their gluon splitting kernel, unmodified (with $\bar{u} \equiv 1-u$ and $\hat{P}_{gg}$ as in \cref{eq:Phat_gg}),
\begin{equation}
\mathcal{B}_{gg}(u) = 2\left[\frac{(1-u\bar{u})^2}{\bar{u}}\right]^{[0,1]}_{(+1)} = \left[u\hat{P}_{gg}(u)\right]^{[0,1]}_{(+1)} = u P_{gg}(u) + \frac{2n_fT_R}{3C_A}\,\delta(1-u)\,,
\end{equation}
that is, the momentum-weighted \DGLAP{} kernel with its momentum integral removed, $\int_0^1\mathrm{d}u\,\mathcal{B}_{gg}(u)=0$, so that the $\LmuP{}$ coefficient integrates to $\gamma_\mathcal{O}$, as in \cref{eq:gammaO}, and does not depend on $n_f$.\footnote{Ref.~\cite{Chen:2024jkb} notes that the $n_f$ part of the infrared pole of the gluon self-energy is missing in ref.~\cite{Balitsky:2019krf}, which evaluates this self-energy for $n_f=0$. Restoring it adds $-\frac{2n_fT_R}{3C_A}\,\delta(1-u)$ both to their ultraviolet coefficient and to their infrared kernel, which becomes $uP_{gg}(u)$; the two cancel in the coefficient of $\LmuP{}$, so that neither the \MSBAR{} gluon-in-gluon kernels of refs.~\cite{Balitsky:2019krf,Balitsky:2021qsr} nor their reduced kernels change at one loop.}
We reproduce all three of their ultraviolet coefficients, which they evaluate for $n_f=0$, where they coincide with $\gamma_\mathcal{O} = 5/6$, $11/6$ and $4/3$ in units of $\alphas C_A/(2\pi)$, for the three combinations in the order listed above. These are the values $\frac{11}{6} - \frac{\lambda+\beta_0}{2C_A}$ of \cref{sec:symmetries}, with $\lambda$ the eigenvalue of \cref{eq:Z1_tilde} in each sector; since the three combinations lie in separate sectors, each of the three tail coefficients in \cref{eq:tail_classification} is tied to a published result. We also agree with the $\delta(1-x)$ contact term and with the coefficients of $[1/(1-x)]^{[0,1]}_{(+1)}$ and $[\log(1-x)/(1-x)]^{[0,1]}_{(+1)}$. Since the combinations differ from one another only in how they weight the auxiliary form factors, $\mathcal{\widetilde M}_{zp}+\mathcal{\widetilde M}_{pz}$ and $\mathcal{\widetilde M}_{ppzz}$ are confirmed separately as well.

The one difference is a regular polynomial in the finite part of the leading-twist component,
\begin{equation}\label{eq:bmr_residual}
	x\tilde{f}^{(1),\text{pseudo,ren}}_{g/g}(x) - x\tilde{f}^{(1),\text{pseudo,ren}}_{g/g}(x)\Big|_{\text{refs.~\cite{Balitsky:2019krf,Balitsky:2021qsr}}} = \frac{\alphas C_A}{2\pi}\,2\left(1-x\right)\left(1+x^2\right)\,, \qquad 0<x<1\,,
\end{equation}
which is identical for all three combinations and is therefore attached to the coefficient of $\mathcal{M}_{pp}(u\nu,z^2)$ that they share. It involves no pole, no logarithm, no plus distribution, and no contact term, and therefore does not affect the anomalous dimensions, the evolution kernel, or the exterior region, which, by \cref{eq:exterior_constraint}, is fixed by the $\LmuP{}$ coefficient alone.

The difference is the finite remainder of an evanescent operator that the kernels of refs.~\cite{Balitsky:2019krf,Balitsky:2021qsr} do not include, a well-known effect of dimensional regularization~\cite{Buras:1989xd,Dugan:1990df,Herrlich:1994kh}, and it can be read off from eq.~(A.1) of ref.~\cite{Balitsky:2021qsr}. That equation gives the one-loop box diagram for forward matrix elements and arbitrary operator indices as an integral over $u$ of bilocal operators $G(uz)[uz,0]G(0)$ with $u$-dependent coefficients, the operator-level form of the convolution with $K(u)$ above: the one-loop correction to $G_{\mu\alpha}(z)[z,0]G_{\nu\beta}(0)$ is itself a sum of bilocal operators, with the first field strength at $uz$ and $u$ integrated from $0$ to $1$. Among the structures that multiply the collinear pole $\Gamma(d/2-2)$, some keep all four operator indices on the field strengths, as in $G_{\mu\alpha}(uz)G_{\nu\beta}(0)$, and others carry two of them on a metric tensor while the field strengths are contracted with each other over a Lorentz index $\xi$ from the loop integration, as in $g_{\alpha\nu}\,G_{\mu\xi}(uz)\,G_\beta^{\ \xi}(0)$. For the operator $\mathcal{O}^+_{03}(z,0) = G_{0i}(z)[z,0]G_{i3}(0) + G_{3i}(z)[z,0]G_{i0}(0)$ of their $M_{03}^+$ combination, with $i$ summed over $i=1,2$, this second kind of structure becomes $g_{ii}\,\big[G_{0\xi}(uz)[uz,0]G_3^{\ \xi}(0) + G_{3\xi}(uz)[uz,0]G_0^{\ \xi}(0)\big]$: the operator's own transverse sum closes into the trace $g_{ii}=-2$, and its place between the two field strengths is taken by $\xi$. The values $\xi=0,3$ drop out by antisymmetry, so $\xi$ runs over the $d-2$ transverse directions of dimensional regularization, and the bracket is the sum of the same operator at the rescaled separation, from $\xi=1,2$, and of an evanescent operator $E_{03}$, from the remaining $d-4$ directions:
\begin{equation}\label{eq:bmr_evanescent}
    G_{0\xi}(uz)\,[uz,0]\,G_3^{\ \xi}(0) + G_{3\xi}(uz)\,[uz,0]\,G_0^{\ \xi}(0)
    = \mathcal{O}^+_{03}(uz,0) + E_{03}(uz,0)\,.
\end{equation}
The kernels quoted in ref.~\cite{Balitsky:2021qsr} follow from eq.~(A.1) only if the sum over $\xi$ is restricted to $\xi=1,2$, that is, only if $E_{03}$ is dropped.\footnote{We thank A.~Radyushkin for confirming this step in private communication.}

Dropping $E_{03}$ costs the finite term \cref{eq:bmr_residual}, because $E_{03}$ multiplies the collinear pole. Per unit tree-level matrix element of $\mathcal{O}^+_{03}$, the terms of eq.~(A.1) that multiply the pole contribute to the kernel
\begin{equation}\label{eq:bmr_nxi}
K^{\text{box}}_{\Gamma(d/2-2)}(u) = -2u + (-g_{ii})\,(1-u)(1+u^2)\,\frac{n_\xi}{n_i} = n_\xi\,(1-u)(1+u^2) - 2u\,.
\end{equation}
Here $-2u$ comes from the terms of eq.~(A.1) without a loop index, and the second term is the bracket of \cref{eq:bmr_evanescent}. Between $d$-dimensional on-shell gluon states each transverse direction of $\xi$ contributes equally, so the bracket, with its $n_\xi$ directions, counts $n_\xi/n_i$ relative to $\mathcal{O}^+_{03}$, with its $n_i=2$ directions. The trace $-g_{ii}=2$ over the operator's own index pair is that same count $n_i$, so the two cancel and only the number $n_\xi$ of loop-index directions remains. Dimensional regularization requires $n_\xi=d-2$, while the coefficient $2\left[(1-u)(1+u^2)-u\right]$ that refs.~\cite{Balitsky:2019krf,Balitsky:2021qsr} quote for $\mathcal{M}_{pp}$ in the $\Gamma(d/2-2)$ part of the box diagram corresponds to $n_\xi=2$. The difference, $(d-4)(1-u)(1+u^2)$, is the contribution of $E_{03}$. Since $(d-4)\Gamma(d/2-2)$ is finite and equal to $2$ at $d=4$, it leaves $2(1-u)(1+u^2)$ on the pole, which is \cref{eq:bmr_residual}.

The count $n_\xi$ is not a normalization convention of the kind discussed in \cref{sec:tree}. There, with the external polarizations summed in $d$ dimensions, the transverse sum over the operator's own indices and the polarization average of the external gluons both drop out of the tree-normalized gluon-in-gluon kernel of a multiplicatively renormalizable combination. The index $\xi$, by contrast, is contracted inside the loop: its count multiplies the collinear pole and is not matched by the tree normalization, which contains only the count of $i$.\footnote{This attribution to $\xi$ refers to $d$-dimensional polarization sums. If the external polarizations are instead kept in four dimensions, the count of $\xi$ is automatically two and the operator's own sum has to be continued to $d-2$ directions. In either convention the kernels of refs.~\cite{Balitsky:2019krf,Balitsky:2021qsr} are reproduced only if both sums are four-dimensional.}

The value $n_\xi=d-2$ is the one the light-cone side requires: the \MSBAR{} splitting functions, the coefficient functions and the global fits, and with them $f_g^{\MSBAR}$, are defined with every such index in $d$ dimensions, so the collinear poles on the two sides of the matching relation cancel with the correct finite remainder only if the equal-time side is regularized in the same way. At one loop the kernel of refs.~\cite{Balitsky:2019krf,Balitsky:2021qsr} is therefore the matching onto
\begin{equation}
x f_4(x) = x f_g^{\MSBAR}(x) + \frac{\alphas C_A}{2\pi}\,\big(R \otimes x f_g^{\MSBAR}\big)(x)\,,\qquad R(u) = 2(1-u)(1+u^2)\,,
\end{equation}
where the Mellin convolution of \cref{eq:mellinconv} acts on the momentum-weighted distribution, as their kernel does. The distribution $f_4$ differs from the \MSBAR{} distribution that lattice pseudo-distribution extractions target by the finite contribution of the dropped evanescent operator, and no conversion between the two is currently applied in those determinations. Since \cref{eq:bmr_residual} is common to the three index combinations, $f_4$ is the same object for each of them at this order. Whether this prescription can be organized consistently beyond one loop we do not address.

\paragraph{Comparison with ref.~\cite{Yao:2022vtp}: coordinate-space kernel.}

The independent calculation of ref.~\cite{Yao:2022vtp}, which computes the complete set of one-loop matching kernels for spacelike correlators onto light-cone distributions, gives a second test of our results. Their gluon correlator is the $z$-weighted field-strength bilinear $\mathbf{F}_{\mu\nu}(z_1,z_2) = z_{12}^\rho G_{\rho\mu}(z_1)[z_1,z_2]G_{\nu\sigma}(z_2) z_{12}^\sigma$ (using our symbol $G_{\mu\nu}$ for the field strength), and their unpolarized distribution is defined by the transverse trace $O_{g,u} = g_\perp^{\mu\nu}\mathbf{F}_{\mu\nu}$. In our frame, $O_{g,u}$ corresponds to the index combination $\mathcal{G}^{ziiz}_{\text{ET}}$ of \cref{eq:Mziiz} with the transverse sum taken over the $d-2$ directions of dimensional regularization, which is $(d-2)/2$ times the displayed $\mathcal{G}^{ziiz}$. Because we compare tree-normalized kernels, this overall factor divides out. The operator $O_{g,u}$ does not isolate $\mathcal{\widetilde M}_{pp}$, as it also contains the four auxiliary form factors $\mathcal{\widetilde M}_{zz}$, $\mathcal{\widetilde M}_{zp}$, $\mathcal{\widetilde M}_{pz}$ and $\mathcal{\widetilde M}_{gg}$. It is, however, multiplicatively renormalizable and belongs to the $Z_{\parallel\perp}^2$ eigenvalue sector.

Ref.~\cite{Yao:2022vtp} presents the gluon-in-gluon kernel for $O_{g,u}$ in coordinate space, both for off-forward kinematics and in the forward limit. In that limit, their factorization reads
\begin{equation}
\tilde{h}(z_{12},p^z,\mu) = \int_0^1\mathrm{d}\alpha\,\boldsymbol{C}(\alpha,\mu^2z_{12}^2)\,h^{l.t.}(\bar{\alpha}\,z_{12}p^z,\mu)\,,
\end{equation}
where we write out the Ioffe-time argument $\bar{\alpha}\,z_{12}p^z$ of the light-cone correlator, which ref.~\cite{Yao:2022vtp} abbreviates as $\bar{\alpha}$. Their $\boldsymbol{C}(\alpha)$ is precisely a pseudo matching kernel at $x=\bar{\alpha}$ (with $\bar{\alpha} \equiv 1-\alpha$), where $L_z = \ln\left[4e^{-2\gamma_E}/(-\mu^2z_{12}^2)\right] = -\LmuP$. When we determine the renormalized pseudo-\PDF{} kernel for $\mathcal{G}^{ziiz}$ from the results above, we reproduce their forward coordinate-space kernel in its entirety. This includes the $\LmuP$ coefficient with $\gamma_\mathcal{O} = 11/6$ in units of $\alphas C_A/(2\pi)$ (their $E_{3,u}$), every plus distribution and every endpoint contact term (which their coordinate-space displays include), and the entire finite contribution.

In addition, ref.~\cite{Yao:2022vtp} presents a second operator projection, and by comparing with both operator choices, we obtain an overdetermined check. Their second operator is the tree-normalized combination
\begin{equation}
\mathcal{\widetilde M}_{pp} + \frac{3}{2}(\mathcal{\widetilde M}_{zz} + \mathcal{\widetilde M}_{zp} + \mathcal{\widetilde M}_{pz} - \mathcal{\widetilde M}_{gg}) + \frac{1}{2}\mathcal{\widetilde M}_{ppzz}\,,
\end{equation}
which weights the auxiliary form factors differently from $O_{g,u}$.\footnote{The identification holds away from four dimensions: projecting $g^{\mu\nu}\mathbf{F}_{\mu\nu}$ onto the tensor decomposition \eqref{eq:Manb} exactly in $d$ dimensions promotes the coefficients $3/2$ and $1/2$ to $(d-1)/(d-2)$ and $1/(d-2)$ respectively, while $O_{g,u}$, with its transverse sum taken over the $d-2$ transverse directions, carries no $\epsilon$ dependence after tree normalization.}%
 We also reproduce their appendix coordinate-space kernel for this operator exactly. The difference between their two operator kernels, the pure polynomial
\begin{equation}
\frac{\alphas C_A}{2\pi}\,\frac{2}{3}\left(2-3x+3x^2-2x^3\right)
\end{equation}
with identical $\LmuP$ coefficient and no contact terms, matches our auxiliary form factors identically.

From the two comparisons together, we therefore pin down $\mathcal{M}_{pp}$ and $\mathcal{M}_{zz}-\mathcal{M}_{gg}$ separately, given that the other auxiliary form factors agree with refs.~\cite{Balitsky:2019krf,Balitsky:2021qsr}. Our agreement with ref.~\cite{Yao:2022vtp} on the coefficient of $\mathcal{M}_{pp}$, which is where \cref{eq:bmr_residual} appears, cannot be an artifact of compensating errors among the auxiliary components, and the gluon-in-gluon kernels of ref.~\cite{Yao:2022vtp} and of refs.~\cite{Balitsky:2019krf,Balitsky:2021qsr} are therefore mutually inconsistent in the $Z_{\parallel\perp}^2$ sector.\footnote{Ref.~\cite{Ji:2025cbb} reports that the differences between published gluon-in-gluon quasi kernels, including those of refs.~\cite{Balitsky:2019krf,Yao:2022vtp}, vanish in the momentum-space factorization theorem; the comparison itself is not shown. The difference \cref{eq:bmr_residual} is not of this kind: its Mellin moments do not vanish, and it shifts the extracted gluon moments.} Both $O_{g,u}$ and the minimal-contamination combination \cref{eq:Gcombination_ziiz_tilde} are tree-normalized members of this sector, so the difference of the two operators is free of $\mathcal{\widetilde M}_{pp}$, and when we translate between the two operators using our auxiliary form factors, exactly \cref{eq:bmr_residual} remains. To our knowledge, this comparison has not been made in the literature.

The agreement with ref.~\cite{Balitsky:2019krf} reported in ref.~\cite{Yao:2022vtp} refers to the ratio scheme, in which the coordinate-space evolution kernel, and consequently the quasi exterior region (\cref{eq:exterior_constraint}), is operator-independent and fixed by the light-cone \DGLAP{} kernel alone. All three calculations agree on the evolution kernel. The finite, scheme-dependent terms, which populate the interior of the momentum-space distribution, can by contrast be compared only through our form-factor analysis.

\paragraph{Numerical impact.}

All published extractions of the unpolarized gluon pseudo-distribution~\cite{Fan:2020cpa,Fan:2021bcr,HadStruc:2021wmh,Salas-Chavira:2021wui,Fan:2022kcb,Good:2023gai,Delmar:2023agv,Good:2023ecp,NieMiera:2025inn,Delmar:2026ltr} match their lattice data with the kernel of refs.~\cite{Balitsky:2019krf,Balitsky:2021qsr}, and the finite difference \cref{eq:bmr_residual} therefore enters each of these results. We use published values of the low moments of the unpolarized gluon \PDF{} of the proton to estimate the numerical impact of this kernel difference.

In the ratio scheme, dividing the bare correlator by its zero-momentum value $\mathcal{M}(0, z^2)$ normalizes the target distribution to $x f_{g/H}(x)/\langle x\rangle_g$ and forces the reduced matching kernel to integrate to $1$. At one loop, this normalization subtracts the $\nu=0$ value of the kernel correction $R(u) \equiv 2(1-u)(1+u^2)$ of \cref{eq:bmr_residual}, $\int_0^1 \mathrm{d}u\,R(u) = 7/6$, and so replaces $R$ by the plus distribution $\left[R\right]^{[0,1]}_{(+1)} = R(u) - \frac{7}{6}\delta(1-u)$, the form in which the difference appears in the reduced Ioffe-time matching relation of ref.~\cite{Balitsky:2019krf}. Matching with the kernel of refs.~\cite{Balitsky:2019krf,Balitsky:2021qsr} therefore yields an extracted distribution that differs from the true \MSBAR{} distribution at one loop by
\begin{equation*}
	\frac{x f_{g/H}(x)}{\langle x\rangle_g}\bigg|_{\text{extracted}} = \frac{x f_{g/H}(x)}{\langle x \rangle_g}\bigg|_{\MSBAR} + \frac{\alphas C_A}{2\pi} \left(\left[R\right]^{[0,1]}_{(+1)} \otimes \frac{x f_{g/H}}{\langle x\rangle_g}\bigg|_{\MSBAR}\right)\!(x) + \mathcal{O}(\alphas^2)\,,
\end{equation*}
with the Mellin convolution of \cref{eq:mellinconv}. Under the Mellin transform, the convolution factors into a simple numerical rescaling of each moment, so that any published moment can be corrected without access to the underlying lattice data. Because \cref{eq:bmr_residual} contains no logarithm of $z^2\mu^2$, the difference is independent of the matching scale; the scale-variation bands shown in lattice publications to estimate perturbative truncation uncertainties therefore do not cover this effect.

Because the plus distribution integrates to zero, the zeroth moment of the correction vanishes identically, and the total momentum fraction $\langle x\rangle_g$ remains unchanged. For the next two moments, the relation yields
\begin{align*}
	\frac{\langle x^2\rangle_g}{\langle x\rangle_g}\bigg|_{\text{extracted}} {} & = \left[1-\frac{11}{15}\frac{\alphas C_A}{2\pi}\right]\frac{\langle x^2\rangle_g}{\langle x \rangle_g}\bigg|_{\MSBAR} + \mathcal{O}(\alphas^2)\,,\\
	\frac{\langle x^3\rangle_g}{\langle x\rangle_g}\bigg|_{\text{extracted}} {} & = \left[1-\frac{14}{15}\frac{\alphas C_A}{2\pi}\right]\frac{\langle x^3\rangle_g}{\langle x \rangle_g}\bigg|_{\MSBAR} + \mathcal{O}(\alphas^2)\,.
\end{align*}
At $\alphas = 0.3$, the extracted moments are suppressed by approximately $11\%$ and $13\%$ relative to the \MSBAR{} values, meaning that the true \MSBAR{} gluon is systematically harder than reported. Published moments are corrected by dividing by these factors, and the corresponding statistical errors scale by the same division.

\begin{figure}[t]
	\centering
	\includegraphics[width=0.62\linewidth]{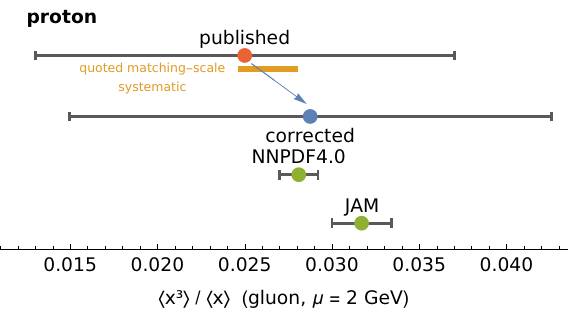}
	\caption{The gluon moment $\langle x^3\rangle_g/\langle x\rangle_g$ at $\mu=2$~GeV, before and after correcting the proton result of ref.~\cite{Delmar:2026ltr} for the matching kernel difference in \cref{eq:bmr_residual}. The thin bars show the statistical errors, which scale with the central value, and the short thick bar shows their quoted matching-scale systematic. The global-fit values are those compiled in ref.~\cite{Delmar:2026ltr}.}
	\label{fig:bmrimpact_moments}
\end{figure}

In \cref{fig:bmrimpact_moments}, we show the moment $\langle x^3\rangle_g/\langle x\rangle_g$ for the proton from ref.~\cite{Delmar:2026ltr}, before and after the correction. Ref.~\cite{Delmar:2026ltr} quotes $\langle x^3\rangle_g/\langle x\rangle_g = 0.0250 \pm 0.0120\,^{+0.0029}_{-0.0002}$ at $\mu = 2$~GeV (statistical and matching-scale uncertainties, respectively) and uses $\alphas(2~\text{GeV}) = 0.293$. Dividing by the transfer factor yields a corrected value of $0.0288$. The resulting shift of $+0.0038$ exceeds their quoted matching-scale systematic, corresponds to $0.3$ of their statistical error, and brings the central value into close agreement with the NNPDF4.0 result compiled in ref.~\cite{Delmar:2026ltr}.

\paragraph{Quasi-\PDF{}.}

We Fourier transform the tilde form factors with respect to the spatial separation $\zt$ at fixed hadron momentum (\cref{eq:fgquasi}, applied to each form factor, as in \cref{eq:pseudo_tilde_expansion}) to yield the bare quasi-\PDF{} components $x\tilde{f}^{(1),\text{quasi,bare}}_{g/g,\mathcal{M}_k}$.
We quote the region $x>0$, and the results for $x<0$ follow from $x\to-x$.
With $\LmuQ$ as defined in \cref{eq:LmuQ} we obtain
\begingroup\allowdisplaybreaks
\begin{align}
	x\tilde{f}^{(1),\text{quasi,bare}}_{g/g,\mathcal{M}_{pp}} = {} & \frac{\alphas C_A}{2\pi} \Bigg\{ \left(\frac{1}{\epsilon} + \LmuQ\right) \left( -\frac{5}{12}\,\delta(1-x) + \frac{1}{2}\left[x\hat{P}_{gg}(x)\right]^{[0,1]}_{(+1)} + \frac{5}{12}\,\frac{\delta^+(1/x)}{x^2} \right) \nonumber \\
	& - \frac{4}{3}\,\delta(1-x) + \frac{4}{3}\,\frac{\delta^+(1/x)}{x^2} + \frac{5}{6}\left[\frac{1}{x}\right]^{[1,\infty)}_{(+\infty)} \nonumber \\
	& + \frac{1}{2}\Bigg[ x\hat{P}_{gg}(x)\left(\frac{3}{2}-\log\big(x(1-x)\big)\right) + x\hat{P}_{gg}(-x)\left(\frac{1}{2}+\log\frac{1+x}{x}\right) \nonumber \\
	& \qquad + 2\left(1+x+3x^2-x^3\right) \Bigg]^{[0,1]}_{(+1)} \nonumber \\
	& + \frac{1}{2}\Bigg[ x\hat{P}_{gg}(x)\left(\frac{1}{2}-\log\frac{x}{x-1}\right) + x\hat{P}_{gg}(-x)\left(\frac{1}{2}+\log\frac{x+1}{x}\right) \nonumber \\
	& \qquad + 2\left(3x+x^3\right) - \frac{5}{3x} \Bigg]^{[1,\infty)}_{(+1)} \Bigg\} \,, \label{eq:fquasi_gg_pp_bare} \\
	x\tilde{f}^{(1),\text{quasi,bare}}_{g/g,\mathcal{M}_{zp}} = {} & x\tilde{f}^{(1),\text{quasi,bare}}_{g/g,\mathcal{M}_{pz}} \nonumber \\
	 = {} & \frac{\alphas C_A}{2\pi} \Bigg\{ -\frac{1}{4} \left(\frac{1}{\epsilon} + \LmuQ\right) \left( \delta(1-x) - \frac{\delta^+(1/x)}{x^2} \right) - \frac{7}{12}\,\delta(1-x) + \frac{7}{12}\,\frac{\delta^+(1/x)}{x^2} \nonumber \\
	& + \frac{1}{2}\left[\frac{1}{x}\right]^{[1,\infty)}_{(+\infty)} + \left[\frac{1+x^2-x^3-2x^4+2x^5}{2\left(x^2-1\right)}\right]^{[0,1]}_{(+1)} + \left[\frac{1}{2x\left(x^2-1\right)}\right]^{[1,\infty)}_{(+1)} \Bigg\} \,, \label{eq:fquasi_gg_zp_bare} \\
	x\tilde{f}^{(1),\text{quasi,bare}}_{g/g,\mathcal{M}_{zz}} = {} & \frac{\alphas C_A}{2\pi} \Bigg\{ \frac{2\left(x-1\right)^3}{3}\,\theta(0<x<1) + \frac{1}{6}\,\frac{\delta^+(1/x)}{x^2} \Bigg\} \,, \label{eq:fquasi_gg_zz_bare} \\
	x\tilde{f}^{(1),\text{quasi,bare}}_{g/g,\mathcal{M}_{ppzz}} = {} & \frac{\alphas C_A}{2\pi} \Bigg\{ \frac{1}{2} \left(\frac{1}{\epsilon} + \LmuQ\right) \left( \delta(1-x) - \frac{\delta^+(1/x)}{x^2} \right) \nonumber \\
	& + \left[\frac{x^3-x^2-2x+1}{x^2-1}\right]^{[0,1]}_{(+1)} - \left[\frac{1}{x\left(x^2-1\right)}\right]^{[1,\infty)}_{(+1)} - \left[\frac{1}{x}\right]^{[1,\infty)}_{(+\infty)} \Bigg\} \,, \label{eq:fquasi_gg_ppzz_bare} \\
	x\tilde{f}^{(1),\text{quasi,bare}}_{g/g,\mathcal{M}_{gg}} = {} & \frac{\alphas C_A}{2\pi} \Bigg\{ \frac{2\left(2x^3-3x^2+3x-2\right)}{3}\,\theta(0<x<1) + \frac{2}{3}\,\frac{\delta^+(1/x)}{x^2} \Bigg\} \,, \label{eq:fquasi_gg_gg_bare}
\end{align}\endgroup
for $x>0$. Here
\begin{equation}\label{eq:Phat_gg}
\hat{P}_{gg}(x) = 2\left[\frac{x}{1-x} + \frac{1-x}{x} + x(1-x)\right]
\end{equation}
is the unregularized bracket of \cref{eq:DGLAP_gg}, so that $\frac{1}{2}x\hat{P}_{gg}(x) = (1-x+x^2)^2/(1-x)$. As in the gluon-in-quark channel, \cref{eq:fquasi_gq_pp_bare}, the logarithms multiply the splitting kernel evaluated at $x$ and at $-x$.

Where a plus bracket appears, it carries the entire contribution within its region, including the terms regular in $x$. The components $\mathcal{\widetilde M}_{zz}$ and $\mathcal{\widetilde M}_{gg}$ coincide on $0<x<1$ with their pseudo-\PDF{} counterparts, \cref{eq:fpseudo_gg_zz_bare,eq:fpseudo_gg_gg_bare}, and differ from them only by a finite boundary term at $x=+\infty$. Its coefficient is minus the integral of the interior, as it must be, since the bare one-loop quasi-distribution integrates to zero (\cref{sec:boundary}); the balanced pairs of contact terms at $x=1$ and boundary terms at $x=+\infty$ in the other components have the same origin.
We find agreement of the quasi- and pseudo-\PDF{} pole structure within the interior region.

The poles in the interior region renormalize as in the pseudo-\PDF{}. The counterterms of \cref{eq:counterterm} cancel the $\delta(1\mp x)/\epsilon$ contact poles of the auxiliary form factors identically, while the $\LmuQ\,\delta(1\mp x)$ terms survive as the scale dependence required by the renormalization group. The boundary poles at $x=\pm\infty$, however, survive renormalization, leaving
\begin{equation}\label{eq:quasi_gg_towers}
	x\tilde{f}^{(1),\text{quasi,ren}}_{g/g,\mathcal{M}_k}\Big|_{\text{boundary pole}} = \frac{\alphas C_A}{2\pi}\, \frac{b_k}{\epsilon} \left[ \frac{\delta^+(1/x)}{x^2} + \frac{\delta^-(1/x)}{x^2} \right]\,,
\end{equation}
where
\begin{equation}
b_{pp} = \frac{5}{12}\,,\qquad b_{zp} = b_{pz} = \frac{1}{4}\,,\qquad \mathrm{and}\qquad b_{ppzz} = -\frac{1}{2}\,.
\end{equation}
These poles drop out of the matching convolution (\cref{sec:boundary}), and we keep them
explicitly.

We construct the physical distribution using $c_{pp}=2$ (\cref{eq:Gcombination_Mpp_tilde}) and the matching minus sign, so that the renormalized quasi-\PDF{}, again quoted for $x>0$, reads
\begin{align}\label{eq:fquasi_gg_ren}
	x\tilde{f}^{(1),\text{quasi,ren}}_{g/g}(x,p^z) = {} & -\frac{\alphas}{2\pi}\, \frac{1}{\epsilon}\, x\, C_A P_{gg}(x) - \frac{\alphas C_A}{2\pi}\, \frac{5}{6\epsilon}\, \frac{\delta^+(1/x)}{x^2} \nonumber \\
	& + \frac{\alphas C_A}{2\pi} \Bigg\{ \frac{8}{3}\,\delta(1-x) - \frac{8}{3}\,\frac{\delta^+(1/x)}{x^2} - \frac{5}{3}\left[\frac{1}{x}\right]^{[1,\infty)}_{(+\infty)} \nonumber \\
	& - \Bigg[ x\hat{P}_{gg}(x)\left(\frac{3}{2}-\log\big(x(1-x)\big)\right) + x\hat{P}_{gg}(-x)\left(\frac{1}{2}+\log\frac{1+x}{x}\right) \nonumber \\
	& \qquad + 2\left(1+x+3x^2-x^3\right) \Bigg]^{[0,1]}_{(+1)} \nonumber \\
	& - \Bigg[ x\hat{P}_{gg}(x)\left(\frac{1}{2}-\log\frac{x}{x-1}\right) + x\hat{P}_{gg}(-x)\left(\frac{1}{2}+\log\frac{x+1}{x}\right) \nonumber \\
	& \qquad + 2\left(3x+x^3\right) - \frac{5}{3x} \Bigg]^{[1,\infty)}_{(+1)} \nonumber \\
	& + \LmuQ \left( \frac{5}{6}\,\delta(1-x) - \left[x\hat{P}_{gg}(x)\right]^{[0,1]}_{(+1)} - \frac{5}{6}\,\frac{\delta^+(1/x)}{x^2} \right) \Bigg\} \,.
\end{align}
Here we write the collinear pole, which is \IR{} and removed by the matching procedure, in the form of \cref{eq:DGLAP}, with $P_{gg}(x)$ supported on $0<x\leq1$.
On $0<x\leq1$, the coefficient of $\LmuQ$ can be rewritten as
\begin{equation}
2x(2-x+x^2) - \delta(1-x) - 2[1/(1-x)]^{[0,1]}_{(+1)}\,,
\end{equation}
in agreement with the $\LmuP$ coefficient of the pseudo-\PDF{} matching contribution \cref{eq:fpseudo_gg_ren}. %

\paragraph{Comparison with ref.~\cite{Yao:2022vtp}: quasi kernel.}

Ref.~\cite{Yao:2022vtp} also gives the forward (\PDF{}) limit of the gluon-in-gluon quasi matching kernel for the operator $O_{g,u}$ introduced above. Their conventions are related to ours by $2 a_s C_A = \frac{\alphas C_A}{2\pi}$ (with $a_s = \frac{\alphas}{4\pi}$) and $l_x \equiv \log(4(p^z)^2 x^2/\mu^2) = \log(x^2) - \LmuQ$. We reconstruct $O_{g,u}$ from our form factors and reproduce their forward quasi matching kernel exactly in all four regions of $x$, excluding the endpoint and boundary contact terms, which they do not display.

\paragraph{Operator dependence and the $|x|>1$ tails.}

The operator of \cref{eq:Gcombination_Mpp_tilde}, which isolates $\mathcal{\widetilde M}_{pp}$ and yields \cref{eq:fquasi_gg_ren}, and the transverse-trace operator $\mathcal{G}^{ziiz}$ of \cref{eq:Mziiz} are both valid definitions of a gluon quasi-\PDF{}, although only the latter contains auxiliary form factors. Both choices match onto the same light-cone \PDF{}, so the operator dependence resides entirely in the finite part of the matching kernel. Their difference provides an explicit conversion dictionary between the two schemes: subtracting our distributions gives
\begin{equation}\label{eq:quasi_gg_dict}
	x\tilde{f}^{(1),\text{quasi,ren}}_{g/g}(x) - x\tilde{f}^{(1),ziiz}_{g/g}(x)
	= \frac{\alphas C_A}{2\pi}
	\left[\frac{1}{\lvert x\rvert-1} + \frac{1}{\lvert x\rvert+1} + \frac{2}{3}\left(\lvert x\rvert - 1\right)\left(4x^2 - 2\lvert x\rvert + 1\right)\theta(1-x^2)\right],
\end{equation}
for the densities away from $\lvert x\rvert=1$.\footnote{The endpoint contacts and the boundary deltas at infinity also differ between the operator choices. They are not displayed here.} This difference is even in $x$, free of logarithms, and independent of $\mu$ and $p^z$, as an operator-choice difference between distributions with identical collinear pole structure must be. It consists of two parts. The first, $1/(\lvert x\rvert-1)+1/(\lvert x\rvert+1) = 2\lvert x\rvert/(x^2-1)$, is a single expression on the whole real line, which the plus prescriptions of the full displays regulate at $\lvert x\rvert=1$. It is the only part that survives in the exterior region, where it is the Cauchy transform, \cref{eq:exterior_constraint}, of the difference $-\delta(1\mp u)$ between the endpoint contact terms of the two scale-logarithm coefficients $\mathcal{L}$; the two operators share the \DGLAP{} part of $\mathcal{L}$. It falls off as $+2/\lvert x\rvert$ at large $\lvert x\rvert$. The second part is a polynomial confined to $\lvert x\rvert\leq1$, and it coincides with the polynomial part of the corresponding difference of the pseudo-distributions, $2(\mathcal{\widetilde M}_{zp}+\mathcal{\widetilde M}_{pz}+\mathcal{\widetilde M}_{zz}-\mathcal{\widetilde M}_{gg})$ in \cref{eq:fpseudo_gg_zp_bare,eq:fpseudo_gg_zz_bare,eq:fpseudo_gg_gg_bare}.

Of the bare form factor components, only $\mathcal{\widetilde M}_{pp}$, $\mathcal{\widetilde M}_{zp}$, $\mathcal{\widetilde M}_{pz}$ and $\mathcal{\widetilde M}_{ppzz}$ carry an asymptotic $[1/x]^{[1,\infty)}_{(+\infty)}$ tail, with coefficients $5/6$, $1/2$, $1/2$ and $-1$. These are twice the boundary-pole coefficients $b_k$ of \cref{eq:quasi_gg_towers}, as required by the dimensional regularization of the $1/|x|$ tail into the boundary deltas at infinity (\cref{eq:plusInfinity}).

When assembling a physical, tree-normalized quasi-\PDF{} $\sum_k c_k \mathcal{\widetilde M}_k$ ($c_{pp}=2$, with the matching minus sign), projecting these component tails onto the three left eigenspaces of the renormalization mixing matrix (\cref{sec:gluon_operator_renorm}) leaves exactly three universal asymptotic tails:
\begin{equation}\label{eq:tail_classification}
	x\tilde{f}^{(1),c}_{g/g}(x)\;\xrightarrow{\;|x|\to\infty\;}\; -\frac{\alphas C_A}{2\pi}\,\frac{1}{|x|} \times
	\begin{cases}
		5/3\,, & Z_{\perp\perp}^2 \quad (\mathcal{\widetilde M}_{pp})\,, \\[0.4ex]
		8/3\,, & Z_{\perp\perp}Z_{\parallel\perp} \quad (\mathcal{G}^{0ii3}+\mathcal{G}^{3ii0})\,, \\[0.4ex]
		11/3\,, & Z_{\parallel\perp}^2 \quad (\mathcal{G}^{ziiz},\; \mathcal{\widetilde M}_{pp}-\mathcal{\widetilde M}_{ppzz})\,,
	\end{cases}
\end{equation}
where each row lists $2\gamma_\mathcal{O} = \frac{11}{3} - \frac{\lambda+\beta_0}{C_A}$ of \cref{sec:symmetries} in units of $\alphas C_A/(2\pi)$, the corresponding eigenvalue sector, and representative operators in parentheses (given either in our form factor basis or as field-strength correlators $\mathcal{G}^{\mu\nu\rho\sigma}$).

This tail behavior is universal across each eigenspace, independently of which specific operator combination is chosen. The additional degrees of freedom allowed within each two-dimensional sector either carry no tail themselves ($\mathcal{\widetilde M}_{gg}$ in the first sector, and the orthogonal combination without $\mathcal{\widetilde M}_{pp}$ in the third, cf.~\cref{eq:eigenspace_parpar}) or are constrained by the tree-level normalization ($c_{zp}+c_{pz}=c_{pp}$ in the second, which fixes the tail since both components carry the same tail coefficient $1/2$).

The $+2/\lvert x\rvert$ fall-off of the dictionary \cref{eq:quasi_gg_dict} is the difference between the outer two rows, the eigenvalue gap $2C_A$ of \cref{eq:Z1_tilde}. Because $\mathcal{L}$ depends only on the eigenvalue (\cref{eq:log_coefficient}), two operators in the same sector share their entire exterior pointwise, not only the tail; the minimal-contamination combination \cref{eq:Gcombination_ziiz_tilde} and $\mathcal{G}^{ziiz}$ are an example. A lattice calculation can therefore read the large-$\lvert x\rvert$ behavior of its quasi-\PDF{} directly from the renormalization of its operator, without a matching calculation. Beyond one loop this holds only for the leading logarithms, because at order $\alphas^2$ the coefficient of the single logarithm also contains the finite one-loop terms, which differ between operators of the same sector.

\subsubsection{Quark in gluon}

\begin{figure}[t]
	\centering
	\begin{subfigure}[b]{0.19\textwidth}
		\centering\includegraphics[width=\linewidth]{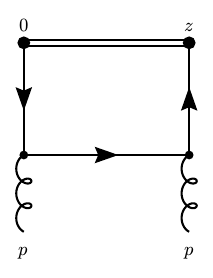}
		\caption{uncrossed}\label{fig:qg-direct}
	\end{subfigure}
	\hspace{0.06\textwidth}
	\begin{subfigure}[b]{0.19\textwidth}
		\centering\includegraphics[width=\linewidth]{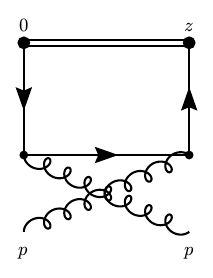}
		\caption{crossed}\label{fig:qg-crossed}
	\end{subfigure}
	\caption{The two one-loop diagrams contributing to the quark-in-gluon matrix element, related by
	crossing the external gluon legs (each carrying momentum $p$). The internal quark line runs from
	the operator insertion at $0$ around to the one at $z$, closing the loop with the gauge link
	$W(z,0)$ (double line); the quark bilocal operator measures the quark content of the gluon. Both
	diagrams must be summed for gauge invariance.}
	\label{fig:qg-diagrams}
\end{figure}

Two diagrams contribute to the quark-in-gluon matrix element (\cref{fig:qg-diagrams}), the uncrossed (\subref{fig:qg-direct}) and crossed (\subref{fig:qg-crossed}) diagrams, related by crossing the external gluon legs. We sum both diagrams, average over colors with $1/(N_c^2-1)$ and over the $n_\text{pol} = d-2$ physical gluon polarizations as required in \MSBAR{} (\cref{sec:tree}), and denote the result $\mathcal{Q}_{qg}$. We find:
\begin{align}
		\mathcal{Q}_{qg}^{(1)} = {} &  -\frac{d}{2(d-4)} g_s^2 T_R
		\frac{1}{(p\cdot z)^2}
		\zt 
		I_z(1,0,0,0) \nonumber \\
		{} & \quad
		 \cdot \Big[
			e^{i p\cdot z}\big((p\cdot z)^2 + 2i p\cdot z - 4\big) + e^{-i p\cdot z}\big((p\cdot z)^2 - 2i p\cdot z - 4\big) - 2\big((p\cdot z)^2 - 4\big)
		\Big]\,. \label{eq:Qqg_bare}
\end{align}
The one-loop matrix element of the quark operator in a gluon state is given as a Feynman-parameter integral in ref.~\cite{Yao:2022vtp}, in off-forward kinematics with $\Gamma=\gamma^t$, and in ref.~\cite{Chen:2024jkb}, in forward kinematics with $\Gamma=\gamma^z$. Ref.~\cite{Balitsky:2019krf} left the quark--gluon terms for future work. As in the other channels, our calculations in $R_\xi$ and axial gauges agree.

Because the quark bilocal operator has no tree-level matrix element in an external gluon state, we normalize by its tree-level matrix element in external quark states (\cref{eq:Qtree_quark}): we divide by $\eta_{pp}p^z = d p^z/2$ for the quasi-\PDF{} and by $-\frac{d}{2}\frac{p\cdot z}{\zt}$ for the pseudo-\PDF{}, factoring this kinematic term out before the Fourier transform. This isolates the amplitude $\mathcal{M}_{pp}$ and the splitting kernel $P_{qg}(x)$. The remaining ratio to the four-dimensional normalization $2p^z$ is the factor $d/4$ displayed below; because the target is a gluon, this factor does not cancel in matching, and setting $d/4\to1$ recovers the standard \MSBAR{} distribution.

\paragraph{Quasi-\PDF{}.}

Inserting the master integral and Fourier transforming, we obtain the exact bare quasi-\PDF{} as the sum of the convergent bulk plus-distributions and the crossing-antisymmetric threshold contact terms. The full result is
\begin{align}
	\label{eq:fquasi_qg_Assembled_Rules}
	& \tilde{f}_{q/g}^{(1),\text{quasi,bare}}\!\left(x,\frac{\mu^2}{(p^z)^2}\right) \nonumber \\
	& \quad = -\frac{\alphas}{2\pi} T_R \frac{d}{4} \Bigg[
	\frac{1}{\epsilon} \Big( P_{qg}(x)\,\theta(0<x<1) - P_{qg}(-x)\,\theta(-1<x<0) \Big) \nonumber \\
	& \qquad + \left( \frac{2}{3}L_\mu^\mathcal{Q} + \frac{13}{9} \right)\big( \delta(1-x) - \delta(1+x) \big) \nonumber \\
	& \qquad + \begin{dcases}
		\left( -2 + P_{qg}(x)\log\frac{x-1}{x} + P_{qg}(-x)\log\frac{x+1}{x} \right)_{\!(+1)}^{[1,\infty)} & x>1 \\[1.4ex]
		\left( \begin{aligned} &P_{qg}(x)\big[L_\mu^\mathcal{Q} - \log\!\big(x(1-x)\big)\big] + P_{qg}(-x)\log\frac{1+x}{x} \\ &\qquad - 8x + 6x^2 \end{aligned} \right)_{\!(+1)}^{[0,1]} & 0<x<1 \\[1.4ex]
		\left( \begin{aligned} &-P_{qg}(-x)\big[L_\mu^\mathcal{Q} - \log\!\big({-}x(1+x)\big)\big] + P_{qg}(x)\log\frac{-x}{1-x} \\ &\qquad - 8x - 6x^2 \end{aligned} \right)_{\!(-1)}^{[-1,0]} & -1<x<0 \\[1.4ex]
		\left( 2 - P_{qg}(x)\log\frac{x-1}{x} - P_{qg}(-x)\log\frac{x+1}{x} \right)_{\!(-1)}^{(-\infty,-1]} & x<-1
	\end{dcases}
	\Bigg]\,,
\end{align}
where $P_{qg}$ is the \DGLAP{} splitting kernel of \cref{eq:DGLAP_qg} and $\LmuQ$ is defined in \cref{eq:LmuQ}.
The individual uncrossed diagram carries a non-integrable $\mathcal{O}(1/x)$ tail, which cancels against the crossed topology so that the full distribution decays as $\mathcal{O}(1/x^2)$ at infinity and requires no boundary counterterm at this order. The full distribution is manifestly odd under $x\to-x$.

The quasi-\PDF{} of this channel appears in refs.~\cite{Wang:2017qyg,Wang:2019tgg}, and the \MSBAR{} quasi kernel in ref.~\cite{Yao:2022vtp}. Ref.~\cite{Wang:2017qyg} evaluated only the uncrossed diagram, whose regions $x>1$ and $x<0$ are \IR{} and \UV{} finite. We can therefore compare this component directly, and we find full agreement. When we set $d/4\to1$ and subtract the collinear pole, we obtain from \cref{eq:fquasi_qg_Assembled_Rules} the one-loop \MSBAR{} matching kernel of this channel, in the combination $C_{qg}(x)-C_{qg}(-x)$ that enters the convolution with the crossing-odd gluon distribution $f_g(y) = -f_g(-y)$. This kernel agrees with the forward kernels of refs.~\cite{Wang:2019tgg,Yao:2022vtp}, which are displayed without the endpoint contact terms.

\paragraph{Pseudo-\PDF{}.}
We insert the master integrals and Fourier transform to find the bare pseudo-\PDF{}. For $0\leq x\leq1$ we obtain:
\begin{equation}
	\tilde{f}_{q/g}^{(1),\text{pseudo,bare}}(x,\mu^2 \tilde{z}^2) = -\frac{\alphas}{2\pi}
	T_R \frac{d}{4} \Big[
		\frac{1}{\epsilon} P_{qg}(x) + \LmuP{}\, P_{qg}(x)
	\Big] \theta(0\leq x \leq 1) \,, \label{eq:fpseudo_qg_bare}
\end{equation}
with the distribution on $-1\leq x<0$ following from the antisymmetry under $x\to-x$.
The collinear $1/\epsilon$ pole correctly reproduces the expected \DGLAP{} splitting kernel, and, as in the quasi-\PDF{}, is removed by matching onto the light-cone \PDF{}. The finite part is a pure logarithm, which matches the structure of the coordinate-space kernel of ref.~\cite{Yao:2022vtp}, itself proportional to $L_z$ alone in this channel.

	\section{Conclusions and Outlook\label{sec:conclusion}}

	We have computed the one-loop matrix elements of the unpolarized quark and gluon quasi- and pseudo-distributions in the \MSBAR{} scheme, together with their matching onto the flavor-singlet quark and gluon light-cone \PDF{}s. For the gluon correlator we worked throughout in the covariant six-form-factor decomposition of \cref{sec:formfactors}. Only the leading-twist form factor $\mathcal{M}_{pp}$ enters the definition of the light-cone gluon \PDF{}, and it is also the only form factor with a tree-level matrix element. The remaining five, which we call the auxiliary form factors, first appear at one loop, where they contribute to the matching kernel of any operator that does not isolate $\mathcal{M}_{pp}$. We derived the $6\times6$ matrix that mixes the form factors under renormalization (\cref{sec:gluon_operator_renorm}). Since every choice of field-strength Lorentz indices is a fixed numerical combination of these form factors, any operator used on the lattice can be assembled from our results without a further loop calculation, and the eigenvectors of the mixing matrix classify the multiplicatively renormalizable combinations completely. We recover the specific combinations established in refs.~\cite{Zhang:2018diq,Balitsky:2019krf,Balitsky:2021qsr} and extend them to the full form-factor space.

	As part of our calculation, we reproduce the known quark \MSBAR{} results~\cite{Izubuchi:2018srq,Chou:2022drv} and the quark-in-gluon matching kernel of refs.~\cite{Wang:2019tgg,Yao:2022vtp}. In the gluon-in-gluon channel we reproduce the \MSBAR{} results of refs.~\cite{Balitsky:2019krf,Balitsky:2021qsr}: the splitting kernel, all three ultraviolet coefficients, and every distributional structure. However, we find a difference of a finite polynomial in the coefficient of the leading-twist form factor $\mathcal{M}_{pp}$, \cref{eq:bmr_residual}, common to all three of their index combinations. With our form-factor decomposition we compare directly with the independent calculation of ref.~\cite{Yao:2022vtp}, with which we agree exactly, in the quasi and pseudo sectors, and in the pseudo sector for both of the gluon operator projections given there. Since the two projections weight the auxiliary form factors differently, the two comparisons together confirm our coefficient of $\mathcal{M}_{pp}$, the one in which we differ from refs.~\cite{Balitsky:2019krf,Balitsky:2021qsr}.

	We trace the difference to the appendix of ref.~\cite{Balitsky:2021qsr}, where it can be read off directly from their operator-level box result. In the gluon-exchange structures of that all-index box operator, a loop-generated index contracted between the two field strengths is summed over two transverse directions instead of $d-2$. In this way an evanescent operator is dropped, and since it multiplies the collinear pole, its omission leaves the finite polynomial. At one loop their kernel therefore matches onto a distribution that differs from $f_g^{\MSBAR}$ by the finite renormalization \cref{eq:bmr_residual}. Note that the transverse count of the operator's own indices and the polarization average of the external gluons drop out of the gluon-in-gluon kernel of a multiplicatively renormalizable combination (\cref{sec:tree}), while the index generated inside the loop, which dimensional regularization continues to $d$ dimensions, multiplies the collinear pole. The current unpolarized gluon pseudo-distribution extractions from lattice data use the kernel of refs.~\cite{Balitsky:2019krf,Balitsky:2021qsr} without this conversion, so that \cref{eq:bmr_residual} enters each of those results. In the ratio scheme, the normalization at $\nu=0$ subtracts the zeroth moment of the correction, so that the total momentum fraction $\langle x\rangle_g$ remains unchanged. For the higher moments, however, we find that at $\alphas=0.3$ the extracted normalized moments $\langle x^2\rangle_g/\langle x\rangle_g$ and $\langle x^3\rangle_g/\langle x\rangle_g$ are suppressed by approximately $11\%$ and $13\%$, respectively, relative to the \MSBAR{} values, so that the true \MSBAR{} gluon is systematically harder than reported in current lattice determinations. In proton extractions such as ref.~\cite{Delmar:2026ltr}, where statistical uncertainties remain dominant but will shrink as statistics improve, the upward correction brings the central value into close agreement with global phenomenological fits.

	We carried out the entire calculation in both a general covariant gauge and an axial gauge. The content of the individual diagrams differs between the two gauges, as does the distinction between real and virtual corrections. Only the sum of contributions is gauge invariant and the agreement between the two gauges is a stringent check of our results. The sum of contributions is also independent of the prescription chosen for the linear Wilson-line propagator, because the eikonal singularity cancels between the real and virtual corrections. This result is the equal-time counterpart of the rapidity-singularity cancellation familiar from light-cone factorization~\cite{Collins:2011zzd}. We generated the amplitudes in an automated pipeline from \qgraf{} to master integrals and reduced them by \IBP{} to three master integrals. We reproduce the one-loop gluon-operator renormalization constants of ref.~\cite{Braun:2020ymy} (\cref{app:vacuum}) through an independent two-loop vacuum calculation.

	By evaluating all Fourier transforms in general dimension $d$ before expanding in $\epsilon$, we resolve the structure of the exterior region $|x|>1$, where coordinate-space and momentum-space determinations in the literature had appeared to disagree. In particular, our gluon-in-quark results implement by construction the regularization prescription of ref.~\cite{Ji:2025cbb}, and we reproduce their kernel for the transverse-trace operator. In the gauge-invariant sum of diagrams, the Fourier transform of the correlator, $\tilde{f}$ for the quark and $x\tilde{f}$ for the gluon operator, approaches a non-integrable $1/\lvert x\rvert$ tail in every channel except the quark in a gluon, where it falls off as $1/x^2$. For the gluon in a gluon, this one-loop tail has only three possible coefficients, determined entirely by the operator's eigenvalue sector (\cref{eq:tail_classification}). More generally, by the Paley--Wiener theorem all analytic contributions to the correlator transform into distributions supported within $\lvert x\rvert\leq1$, so that at leading power the exterior of a hadronic quasi-distribution is generated entirely by the short-distance scale logarithm, whose coefficients are fixed by the renormalization group, and contains no independent non-perturbative physics. At one loop, the exterior is the integral of the scale evolution of the light-cone distribution, minus the running of the operator, against the kernel $1/\lvert x-w\rvert$ (\cref{eq:exterior_hadron}). Consequently, the exterior of a quasi-distribution reconstructed from lattice data has to agree with this relation up to power corrections, and the difference measures those corrections together with the systematics of the Fourier transform.

	Subtracting the $1/|x|$ tail leaves residual boundary distributions $\delta^\pm(1/x)/x^2$ at $x=\pm\infty$ that carry ultraviolet poles. In the matching convolution these terms evaluate the light-cone \PDF{}s at vanishing momentum fraction $y\to0$. For the isovector distribution, ref.~\cite{Izubuchi:2018srq} showed that they drop out of the convolution because the distribution grows more slowly than $1/y$ at small $y$, which is not the case for sea quarks~\cite{Chou:2022drv,Ball:2016spl} or for the gluon at $\mu^2\approx10\,\text{GeV}^2$~\cite{Ball:2016spl}. However, in both channels the boundary terms are controlled by sum rules: for quarks the two boundaries enter with equal weight, so the sea cancels and only the finite net quark number remains; for gluons the correlator carries an extra power of $y$, so the boundary terms involve momentum fractions, which the momentum sum rule bounds. In both channels the boundary terms therefore drop out of the matching convolution, poles included. Furthermore, the boundary terms probe only the $y\to0$ region, where the parton momentum is no longer a hard scale and which lies outside the reach of the perturbative matching in any case (\cref{sec:boundary}).

	\paragraph{Outlook.} Our calculation extends in several directions. In particular, the projector setup and the \IBP{} reduction are not tied to one loop and carry over to two loops, although the Fourier transforms become considerably more involved there. At higher orders it may be advantageous to combine the loop integration with the Fourier transform instead of reducing first and transforming afterwards, as we do here; this has been used in the literature~\cite{Chen:2020iqi}.

	Coulomb-gauge correlators give a Wilson-line-free definition in the same universality class~\cite{Gao:2023lny,Zhao:2023ptv}, free of the linear divergence of the Wilson-line self-energy and of its renormalon, and Coulomb-gauge fixing has already been explored for the gluon correlator on the lattice~\cite{Good:2024iur}. Since the renormalization of Coulomb-gauge correlators does not depend on the separation, the corresponding quasi-distributions can also be computed directly in momentum space, by a discrete Fourier transform over all separations in the periodic lattice volume. This has been done for the valence quark distribution of the pion~\cite{Grebe:2026qmt}. It will be interesting to work out the gluon matching in Coulomb gauge in the same form-factor language. We expect the boundary-term analysis and the exterior-region results to carry over, since by the Paley--Wiener theorem both come from the non-analytic short-distance scale logarithm at vanishing Ioffe time, and any correlator that matches onto the light-cone \PDF{}s carries this logarithm, as the renormalization group requires, with or without a Wilson line. A momentum-space calculation of this kind gives the exterior of the quasi-distribution directly and could therefore test this expectation.

	The form-factor decomposition separates the leading-twist form factor $\mathcal{M}_{pp}$ from the auxiliary form factors within the same correlator, so the present calculation is a natural starting point for studying higher-twist gluon contributions and, through the analogous decomposition, the polarized distributions. The polarized correlator $G^{\mu\nu}(z)\widetilde G^{\rho\sigma}(0)$, whose one-loop corrections have been given by Balitsky, Morris and Radyushkin~\cite{Balitsky:2021cwr}, carries a second scheme choice alongside the transverse trace, namely the treatment of $\epsilon^{\mu\nu\rho\sigma}$ away from four dimensions~\cite{tHooft:1972tcz,Breitenlohner:1977hr,Larin:1993tq}. Both prescriptions act on the operators that multiply the collinear pole, so the finite terms of a polarized matching kernel depend on the prescription chosen for each, and two kernels computed with different prescriptions differ by a finite renormalization. Neither prescription is stated in ref.~\cite{Balitsky:2021cwr}, and we have not repeated that calculation, so the convention of the polarized kernel, and the conversion it requires to reach the polarized gluon distribution in \MSBAR{}, are left open.

	A further step is to smear the fields along the gradient flow. Flowed operators replace the linear power divergence of the Wilson line with a controlled flow-time dependence and are better suited to the lattice, while our tensor decomposition and much of the machinery developed here apply unchanged.

	Since our form-factor results are operator independent, a lattice calculation can use the index combination with the best signal and discretization, and every tree-normalized choice matches onto the same continuum \PDF{}. For any such operator, the region $|x|>1$ of a one-loop kernel is fixed by the coefficient of its coordinate-space logarithm (\cref{eq:exterior_constraint}), and for a multiplicatively renormalizable operator this coefficient depends only on the eigenvalue sector (\cref{eq:log_coefficient}). In future calculations, for new operators or in other schemes, this relation can be used as a simple check of the Fourier transform to momentum space.

	\paragraph{Acknowledgments.} We would like to thank Yong Zhao for information on
	the $i\eta$-regularization of the linear propagator integrals in 
	ref.~\cite{Izubuchi:2018srq} and Robert
	Szafron 
	for discussions about Feynman rules for Wilson lines. We would also like to thank Rui-Lin Zhu, 
	Wei Wang and Shuai Zhao for 
	comments on 
	the self-energy corrections in ref.~\cite{Wang:2017qyg} and Jiunn-Wei
	Chen for comments about ref.~\cite{Chou:2022drv}. We also thank Manfred Kraus for
	an inspiring suggestion on
	\IBP{} reduction of Fourier integrals. We are grateful to Anatoly Radyushkin for
	correspondence on the gluon-in-gluon matching kernel and for sharing a detailed
	hand calculation of the box contribution of
	refs.~\cite{Balitsky:2019krf,Balitsky:2021qsr}.

	Tobias Neumann acknowledges extensive assistance from Anthropic's Claude models
	throughout the later stages of this work: the automated
	consistency suites, the exact-$d$ recomputation of the classes of
	refs.~\cite{Balitsky:2019krf,Balitsky:2021qsr}, the display-level
	comparisons with refs.~\cite{Balitsky:2019krf,Balitsky:2021qsr,Yao:2022vtp}, and
	the cross-checking of displayed results in the manuscript, as well as
	additional assistance from Google's Gemini models. All results were
	verified by the authors, who take full responsibility for the content.

	Tobias Neumann was supported by
	the United States Department of Energy under Grant Contract DE-SC0012704 for part of this work. 
	
	Christopher Monahan 
	and Tobias Neumann are supported 
	in part by USDOE grant No.~DE-SC0023047.
	Christopher Monahan is supported in part by USDOE grant No.~DE-SC0025908. This work has benefited from the collaboration enabled by the Quark-Gluon
	Tomography (QGT) Topical Collaboration, U.S. DOE
	Award No.~DE-SC0023646.
    This research used resources of the National Energy Research Scientific Computing Center (NERSC), a U.S. Department of Energy Office of Science User Facility located at Lawrence Berkeley National Laboratory, operated under Contract No. DE-AC02-05CH11231 using NERSC awards ERCAP0033748 and HEP-ERCAP0036607 \enquote{Higher-order calculations for precision collider phenomenology}.

\appendix

\crefalias{section}{appendix}
\crefalias{subsection}{appendix}
\crefalias{subsubsection}{appendix}

	\section{Feynman rules\label{app:feynman}}

	Here we collect the Feynman rules used throughout this work: the standard \QCD{} vertices and
	propagators, the eikonal rules for the Wilson line, and the vertices generated by the composite
	operator. For the \QCD{} sector we adopt the conventions of ref.~\cite{Romao:2012pq}. All external
	momenta are defined as incoming, with the exception of the ghost vertex where the incoming ghost
	momentum $p_1$ is indicated explicitly. In our diagrammatic figures, gluons appear as coils,
	quarks as solid arrowed lines, ghosts as dashed arrowed lines, the Wilson line as a double line,
	and the composite-operator insertion as a crossed circle ($\otimes$) carrying the open Lorentz
	indices $\mu\nu$. The sign factors $\eta_s,\eta_G=\pm1$ track convention choices in the sense of
	ref.~\cite{Romao:2012pq} for the signs of the gauge coupling and gauge-fixing terms; they always
	appear squared in physical quantities and cancel from all final matrix elements. In addition, $T^a$
	denote the $\SU(3)$ generators in the fundamental representation. Throughout this work, $z$ is the
	spacelike separation four-vector, $\tilde z\equiv\sqrt{-z^2}$ its length (such that $\tilde z^2=-z^2>0$,
	distinguished from the Minkowski invariant $z^2<0$), and $r^\mu\equiv z^\mu/\tilde z$ the spacelike unit
	vector along the Wilson line ($r^2=-1$).

	\paragraph{Propagators.}
	We evaluate the gluon propagator in both the general axial gauge with gauge parameter $\lambda$
	(taking the homogeneous limit $\lambda\to0$) and the covariant $R_\xi$ gauge with parameter $\xi_G$.
	Because the gauge condition leaves the spurious $1/(r\cdot p)$ poles of the axial-gauge propagator
	undetermined, we regularize them using the principal-value prescription detailed in
	\cref{sec:conventions}.
	\begin{center}
	\renewcommand{\arraystretch}{1.5}
	\begin{tabular}{@{}>{\centering\arraybackslash}m{0.22\linewidth}@{\hspace{1em}}m{0.70\linewidth}@{}}
		\includegraphics[height=8mm]{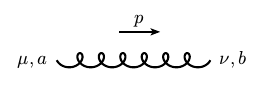} &
		$\displaystyle i\Pi^{\mu\nu,ab}_{\text{axial}}(p)=\frac{i\delta^{ab}}{p^2+i0}\left(-g^{\mu\nu}+\frac{r^\mu p^\nu+r^\nu p^\mu}{r\cdot p}-\frac{(r^2+\tilde z^2\lambda\, p^2)\,p^\mu p^\nu}{(r\cdot p)^2}\right)$ \\
		\includegraphics[height=8mm]{figs/fr-gluon-prop.pdf} &
		$\displaystyle i\Pi^{\mu\nu,ab}_{R_\xi}(p)=\frac{i\delta^{ab}}{p^2+i0}\left(-g^{\mu\nu}+(1-\xi_G)\frac{p^\mu p^\nu}{p^2}\right)$ \\
		\includegraphics[height=8mm]{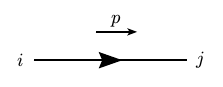} &
		$\displaystyle \frac{i\,\slashed{p}}{p^2+i0}\,\delta_{ij}$ \\
		\includegraphics[height=8mm]{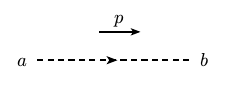} &
		$\displaystyle \frac{i\,\eta_G\,\delta^{ab}}{p^2+i0}$ \\
	\end{tabular}
	\end{center}

	\paragraph{Gauge and matter vertices.}
	\begin{center}
	\renewcommand{\arraystretch}{1.5}
	\begin{tabular}{@{}>{\centering\arraybackslash}m{0.22\linewidth}@{\hspace{1em}}m{0.70\linewidth}@{}}
		\includegraphics[height=19mm]{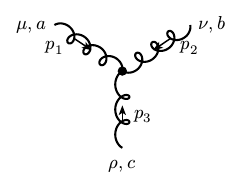} &
		$\displaystyle -\eta_s g_s f^{abc}\big[g^{\mu\nu}(p_1-p_2)^\rho+g^{\nu\rho}(p_2-p_3)^\mu+g^{\rho\mu}(p_3-p_1)^\nu\big]$ \\
		\includegraphics[height=19mm]{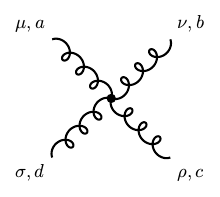} &
		$\displaystyle\begin{aligned}-ig_s^2\big[&f^{eab}f^{ecd}(g^{\mu\rho}g^{\nu\sigma}-g^{\mu\sigma}g^{\nu\rho})+f^{eac}f^{edb}(g^{\mu\sigma}g^{\nu\rho}-g^{\mu\nu}g^{\rho\sigma})\\&{}+f^{ead}f^{ebc}(g^{\mu\nu}g^{\rho\sigma}-g^{\mu\rho}g^{\nu\sigma})\big]\end{aligned}$ \\
		\includegraphics[height=19mm]{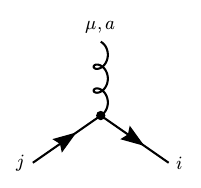} &
		$\displaystyle -i\eta_s g_s\,\gamma^\mu (T^a)_{ij}$ \\
		\includegraphics[height=19mm]{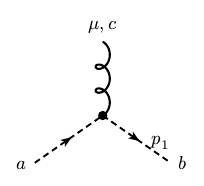} &
		$\displaystyle -\eta_s\eta_G g_s\,f^{abc}\,p_1^\mu$ \\
	\end{tabular}
	\end{center}

	\paragraph{Wilson line.}
	Following ref.~\cite{VanderVeken:2014raa}, we split the finite gauge link $W(z,0)$ at infinity into two
	semi-infinite Wilson lines, $W(z,0)=U(z;\infty)\,U(\infty;0)$, each radiating gluons through the
	eikonal rules displayed below (the segment running off to infinity is trivial and drops out).
	These eikonal rules are direction dependent and uniquely fixed: convergence of the semi-infinite
	path determines the pole prescription, so the propagator carries $+i0$ when the momentum flows
	along the path and is complex-conjugated to $-i0$ when flowing against it. Consequently, the
	vertex sign depends strictly on the path orientation, marked by the arrow along the double line
	in the diagrammatic rules.

	The gauge link sits in the fundamental representation for the quark operator and in the adjoint
	representation for the gluon operator. In the adjoint case, the Wilson-line vertex carries the
	adjoint generator $(T^a_{\mathrm{adj}})_{bc}=-if^{abc}$ in place of $(T^a)_{ij}$, yielding
	$V_{W}^{a,bc}=i\eta_s g_s\,r^\mu (T^a_{\mathrm{adj}})_{bc} =\eta_s g_s\,r^\mu f^{abc}$.
	\begin{center}
	\renewcommand{\arraystretch}{1.5}
	\begin{tabular}{@{}>{\centering\arraybackslash}m{0.22\linewidth}@{\hspace{1em}}m{0.70\linewidth}@{}}
		\includegraphics[height=8mm]{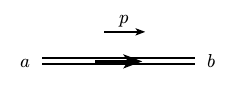} &
		$\displaystyle D_W^{ab}(k)=\frac{i\delta^{ab}}{r\cdot k+i0}\,,\qquad r^\mu=z^\mu/\tilde z$ \\
		\includegraphics[height=17mm]{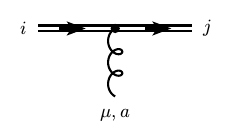} &
		$\displaystyle V_{W,ij}^{a}=i\eta_s g_s\,r^\mu (T^a)_{ij}$ \\
	\end{tabular}
	\end{center}

	\paragraph{Composite operator.}
	The field-strength insertion generates an abelian (one-gluon) and a non-abelian (two-gluon)
	vertex. In each case, an equivalent second vertex is obtained by replacing the open Lorentz
	indices $\mu,\nu$ with $\rho,\sigma$.
	\begin{center}
	\renewcommand{\arraystretch}{1.5}
	\begin{tabular}{@{}>{\centering\arraybackslash}m{0.22\linewidth}@{\hspace{1em}}m{0.70\linewidth}@{}}
		\includegraphics[height=17mm]{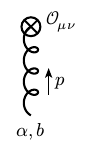} &
		$\displaystyle V_{\cal O,\mu\nu\alpha}^{ab}(p)=-i(p_\mu g_{\nu\alpha}-p_\nu g_{\mu\alpha})\,e^{-ip\cdot z}\delta^{ab}$ \\
		\includegraphics[height=17mm]{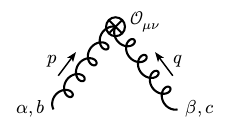} &
		$\displaystyle V_{\cal O,\mu\nu\alpha\beta}^{abc}(p,q)=-\eta_s g_s f^{abc}(g_{\mu\alpha}g_{\nu\beta}-g_{\mu\beta}g_{\nu\alpha})\,e^{-i(p+q)\cdot z}$ \\
	\end{tabular}
	\end{center}

	\subsection{Conventions and gauge fixing}
	\label{sec:feynmanrules}

	We derive the Feynman rules above from the non-abelian field strength tensor
	\begin{gather}
		G_{\mu\nu}^a = \partial_\mu A_\nu^a - \partial_\nu A_\mu^a - \eta_s g_s f^{abc} A_\mu^b
		A_\nu^c\,,
	\end{gather}
	with $\SU(3)$ structure constants $f^{abc}$ defined by $[T^a, T^b] = i f^{abc} T^c$ and
	generators $T^a$ normalized as $\operatorname{Tr}\{T^aT^b\} = T_R \delta^{ab}$, where we set
	$T_R=1/2$. The convention factor $\eta_s=\pm1$ enters the gauge transformation and the covariant
	derivative as $D_\mu = \partial_\mu + i \eta_s g_s A^a_\mu T^a$, and we verify that all physical
	matrix elements are strictly invariant under $\eta_s\to-\eta_s$~\cite{Romao:2012pq}.

	Consistency across all diagrams requires fixing the gauge uniformly to ensure that the complete
	physical matrix elements remain gauge invariant. For this purpose, we perform our loop calculations
	in both covariant $R_\xi$ gauge, following ref.~\cite{Romao:2012pq}, and general axial gauge. The
	subtleties of axial gauges, particularly regarding the pole prescription for linear propagators,
	are examined in ref.~\cite{Leibbrandt:1987qv} and adapted to our equal-time spatial Wilson lines in
	\cref{sec:conventions}.

	\subsubsection{General axial gauge}
	\label{sec:axialgauge}

		In axial gauge, we adopt the gauge-fixing and ghost Lagrangian
	\begin{align}
		\mathcal{L}^\text{axial}_\text{gauge+ghost} & = -\frac{1}{2\lambda} (r^\mu A_\mu^a)^2  + \eta_G\bar{c}^a r^\mu (\delta^{ab} \partial_\mu + \eta_s g_s f^{abc} A_\mu^c) c^b\,,
	\end{align}
	which depends on the gauge parameter $\lambda$ and the spacelike four-vector $r^\mu$. While the gauge-fixing term carries the parameter $\lambda$, the ghost Lagrangian depends on the gauge choice only through the reference vector $r^\mu$ and incorporates the convention sign choice $\eta_G = \pm 1$, which appears as $\eta_G^2$ in all contributions relevant here. Following ref.~\cite{Leibbrandt:1987qv}, we use the term ``general axial gauge'' for the gauge condition $r^2<0$ with $\lambda\neq0$, to distinguish it from homogeneous axial gauge defined by $r^2<0$ and $\lambda=0$. In general axial gauge ghost fields are formally present, but they decouple from gauge-invariant quantities in the limit $\lambda\to0$.

	Specifically, we choose $r$ along the spatial $z$ direction, $r^\mu=(0,0,0,1)$, and take the limit $\lambda\to0$, which imposes $A_z=0$. Under this choice, the spatial Wilson line along the $z$ direction collapses to unity and drops out of all Feynman diagrams, while the axial-gauge propagator reduces to
	\begin{gather}
		i\Pi^{\mu\nu,ab}_\text{axial}(p) = \frac{i\delta^{ab}}{p^2 + i0} \left(
		-g^{\mu\nu} + \frac{r^\mu p^\nu + r^\nu p^\mu}{ p^z }
		+ \frac{p^\mu p^\nu}{(p^z)^2}
		\right) \,.
	\end{gather}
	The ghosts decouple as well, because the ghost-antighost-gluon vertex is proportional to $r_\mu$ and the propagator satisfies $r_\mu \Pi^{\mu\nu,ab}_\text{axial}=0$ in this limit, so that the ghost vertex vanishes when contracted with any internal gluon line.

	For quarks, the direct correspondence between these additional axial-gauge propagator terms and the Wilson-line diagrams is discussed in ref.~\cite{Ji:2015jwa}.\footnote{See
	fig.~1 of ref.~\cite{Ji:2015jwa}, where the second row represents the three terms in the axial
	gauge. Note that for the second diagram in that figure the conjugate diagram must be added to
	account for both contributions to the second term in the axial gauge, providing a useful sign check on the Feynman rules.}

	\subsubsection{Covariant \texorpdfstring{$R_\xi$}{R\_xi} gauge and the Wilson line}

	To check gauge independence, we perform companion calculations in covariant $R_\xi$ gauge using the gauge-fixing Lagrangian
	\begin{equation}
		\mathcal{L}^\text{cov.}_\text{gauge}  = -\frac{1}{2\xi_G} (\partial^\mu A_\mu^a)^2\,.
	\end{equation}
	In this gauge, the Wilson line contributes explicitly to Feynman diagrams. We define a Wilson line connecting spacetime points $a$ and $b$ along a path $\mathcal{C}$ as
	\begin{gather}
		W^\mathcal{C}(b,a) = \mathcal{P}\exp\left( i \eta_s g_s \int_a^b \mathrm{d}\lambda\,
		\frac{\mathrm{d}x^\mu}{\mathrm{d}\lambda} A_\mu(x)
		\right)\,,
	\end{gather}
	where $x^\mu(\lambda)$ parametrizes the contour and $A_\mu(x) = A_\mu^a(x) T^a$. For the straight spacelike path $x^\mu(\lambda)= r^\mu \lambda$ with $\lambda\in[0,\zt]$ and $r^\mu = (0,0,0,1)$ relevant to equal-time lattice correlators, this parametrization yields
	\begin{align}
		W(z,0) & = \mathcal{P}\exp\left(  i \eta_s g_s \int_0^{\zt} \mathrm{d}\lambda\,
		r^\mu A_\mu(\lambda)
		\right) =
		\mathcal{P}\exp\left( - i \eta_s g_s  \int_0^{\zt} \mathrm{d}\lambda\,
		A_z(\lambda)
		\right)\,,
	\end{align}
	where the minus sign in the second form arises because $r^\mu A_\mu = A_3 = -A^3 = -A_z$ in our mostly-minus Minkowski metric signature. Finally, following ref.~\cite{VanderVeken:2014raa}, we incorporate a convergence factor $+i0$ in the semi-infinite path parametrization, which produces the direction-dependent $\pm i0$ prescriptions of the eikonal rules displayed above.
\section{Fourier transforms\label{app:FT}}

Quasi- and pseudo-distributions are obtained from the coordinate-space matrix elements of \cref{sec:matrixelems} by Fourier transformation to the parton momentum fraction $x$, conjugate to the Ioffe time $\nu = \zt P^z = -P\cdot z$ (\cref{sec:definitions}). For quasi-distributions, the defining transform is taken in the spatial separation $\zt$ at fixed hadron momentum $P^z$ with phase $e^{i\zt x P^z}$ (\cref{eq:fgquasi}); substituting $\nu = \zt P^z$ converts this integration into a Fourier transform over $\nu$ with phase $e^{i\nu x}$, while the short-distance factor $(\zt^2)^\epsilon$ becomes $(\nu^2)^\epsilon/(P^z)^{2\epsilon}$, whose momentum scale combines with $\mu^{2\epsilon}$ into the scale logarithm $L_\mu^\mathcal{Q}$. For pseudo-distributions, the transform is taken directly with respect to $\nu$ at fixed spacelike separation $z^2 = -\zt^2$, leaving $(\zt^2)^\epsilon$ as an overall prefactor that factors out of the Fourier integration. Throughout this appendix, all Fourier transforms are therefore formulated uniformly as integrals over the Ioffe time $\nu$ with measure $\int\frac{\mathrm{d}\nu}{2\pi}e^{i\nu x}$. At tree level, the correlator depends on $\zt$ only through the phase $e^{i P\cdot z} = e^{-i\nu}$, which transforms to the tree-level distribution $\delta(1-x)$.

We Fourier transform the assembled matrix elements in general dimension $d=4-2\epsilon$ after reducing the loop integrals to the master integrals of \cref{sec:oneloopints}. Keeping the dimension generic before expanding in $\epsilon$ avoids the scale ambiguities of Fourier-transforming $\log(\zt^2)$ terms directly (\cref{sec:boundary}) and determines the boundary terms at $x=\pm\infty$ uniquely. In \cref{app:FTkernels} we classify the kernel structures and their master power-law transforms. In \cref{app:FTplus} we formulate the regularized plus distributions and boundary terms that isolate the collinear and ultraviolet singularities. Finally, in \cref{app:FTadvanced} we compute the sign-exponential and incomplete-gamma transforms.

\subsection{Kernel inventory and the power-law transform\label{app:FTkernels}}

After reducing the loop amplitudes to the master integrals of \cref{sec:oneloopints}, the $\nu$ dependence of every one-loop matrix element decomposes into a small number of kernel classes. For the quasi-\PDF{} transforms, these are the power-law kernels $(\nu^2)^\epsilon/\nu^n$ ($0\leq n\leq4$), either bare or accompanied by a phase $e^{\mp i\nu}$, together with the sign-exponential kernel $e^{-i\nu}(-i\nu)^{-2\epsilon}(\nu^2)^\epsilon$, the incomplete-gamma kernel $e^{-i\nu}(-i\nu)^{-2\epsilon}(\nu^2)^\epsilon\,\Gamma(2\epsilon,-i\nu)$, and their $e^{+i\nu}$ mirrors. For the pseudo-\PDF{} transforms, the classes are the rational powers $1/\nu^n$, the pure phases $e^{\mp i\nu}$, and the corresponding kernels without the $(\nu^2)^\epsilon$ factor, $e^{-i\nu}(-i\nu)^{-2\epsilon}$ and $e^{-i\nu}(-i\nu)^{-2\epsilon}\,\Gamma(2\epsilon,-i\nu)$. These functions originate from the master integrals through the factors $(\mp i\nu)^{d-3}$ and $\Gamma(3-d,\mp i\nu)$: with the recurrence relation $\Gamma(s+1,x) = s\,\Gamma(s,x) + x^s e^{-x}$, the incomplete gamma functions reduce to index $2\epsilon$, while partial fractioning the rational \IBP{} coefficients in $\nu$ yields the power-law kernels $1/\nu^n$.

The power-law Fourier transforms evaluate according to a single master formula, valid at generic $\epsilon$ for integer $n\geq0$:
\begin{equation}
\int_{-\infty}^\infty \frac{\mathrm{d}\nu}{2\pi}\, e^{i\nu(x-c)}\, \frac{(\nu^2)^\epsilon}{\nu^n} = -(-i)^n\, \frac{\Gamma(1-n+2\epsilon)}{\Gamma(1-\epsilon)\,\Gamma(\epsilon)}\, |x-c|^{n-1-2\epsilon}\, \operatorname{sign}^n(x-c)\,,
\label{eq:ftPowerMaster}
\end{equation}
where any exponential factor of the kernel is absorbed into the centering $c$: kernels accompanied by $e^{\mp i\nu}$ have $c=\pm1$, while kernels without an exponential phase have $c=0$. In particular, for $n\geq1$, individual diagrammatic terms carry kinematic poles $1/\nu^n$ at $\nu=0$. In dimensional regularization, \cref{eq:ftPowerMaster} evaluates their Fourier transforms by standard analytic continuation in $\epsilon$ from the domain $\operatorname{Re}(\epsilon) > (n-1)/2$ where the integral converges. In the complete physical correlator, the short-distance operator product expansion guarantees regularity at $\nu=0$ (\cref{sec:symmetries}, and shown explicitly in \cref{sec:quarkinquark}), so all kinematic $1/\nu^n$ poles cancel in the sum of diagrams.

Consequently, the individual $1/\nu^n$ kernels produce Fourier transforms with non-decaying tails: the $n=1$ instances approach a constant as $|x-c|^{-2\epsilon}$, while the $n\geq2$ instances grow as $|x-c|^{n-1-2\epsilon}$. In \cref{app:FTplus} we regularize these tails term by term.

For pseudo-distributions, because the Fourier transform is evaluated at fixed $z^2$, the factor $(\nu^2)^\epsilon$ is absent from the integration. The power-law kernels therefore reduce to the bare rational powers $1/\nu^n$, whose Fourier transforms evaluate elementarily: for $n=1,\dots,4$ and centerings $c\in\{-1,0,+1\}$,
\begin{equation}
	\int_{-\infty}^\infty \frac{\mathrm{d}\nu}{2\pi}\, e^{i\nu(x+c)}\, \frac{1}{\nu^{n}}
	= \frac{i^{\,n}}{2\,(n-1)!}\,(x+c)^{n-1}\,\operatorname{sign}(x+c)\,,
	\label{eq:ftPseudoPoly}
\end{equation}
where each member of the family is defined by a common symmetric prescription at $\nu=0$ (principal value for odd $n$, Hadamard finite part for even $n$). \Cref{eq:ftPseudoPoly} is the $\epsilon\to0$ limit of the master formula \eqref{eq:ftPowerMaster}, and these kinematic poles cancel in the assembled pseudo-\PDF{} by the same regularity argument. The remaining pseudo-\PDF{} kernels are the pure phases $e^{\mp i\nu}$, which transform to the contact terms $\delta(1\mp x)$, and the $\epsilon$-dependent transforms without the $(\nu^2)^\epsilon$ factor, \cref{eq:ruleGammaPowerExact,eq:rule21exact}, evaluated in \cref{app:FTadvanced}.

The matrix elements also contain each kernel class with the opposite phase factor $e^{+i\nu}$ (and, for the kernels of \cref{app:FTadvanced}, with $(+i\nu)^{-2\epsilon}$ and $\Gamma(2\epsilon,+i\nu)$ in place of their $-i$ counterparts). All mirrored transforms follow from the derived ones through the exact reflection identity
\begin{equation}
	\int_{-\infty}^\infty \frac{\mathrm{d}\nu}{2\pi}\, e^{i\nu x}\, K(-\nu) \;=\; \left(\int_{-\infty}^\infty \frac{\mathrm{d}\nu}{2\pi}\, e^{i\nu x'}\, K(\nu)\right)_{x'\to -x}\,,
	\label{eq:ftReflection}
\end{equation}
so that the transform of the reflected kernel is obtained by replacing $x\to-x$ in the original transform. Under this reflection, the four support regions interchange pairwise ($\{x>1\}\leftrightarrow\{x<-1\}$ and $\{0<x<1\}\leftrightarrow\{-1<x<0\}$), the contact terms reflect as $\delta(x\mp1)\leftrightarrow\delta(x\pm1)$, and the boundary distributions at $+\infty$ and $-\infty$ interchange ($\delta^\pm\leftrightarrow\delta^\mp$). For the power-law kernels, this reflection is already incorporated in the centering $c=\pm1$ of \cref{eq:ftPowerMaster}. For the sign-exponential and incomplete-gamma kernels, we derive the $e^{-i\nu}$ representatives and obtain their mirrors directly from \cref{eq:ftReflection}.

\subsection{Plus distributions and boundary terms\label{app:FTplus}}

\paragraph{Plus distributions.} On a finite interval $[a,b]$, a plus distribution $[g(x)]_{(+c)}^{[a,b]}$ subtracts the value of a test function at $x=c$:
\begin{equation}
\int_a^b \mathrm{d}x\, [g(x)]_{(+c)}^{[a,b]} f(x) = \int_a^b \mathrm{d}x\, g(x)\left(f(x) - f(c)\right)\,.
\end{equation}
For a power divergence $x^{-1+\epsilon}$ on $[0,1]$ with $\epsilon>0$, adding and subtracting $f(0)$ isolates the $1/\epsilon$ pole through the endpoint integral:
\begin{align}
    \int_0^1 \mathrm{d}x\, \frac{f(x)}{x^{1-\epsilon}} & = \int_0^1 \mathrm{d}x\, 
    \frac{f(x)-f(0)}{x^{1-\epsilon}} + f(0) \int_0^1 \mathrm{d}x\, \frac{1}{x^{1-\epsilon}} \nonumber \\
    & \equiv \int_0^1 \mathrm{d}x\, \left[\frac{1}{x^{1-\epsilon}}\right]_{(+0)}^{[0,1]} f(x) + \frac{1}{\epsilon}\, f(0)\,.
\end{align}
Expanding $x^{-1+\epsilon}$ in powers of $\epsilon$ then generates the standard logarithmic plus distributions $[\log^n(x)/x]_{(+0)}^{[0,1]}$ alongside the contact term $\delta(x)/\epsilon$.

\paragraph{Subtractions at $\pm\infty$ and boundary deltas.} On an unbounded domain such as $[1,\infty)$, we regularize the non-integrable tail $x^{-1+\epsilon}$ (convergent for $\epsilon <0$) by mapping to the unit interval through $t=1/x$, where $x\to\infty$ maps to $t\to0^+$. Subtracting the boundary value at $t=0$ and transforming back yields
\begin{equation}
\int_1^\infty \mathrm{d}x\, \frac{\Phi(x)}{x^{1-\epsilon}} = \int_1^\infty\mathrm{d}x\, \left[\frac{1}{x^{1-\epsilon}}\right]_{(+\infty)}^{[1,\infty)} \Phi(x) - \frac{1}{\epsilon} \int_{1}^{\infty}\mathrm{d}x\, \frac{\Phi(x)}{x^2} \delta^+(1/x)\,,
\label{eq:plusInfinity}
\end{equation}
where we adopt the notation of refs.~\cite{Izubuchi:2018srq,Chou:2022drv}, in which $\delta^+(1/x) = \lim_{\beta\to 0^+} \delta(1/x - \beta)$ approaches zero from above, placing its support at $x=+\infty$. The boundary delta pairs with test functions as $\langle \delta^+(1/x)/x^2, \Phi\rangle = \Phi(+\infty)$. Applying the same transformation to $(-\infty,-1]$ gives the mirror identity
\begin{equation}
\int_{-\infty}^{-1} \mathrm{d}x\, \frac{\Phi(x)}{|x|^{1-\epsilon}} = \int_{-\infty}^{-1}\mathrm{d}x\, \left[\frac{1}{|x|^{1-\epsilon}}\right]_{(+(-\infty))}^{(-\infty,-1]} \Phi(x) - \frac{1}{\epsilon} \int_{-\infty}^{-1}\mathrm{d}x\, \frac{\Phi(x)}{x^2} \delta^-(1/x)\,,
\label{eq:plusMinusInfinity}
\end{equation}
where $\delta^-(1/x) = \lim_{\beta\to 0^+} \delta(1/x + \beta)$ has support at $x=-\infty$ and pairs to $\langle \delta^-(1/x)/x^2, \Phi\rangle = \Phi(-\infty)$.

\paragraph{Simultaneous singularities at $1$ and $\infty$.} In the quasi-\PDF{}, matrix elements on $[1,\infty)$ typically carry both a collinear singularity at $x=1$ and an asymptotic $1/x$ tail at infinity. In this case, we disentangle the two singularities by subtracting and adding back the asymptotic tail $x^{-1-\epsilon}$:
\begin{align}
	\int_1^\infty \mathrm{d}x\, \Phi(x) (x-1)^{-1-\epsilon} = {} & \int_1^\infty\mathrm{d}x\, \Phi(x) \Big( (x-1)^{-1-\epsilon} - x^{-1-\epsilon}\Big)_{(+1)}^{[1,\infty)} - \frac{1}{\epsilon}\,\Phi(1) \nonumber \\
	{} & + \int_1^\infty\mathrm{d}x\, \left[\frac{1}{x^{1+\epsilon}}\right]_{(+\infty)}^{[1,\infty)} \Phi(x) + \frac{1}{\epsilon} \int_{1}^{\infty}\mathrm{d}x\, \frac{\Phi(x)}{x^2} \delta^+(1/x)\,.
\end{align}
This decomposition cleanly isolates the collinear divergence at $x=1$ as an infrared pole $-1/\epsilon$ multiplying $\delta(x-1)$, while the large-$x$ behavior is contained entirely in the plus distribution $[\,\cdot\,]_{(+\infty)}^{[1,\infty)}$ and the ultraviolet boundary pole $+1/\epsilon$ multiplying $\delta^+(1/x)/x^2$. We treat all unbounded regions in the quasi-\PDF{} in this manner.

\paragraph{Growing tails and moment subtractions.} For intermediate bookkeeping of individual diagrammatic terms, we regularize the growing tails $|x-c|^{n-1-2\epsilon}$ generated by the kinematic poles $1/\nu^n$ ($n\geq 1$) of \cref{eq:ftPowerMaster}. On unbounded domains, subtracting the boundary value alone is insufficient. We therefore subtract the boundary value together with subsequent moments of the test function, pairing the higher subtractions with derivatives of the boundary delta. Expanding the test function as $\Phi(x) = \Phi(\infty) + \Phi^+_1/x + \mathcal{O}(1/x^2)$ as $x\to\infty$, we define the first-moment boundary distribution by
\begin{equation}
	\left\langle \frac{(\delta^+)'(1/x)}{x^2},\,\Phi\right\rangle = -\Phi^+_1\,,
	\label{eq:boundaryDeltaPrime}
\end{equation}
with the analogous definition at $-\infty$. To illustrate how moment subtractions regularize an unbounded domain, consider the tail $x^{-2\epsilon}$ restricted to the region $[1,\infty)$ (leaving aside the other regions, including the interior $[-1,1]$, where the power law is integrable). Subtracting the value and first moment on $[1,\infty)$ and evaluating the endpoint integrals by analytic continuation yields the distributional identity
\begin{equation}
	x^{-2\epsilon}\,\theta(x-1) = \left[x^{-2\epsilon}\right]_{(+\infty,1)}^{[1,\infty)}
	- \frac{1}{1-2\epsilon}\,\frac{\delta^+(1/x)}{x^2}
	- \frac{1}{2\epsilon}\,\frac{(\delta^+)'(1/x)}{x^2}\,,
	\label{eq:conservativeSubtraction}
\end{equation}
where the order subscript $1$ marks the subtraction of the first moment, and the mirror identity at $-\infty$ involves $\delta^-(1/x)/x^2$ and $(\delta^-)'(1/x)/x^2$. The value delta appears with a finite coefficient, while the $1/\epsilon$ boundary pole multiplies the first-moment derivative alone. For higher powers $1/\nu^n$, moments through order $n$ are subtracted, with the $1/\epsilon$ pole attaching to the highest derivative $(\delta^\pm)^{(n)}(1/x)/x^2$.

We stress that this derivative-delta machinery is purely internal bookkeeping: all growing tails and their associated derivative boundary deltas cancel identically in the complete physical correlator. By locality (\cref{sec:quarkinquark}), the $1/\nu^n$ kernels enter only in combinations regular at $\nu=0$, such as $(1-i\nu-e^{-i\nu})/\nu^2$, whose Fourier transforms decay at least as fast as $1/x$. The derivative boundary deltas therefore drop out of the assembled quasi-distributions, leaving only the contact terms $\delta(1\mp x)$ and the value boundary deltas $\delta^\pm(1/x)/x^2$.

We retain these value boundary deltas throughout because their coefficients carry ultraviolet $1/\epsilon$ poles that survive multiplicative renormalization: in the quark quasi-\PDF{}, these poles combine to $-3/(4\epsilon)$ on each of $\delta^\pm(1/x)/x^2$ (\cref{sec:quarkinquark}), while the corresponding gluon boundary poles are given in \cref{eq:quasi_gg_towers}. As established in \cref{sec:boundary}, these boundary poles cancel the non-vanishing integral of the interior poles so that the bare quasi-distribution integrates to zero, while in the matching convolution they probe the light-cone distribution only at vanishing momentum fraction.

\subsection{Sign-exponential and incomplete-gamma transforms\label{app:FTadvanced}}

We evaluate the sign-exponential and incomplete-gamma transforms without expanding in $\epsilon$.

\paragraph{Sign function exponential.} For $\nu\neq0$, the factor $(-i\nu)^{-2\epsilon}(\nu^2)^\epsilon$ reduces to a pure phase, $e^{i\pi\epsilon\operatorname{sign}(\nu)} = \cos(\pi\epsilon) + i\sin(\pi\epsilon)\operatorname{sign}(\nu)$. The $\epsilon$ dependence therefore factors out of the integration: the constant term transforms to $\delta(x-1)$, while $\operatorname{sign}(\nu)$ generates the Cauchy principal value $\text{P.V.}[1/(x-1)]$. The transform evaluates to
\begin{equation}
\int_{-\infty}^\infty \frac{\mathrm{d}\nu}{2\pi}\, e^{i\nu(x-1)}\, (-i\nu)^{-2\epsilon} (\nu^2)^\epsilon =
	\cos(\pi\epsilon)\,\delta(x-1) - \text{P.V.}\left(\frac{\sin(\pi\epsilon)}{\pi(x-1)}\right)\,,
	\label{eq:rule38exact}
\end{equation}
where the principal value acts on a test function $\Phi(x)$ as
\begin{equation}
	\left\langle \text{P.V.}\,\frac{1}{x-1}, \Phi \right\rangle := \lim_{\eta\to0^+} \int_{|x-1|>\eta}\mathrm{d}x\,\frac{\Phi(x)}{x-1}\,.
\end{equation}
Because the trigonometric coefficients expand as $\cos(\pi\epsilon)=1-\pi^2\epsilon^2/2+\mathcal{O}(\epsilon^4)$ and $\sin(\pi\epsilon)/\pi=\epsilon+\mathcal{O}(\epsilon^3)$, this transform is finite and carries no $1/\epsilon$ pole. In intermediate steps on individual half-line domains, we regularize the unbounded $1/(x-1)$ tails by explicit plus-distribution subtractions at $\pm\infty$. When summed over the entire real line, these boundary evaluations cancel exactly by the antisymmetry of $1/(x-1)$ about $x=1$, leaving no net boundary delta $\delta^\pm(1/x)/x^2$.

\paragraph{Incomplete gamma function with sign exponential.} We next compute
\begin{equation}
\int_{-\infty}^\infty \frac{\mathrm{d}\nu}{2\pi}\, e^{i\nu(x-1)}\, (-i\nu)^{-2\epsilon}(\nu^2)^\epsilon\, \Gamma(2\epsilon,-i\nu)\,.
\label{eq:rule39def}
\end{equation}
For $\nu\neq0$, we rewrite the prefactor as $e^{i\pi\epsilon\operatorname{sign}(\nu)}$ and evaluate this integral for $x\neq 1$ by integration by parts. Integrating the phase $e^{i\nu(x-1)}$ yields $1/[i(x-1)]$, while differentiating $e^{i\pi\epsilon\operatorname{sign}(\nu)}\,\Gamma(2\epsilon,-i\nu)$ produces
\begin{equation}
\frac{\mathrm{d}}{\mathrm{d}\nu}\, e^{i\pi\epsilon\operatorname{sign}(\nu)}\, \Gamma(2\epsilon,-i\nu) = -\frac{e^{i\nu}(\nu^2)^\epsilon}{\nu} + 2 i\,\Gamma(2\epsilon) \sin(\pi\epsilon) \delta(\nu)\,,
\end{equation}
where the second term arises from the jump of the sign exponential at $\nu=0$. The boundary terms at $\nu=\pm\infty$ vanish because $\Gamma(2\epsilon,-i\nu) \sim (-i\nu)^{2\epsilon-1} e^{i\nu}$ for $\epsilon < 1/2$. Evaluating the remaining Fourier transform gives
\begin{align}
\int_{-\infty}^\infty \frac{\mathrm{d}\nu}{2\pi}\, e^{i\nu(x-1)}\, & (-i\nu)^{-2\epsilon}(\nu^2)^\epsilon\, \Gamma(2\epsilon,-i\nu) \nonumber \\
& = \frac{1}{i(x-1)} \int_{-\infty}^\infty \frac{\mathrm{d}\nu}{2\pi}\, e^{i\nu x} \left( \frac{(\nu^2)^\epsilon}{\nu} - 2i\Gamma(2\epsilon) \sin(\pi\epsilon) \delta(\nu) \right) \nonumber \\
& = -\frac{\Gamma(2\epsilon)}{\Gamma(1-\epsilon)\,\Gamma(\epsilon)}\, \frac{1-\operatorname{sign}(x)\,|x|^{-2\epsilon}}{x-1}\,,
\label{eq:rule39exact}
\end{align}
where the last step applies Euler's reflection formula $\sin(\pi\epsilon)/\pi = 1/[\Gamma(1-\epsilon)\,\Gamma(\epsilon)]$.
Since integration by parts assumes $x\neq 1$, contact terms proportional to $\delta(x\mp1)$ could in principle be present. Pairing the defining integral \cref{eq:rule39def} directly with test functions via the representation $\Gamma(2\epsilon,-i\nu) = e^{i\nu}\int_0^\infty \mathrm{d}s\,(s-i\nu)^{2\epsilon-1}\,e^{-s}$ confirms that no contact terms arise, establishing \cref{eq:rule39exact} as the complete distributional result. When cast in the subtracted form of \cref{app:FTplus}, the endpoint integrals accompanying the plus distributions at $x=\pm1$ are finite and of order $\epsilon$, so this kernel carries no $1/\epsilon$ contact pole. Its mirror $(+i\nu)^{-2\epsilon}(\nu^2)^\epsilon\,\Gamma(2\epsilon,+i\nu)\,e^{+i\nu}$ follows directly from \cref{eq:ftReflection}.

\paragraph{Incomplete gamma function with power.} Integration by parts does not yield a closed form for this kernel. Instead, using the integral representation $\Gamma(s,z) = z^s \int_1^\infty\mathrm{d}u\, u^{s-1} e^{-uz}$ with $s=2\epsilon$ and $z=-i\nu$, the prefactor $(-i\nu)^{2\epsilon}$ cancels the $(-i\nu)^{-2\epsilon}$ in the integrand, and the remaining $\nu$ integral generates the Dirac delta $\delta(x-1+u)$, giving
\begin{equation}
	\int_{-\infty}^\infty \frac{\mathrm{d}\nu}{2\pi}\, e^{i\nu(x-1)}\, (-i\nu)^{-2\epsilon}\, \Gamma(2\epsilon,-i\nu) = \int_1^\infty\mathrm{d}u\, u^{2\epsilon-1} \delta(x-1+u) = (1-x)^{2\epsilon-1}\theta(-x)\,.
	\label{eq:ruleGammaPowerExact}
\end{equation}
The support is strictly confined to $(-\infty,0]$, where the tail $(1-x)^{2\epsilon-1}$ is a standard dimensionally regulated power law regularized by a boundary subtraction at $-\infty$.

Finally, the companion pseudo-\PDF{} transform without the incomplete gamma function evaluates to
\begin{equation}
	\int_{-\infty}^\infty \frac{\mathrm{d}\nu}{2\pi}\, e^{i\nu(x-1)}\, (-i\nu)^{-2\epsilon} = \frac{\theta(1-x)\, |x-1|^{-1+2\epsilon}}{\Gamma(2\epsilon)}\,.
	\label{eq:rule21exact}
\end{equation}
Both \cref{eq:ruleGammaPowerExact,eq:rule21exact} have support bounded from above by $x\leq 1$, so that their non-integrable tails and boundary terms lie exclusively at $x=-\infty$, providing explicit realizations of the one-sided boundary terms discussed in \cref{sec:boundary}.
\section{Two-loop vacuum calculation of the gluon renormalization constants\label{app:vacuum}}

The form-factor mixing matrix of \cref{sec:gluon_operator_renorm} is built from the two vacuum eigenvalues $Z_{\perp\perp}$ and $Z_{\parallel\perp}$ of \cref{eq:sector_values}, which we determine from a dedicated two-loop vacuum calculation. Removing the coupling renormalization that distinguishes our operator from that of ref.~\cite{Braun:2020ymy} reproduces their one-loop values exactly. In this appendix we summarize the two-loop vacuum calculation and its results. Beyond determining the two eigenvalues, it tests our diagram generation, color and Lorentz algebra, \IBP{} reduction, and master integrals on an independent set of two-loop topologies.

In on-shell partonic matrix elements, the loop corrections contain collinear divergences alongside the ultraviolet ones. Because our reduction to master integrals uses dimensional regularization with $\epsilon_\IR = \epsilon_\UV$, the partonic amplitudes cannot separate the two types of poles on their own. In the vacuum, by contrast, the operator has no external states, the correlator depends only on the single invariant scale $z^2$, and all loop divergences are strictly \UV{}. We can therefore read off the renormalization constants directly, without needing an auxiliary infrared regulator or risking mixing with gauge-non-invariant operators~\cite{Braun:2020ymy}. However, the one-loop vacuum correlator is \UV{}-finite and only fixes the tree normalization, so the renormalization constants first appear at two loops.

Ref.~\cite{Braun:2020ymy} defines the bilocal operator with an explicit factor of the strong coupling,
\begin{equation}\label{eq:Braundef}
\mathbb{G}_{\mu\nu\alpha\beta}(z)=g_s^2 G_{\mu\nu}(z)\,W(z,0)\,G_{\alpha\beta}(0)\,,
\end{equation}
with the adjoint Wilson line along the spacelike direction $z^\mu$. In the auxiliary-field formalism of that reference, the Wilson line is represented by an auxiliary field $h_v$ along the direction $v^\mu = z^\mu / \zt$, factorizing the bilocal operator into two local endpoint operators:
\begin{equation}
  G^{\perp\perp}_{\mu\nu}=g^\perp_{\mu\alpha}g^\perp_{\nu\beta}\,g_s G^{\alpha\beta}h_v\,,
  \qquad
  G^{\parallel\perp}_{\mu\nu}=\big(g^\parallel_{\mu\alpha}g^\perp_{\nu\beta}
  -g^\parallel_{\nu\alpha}g^\perp_{\mu\beta}\big)\,g_s G^{\alpha\beta}h_v\,.
\end{equation}
These two structures renormalize multiplicatively with constants denoted $Z^{\text{B}}_{\perp\perp}$ and $Z^{\text{B}}_{\parallel\perp}$. In our calculation, we work directly with the bilocal operator without auxiliary fields. We project its vacuum expectation value onto the two sectors using the projectors $\hat{\Pi}^{\perp\perp}$ and $\hat{\Pi}^{\parallel\perp}$ of \cref{sec:gluon_operator_renorm}. Because the operator carries two field strengths, the vacuum correlator in each sector renormalizes with the square of the corresponding renormalization constant, $Z_{\perp\perp}^2$ or $Z_{\parallel\perp}^2$.

We generate the vacuum diagrams with \qgraf{}~\cite{Nogueira:1991ex} and classify them by the number of gluons emitted from the Wilson line: 32 diagrams with zero emissions, five with one emission, and three diagrams with two emissions, including the Wilson-line self-energy. The diagrams are evaluated with the Feynman rules of \cref{app:feynman}, keeping the covariant gauge parameter $\xi_G$ generic to verify the gauge independence of the sum.

We project each diagram onto $\hat{\Pi}^{\perp\perp}$ or $\hat{\Pi}^{\parallel\perp}$, carry out the color and Lorentz algebra, and map the amplitudes onto two-loop vacuum topologies with gluon propagators and linear eikonal propagators from the Wilson line. Following the $R_\xi$ Feynman rules, the sign in the eikonal prescription, $+i0$ or $-i0$, depends on the direction of momentum flow along the line. Using integration-by-parts (\IBP{}) reduction with \texttt{Kira}~\cite{Maierhofer:2017gsa,Klappert:2020nbg}, all diagrams reduce to two master integrals: a product of two one-loop massless bubbles (which is finite) and a single non-trivial two-loop eikonal vacuum integral
\begin{align}
  J^\pm(d, z^2) &= \int \frac{\mathrm{d}^d k}{(2\pi)^d} \frac{\mathrm{d}^d \ell}{(2\pi)^d} \frac{1}{k^2 (k-\ell)^2 (\ell \cdot r \pm i0)} \nonumber \\
  &= \frac{i\,4^{d-5}}{\pi^{d+1/2}} (\zt^2)^{2-d} C^\pm(d) \Gamma(3-d) \Gamma\left(\frac{d}{2}-1\right)^2 \Gamma\left(d - \frac{5}{2}\right) ,
\end{align}
with $r^\mu = z^\mu / \zt$, $\zt^2 = -z^2$, and the eikonal prescription coefficients
\begin{equation}
  C^+(d) = 1 \,, \qquad C^-(d) = 2 \cos(d\pi) - 1 \,.\label{eq:vac_eikonal_coeffs}
\end{equation}
Since $C^-(4-2\epsilon) = 1 + \mathcal{O}(\epsilon^2)$, the choice of eikonal sign does not affect the $1/\epsilon$ pole of the master integral.

Parameterizing the bare correlator in each sector as $\Pi^{\text{bare}}=\Pi^{(0)}\,(1+Z_\Pi^{(1)}\,a/\epsilon+\dots)$ with $a=g_s^2/(16\pi^2)$, we extract the pole coefficients
\begin{equation}\label{eq:ZPi}
  Z_\Pi^{(1),\perp\perp}=\frac{5C_A-4T_Rn_f}{3}\,,\qquad
  Z_\Pi^{(1),\parallel\perp}=\frac{11C_A-4T_Rn_f}{3}=\beta_0\,,
\end{equation}
with $\beta_0=\tfrac{11}{3}C_A-\tfrac{4}{3}T_R n_f$. Since the bare correlator renormalizes as $\Pi^{\text{bare}} = Z^{-2}\Pi$ with finite $\Pi$, the one-loop constants are $Z^{(1)} = -Z_\Pi^{(1)}/2$ in each sector, reproducing \cref{eq:sector_values}, and the corresponding diagonal entries of the form-factor mixing matrix are $(\widetilde{Z}_\mathcal{M}^{(1)})_{pp,pp} = 2Z_{\perp\perp}^{(1)} = -Z_\Pi^{(1),\perp\perp}$ and $(\widetilde{Z}_\mathcal{M}^{(1)})_{zz,zz} = 2Z_{\parallel\perp}^{(1)} = -Z_\Pi^{(1),\parallel\perp}$.

While the sum over all diagrams is gauge independent, individual components depend on $\xi_G$. In particular, the Wilson-line self-energy from the two-emission diagrams, $Z_h=C_A\,[2+(1-\xi_G)]$, is identical in both channels. In Landau gauge ($\xi_G=0$), this evaluates to $3C_A$, matching the adjoint ($C_F\to C_A$) image of the fundamental Wilson-line self-energy $C_F\,[2+(1-\xi_G)]$ quoted in ref.~\cite{Braun:2020ymy} ($Z_h^{(1)}=3C_F=4$ for $N_c=3$). In Feynman gauge ($\xi_G=1$), our expression evaluates to $2C_A$.

To compare the overall renormalization constants with ref.~\cite{Braun:2020ymy}, we account for the explicit coupling factor in their operator definition, \cref{eq:Braundef}, which renormalizes as $\mathbb{G}^{\text{bare}} = (Z^{\text{B}})^{-2}\, \mathbb{G}$. Our operator, $\mathcal{O}^{\text{bare}}(z) = G^{\text{bare}}(z)\,W^{\text{bare}}(z,0)\,G^{\text{bare}}(0)$, omits this coupling prefactor. The two bare operators are related by the bare coupling $g_{s,0}^2 = Z_a g_s^2 \mu^{2\epsilon}$ with $Z_a^{(1)} = -\beta_0$, giving $\mathbb{G}^{\text{bare}} = Z_a g_s^2 \mu^{2\epsilon} \mathcal{O}^{\text{bare}}$, while the renormalized operators satisfy $\mathbb{G} = g_s^2\mu^{2\epsilon}\,\mathcal{O}$. The two sets of renormalization constants therefore differ by the coupling counterterm,
\begin{equation}
  Z^2 = Z_a\,(Z^{\text{B}})^2\,,\qquad Z^{(1)} = Z^{\text{B}(1)} - \frac{\beta_0}{2}\,.\label{eq:vac_Zcombination}
\end{equation}
Removing the coupling renormalization from \cref{eq:sector_values} gives
\begin{equation}\label{eq:vac_reconstruction}
  Z_{\perp\perp}^{\text{B}(1)}=\frac{\beta_0-Z_\Pi^{(1),\perp\perp}}{2}=C_A=3\,,\qquad
  Z_{\parallel\perp}^{\text{B}(1)}=\frac{\beta_0-Z_\Pi^{(1),\parallel\perp}}{2}=0\,,
\end{equation}
in exact agreement with the one-loop values of ref.~\cite{Braun:2020ymy}. Their constants are manifestly $n_f$-independent, because the quark-loop contributions from the gluon vacuum polarization are absorbed by the coupling renormalization $\beta_0$.

With the vacuum eigenvalues established, the mixing matrix $\widetilde{Z}_\mathcal{M}$ provides a complete classification of all multiplicatively renormalizable combinations of the gluon bilocal operator quoted in the literature. We have re-derived each quoted combination by explicit component evaluation of the tensor decomposition in \cref{eq:Manb} and verified that its coefficient vector is a left eigenvector of $\widetilde{Z}_\mathcal{M}$. In the following, transverse indices $i,j\in\{1,2\}$ are summed over, while $t$ and $z$ denote the temporal and longitudinal directions, and $\mu$ runs over all four spacetime directions. As discussed in \cref{sec:formfactors,sec:tree}, continuing the transverse sums to $d-2$ directions introduces at most $\mathcal{O}(\epsilon)$ admixtures of auxiliary form factors that remain within the same eigenvalue sector.

Each quoted combination falls into one of the three eigenvalue sectors:
First, the sector $Z_{\perp\perp}^2$ contains the lattice combination $\mathcal{G}^{0ii0}+\mathcal{G}^{jiij}$ of ref.~\cite{Balitsky:2019krf} (\cref{eq:Gcombination_Mpp}), the individual components $\mathcal{G}^{0ii0}$ and $\mathcal{G}^{jiij}$, and the operator $O_1=F^{ti}\,W\,F_{i}{}^{t}$ of refs.~\cite{Zhang:2018diq,Wang:2019tgg}.
Second, the mixed sector $Z_{\perp\perp}Z_{\parallel\perp}$ contains the combinations $\mathcal{G}^{0ii3}\pm\mathcal{G}^{3ii0}$ of ref.~\cite{Balitsky:2019krf} and the operator $O_3=F^{ti}\,W\,F_{i}{}^{z}$ of refs.~\cite{Zhang:2018diq,Wang:2019tgg}.
Finally, the sector $Z_{\parallel\perp}^2$ contains the minimal-contamination combination $\mathcal{G}^{3ii3}+2\,\mathcal{G}^{3003}$ of ref.~\cite{Balitsky:2021qsr} (\cref{eq:Gcombination_ziiz}), the individual components $\mathcal{G}^{3ii3}$ (our $\mathcal{G}^{ziiz}$) and $\mathcal{G}^{3003}$, the covariant contraction $g_{\alpha\beta}\,\mathcal{G}^{3\alpha3\beta}$~\cite{Balitsky:2019krf,Balitsky:2021qsr}, the operators $O_2=F^{zi}\,W\,F_{i}{}^{z}$ and $O_4=F^{z\mu}\,W\,F_{\mu}{}^{z}$ of refs.~\cite{Zhang:2018diq,Wang:2019tgg}, and the operator $\sum_{i}G^{z}{}_{i}(z)\,W\,G^{iz}(0)$ of ref.~\cite{Wang:2017qyg} (which coincides with $O_2$).

Conversely, the operator $O_5=F^{t\mu}\,W\,F_{\mu}{}^{t}$ of refs.~\cite{Zhang:2018diq,Wang:2019tgg}, used for the gluon quasi-\PDF{} in ref.~\cite{Fan:2018dxu}, is \emph{not} a left eigenvector of $\widetilde{Z}_\mathcal{M}$, confirming the observation of ref.~\cite{Zhang:2018diq} that this operator does not renormalize multiplicatively.

	\newpage
	
\providecommand{\href}[2]{#2}\begingroup\raggedright\endgroup
	
\end{document}